\documentclass[aps,prd,amsmath,amssymb,amsfonts,showpacs,superscriptaddress,floatfix,twocolumn, longbibliography]{revtex4-1}
\usepackage{amsmath}
\usepackage{amssymb}
\usepackage{graphicx}
\usepackage{caption}
\usepackage{subcaption}
\usepackage{epsfig}
\usepackage{braket}
\usepackage{slashed}
\usepackage{bm}
\usepackage[usenames,dvipsnames]{color}
\usepackage{color}
\usepackage{float}
\usepackage{hyperref}
\usepackage{float}

\newcommand{\idm}{\openone}

\newcommand{\physrep}{Phys. Rep.}

\def\be{\begin{equation}}
\def\ee{\end{equation}}

\def\bea{\begin{eqnarray}}
\def\eea{\end{eqnarray}}
\numberwithin{equation}{section}
\renewcommand\theequation{\arabic{section}.\arabic{equation}}
\begin{document}

\title{Confinement and string breaking for QED$_2$ in the Hamiltonian picture} 

\author{Boye Buyens}
\affiliation{Department of Physics and Astronomy,
Ghent
 University,
  Krijgslaan 281, S9, 9000 Gent, Belgium}

\author{Jutho Haegeman}
\affiliation{Department of Physics and Astronomy,
Ghent
 University,
  Krijgslaan 281, S9, 9000 Gent, Belgium}
  
  \author{Henri Verschelde}
\affiliation{Department of Physics and Astronomy,
Ghent
 University,
  Krijgslaan 281, S9, 9000 Gent, Belgium}

\author{Frank Verstraete}
  \affiliation{Department of Physics and Astronomy,
Ghent
 University,
  Krijgslaan 281, S9, 9000 Gent, Belgium}
  \affiliation{Vienna Center for Quantum Science and Technology, Faculty of Physics, University of Vienna, Boltzmanngasse 5, 1090 Vienna, Austria}

 \author{Karel Van Acoleyen}
\affiliation{Department of Physics and Astronomy,
Ghent
 University,
  Krijgslaan 281, S9, 9000 Gent, Belgium}

\begin{abstract}
The formalism of matrix product states is used to perform a numerical study of (1+1)-dimensional QED -- also known as the (massive) Schwinger model -- in the presence of an external static `quark' and `antiquark'. 
We obtain a detailed picture of the transition from the confining state at short interquark distances to the broken-string `hadronized' state at large distances and this for a wide range of couplings, recovering the predicted behavior in both the weak- and strong-coupling limit of the continuum theory. In addition to the relevant local observables like charge and electric field, we compute the (bipartite) entanglement entropy and show that subtraction of its vacuum value results in a UV-finite quantity. We find that both string formation and string breaking leave a clear imprint on the resulting entropy profile. Finally, we also study the case of fractional probe charges, simulating for the first time the phenomenon of partial string breaking.  

\end{abstract}

\maketitle
\section{Introduction}

\noindent
The confinement of color charge in quantum chromodynamics (QCD)  is one of the beautiful key mechanisms of the Standard Model. Focusing on the static aspect of confinement, one can probe the theory with a heavy quark-antiquark ($q\bar{q}$) pair and examine how the modified ground state evolves as a function of the interquark distance \cite{Bali}. For small distances a color electric flux tube forms between the pair, resulting in a static potential (i.e. the surplus energy of the modified ground state) that grows linearly with the distance. This flux tube can therefore be conveniently modeled by an interquark string with a certain string tension. One can then describe a heavy quarkonium state as a $q\bar{q}$ pair that is kept together by this confining string. However, there exists a critical distance at which the string breaks. Beyond this distance the flux tube disappears and the potential flattens out to a constant. At this point it has become energetically favorable to excite light particles out of the vacuum that completely screen both the probe quark and antiquark, leading to two isolated color singlets. In a dynamical setting these would then be the two freely propagating jets of hadrons that emerge as the final product of some particle collision.

This phenomenological picture is corroborated both by experiment and theoretical work. At the computational level, the static potential has been studied extensively over the years with lattice QCD. The linearly rising confining interquark potential has been obtained, both in the quenched case \cite{Bali1992,Lang:1982tj,Stack:1983wb,Griffiths:1983ah,Otto:1984qr,Hasenfratz:1984gc,Barkai:1984ca,{Sommer:1985du,Huntley:1986ts,Hoek:1987uy,Ford:1988ki}} that excludes dynamical light quark degrees of freedom and in the unquenched case \cite{Born:1994cq,Heller:1994rz,Glassner:1996xi,Aoki:1998sb,Bali:2000vr,Bernard:2000gd} that includes these degrees of freedom. In the latter case, where the dynamical quarks can screen the heavy probe charges, the phenomenon of string breaking has also been observed \cite{Bali2} as an asymptotic flattening of the calculated potential. Nevertheless our understanding of confinement is incomplete: the Euclidean space-time lattice Monte Carlo simulations cannot access the real-time aspects of the dynamical string formation and string breaking. Furthermore, even in the static case, it is not settled yet if one can fully describe the confinement mechanism - specifically the nonperturbative string formation - in terms of (semi-) local degrees of freedom (e.g., center vortices, magnetic monopoles) \cite{Alkofer:2010ue,Alkofer:2006fu,Greensite}.

In this paper, we study how confinement and string breaking show up in the Hamiltonian setup, as opposed to the Euclidean path integral setup of lattice Monte Carlo. We do this for the simplest nontrivial quantum gauge field theory: (1+1)-dimensional quantum electrodynamics (QED$_2$), also known as the Schwinger model \cite{Schwinger1962}. The Schwinger model has a long-standing tradition as toy model for QCD, sharing its confining and chiral symmetry-breaking properties. (We will therefore often refer to `quark' and `antiquark' both for the external
probe charges and for the light dynamical fermions.) But notice that in the future the significance of QED$_2$ could go well beyond being a toy model, as QED$_2$ or QED$_2$-like theories might be realized effectively by quantum simulators \cite{Cirac2010,Zohar2011,Zohar2012,Banerjee2012,Banerjee2013,Hauke2013,Stannigel2013,
Tagliacozzo2013a,Wiese2013,Zohar2013,Zohar2013a,Zohar2013b,Zohar2013c,Kosior2014,Kuehn2014,
Marcos2014,Wiese2014,Mezzacapo2015,Notarnicola2015,Zohar2015a}. 

An important difference with QCD is that QED$_2$ already exhibits confinement at the perturbative level, as the Coulomb potential is linear in (1+1) dimensions. Furthermore, the theory can also be solved in a strong-coupling expansion, via bosonization \cite{Coleman}. We make extensive use of both the strong- and weak-coupling results in the analysis of our numerical results. Our simulations of the lattice Hamiltonian are performed close to the continuum limit, indeed allowing for a quantitative check against these analytic continuum results in the appropriate regimes.

Specifically, we simulate the modified vacuum structure in the presence of two probe charges and this for different distances and values of the charges. As we will show, already in this static case the Hamiltonian simulations give a complementary view on the confining properties of the theory. At the practical level, the direct access to the quantum state allows for a relatively easy calculation of all local observables. In this way we could not only extract the static interquark potential, but also for instance determine the detailed spatial profile of the electric string or the precise charge distribution of the light fermions around the probe charges. At a more fundamental level, our tensor network state simulations (see below) allow for a direct calculation of the entanglement entropy between different regions. In the past decade it has become clear that entanglement entropy is a very useful quantity for the characterization of quantum many-body systems and quantum field theories \cite{introreview}, in particular also for the investigation of the confining properties of gauge theories \cite{Klebanov:2007ws}. In this context the entanglement entropy is typically calculated either from the dual geometry in the AdS/CFT approach \cite{Klebanov:2007ws,Kol:2014nqa} through the Ryu-Takayanagi conjecture \cite{Ryu:2006bv}, or from lattice Monte Carlo simulations \cite{Buividovich:2008kq,Buividovich:2008gq,Nakagawa:2009jk} through the replica trick, allowing for calculation of the discrete Renyi entropies. In contrast, tensor network state simulations give access to the full Schmidt spectrum of the state.  The Schmidt spectrum $\{\lambda_\alpha\}$ follows from the Schmidt decomposition: if $\ket{\Psi} \in \mathcal{H}_A \otimes \mathcal{H}_B$ is a state belonging to the tensor product of the two Hilbert spaces $\mathcal{H}_A$ and $\mathcal{H}_B$, then one can write 
\be \label{eq:SchmidtGeneral} \ket{\Psi} = \sum_{\alpha = 1}^{d} \sqrt{\lambda_\alpha} \Ket{\Psi_\alpha^{(A)}} \otimes \Ket{\Psi_\alpha^{(B)}}, \ee
with $d \leq \max(\mbox{dim}\bigl(\mathcal{H}_A), \mbox{dim}(\mathcal{H}_B)\bigl)$, $\Ket{\Psi_\alpha^{(A)}} \in \mathcal{H}_A$ and $\Ket{\Psi_\alpha^{(B)}} \in \mathcal{H}_B$ orthonormal unit vectors and $\lambda_\alpha$, called the Schmidt values, non-negative numbers that sum to one. From the Schmidt values one can calculate all Renyi entropies, including the von Neumann entropy. In our simulations we find that subtraction of the vacuum entropy results in a UV-finite entanglement (von Neumann) entropy and that both the string formation and string breaking leave characteristic imprints on this renormalized entropy. 

As was mentioned in the previous paragraph, we use the general formalism of tensor network states (TNS) \cite{TNS1, TNS2} for our simulations. Although the TNS formalism has been mainly developed in the context of condensed matter physics, it is actually a universal method in the same way that the Feynman diagrammatic approach has a universal character. The latter applies whenever the interactions are weak, whereas the TNS method applies whenever the interactions are local. It is in fact precisely the entanglement structure of low-energy states for local systems, captured by the so-called area law \cite{area}, which lies at the root of the TNS description. 

For one spatial dimension, the most widely used TNS go by the name of Matrix product states (MPS) \cite{Fannes,Schollwoeck2011}. Recently, different applications of MPS on (1+1)-dimensional gauge theories, have demonstrated its potential in the context of gauge theories. In \cite{Buyens, Montangero, BanulsSB} the MPS formalism was used for the numerical simulation of nonequilibrium physics, but static properties \cite{Byrnes, BanulsSB,Banuls,BanulsProc,Sugihara,Rico,Buyens,Kuehn2014,Silvi,Buyens2, Milsted2015} and finite temperature properties \cite{Saito2014,Saito} have also been studied. Notably \cite{BanulsSB} simulated string breaking for probe charges in a $SU(2)$ quantum link lattice model. In higher dimensions, the TNS formalism is at present less developed, nevertheless some first promising results have appeared for (2+1)-dimensional gauge theories \cite{Luca,Luca2,Milsted2014,gaugingStates,2015arXiv150708837Z}.

In the next section we discuss the starting point of our simulations, introducing both the relevant lattice Hamiltonian in the presence of probe charges and the appropriate form of MPS that is dictated by gauge invariance. We then first consider the asymptotic case of two (fractional) probe charges at infinity in section \ref{section:String tension}. In section \ref{sec:potential} we consider finite interquark distances and study how the ground state evolves as a function of this distance. We distinguish three different cases. First we consider the strong-coupling limit. This is a special case, since in this limit the interquark string never forms and all probe charges, fractional or integer, are screened asymptotically. We then go away from the strong-coupling limit, considering first the case of unit probe charges. In this case we clearly observe the transition from a string state at short interquark distances to a broken-string two meson state at large distances. Then, in addition to unit probe charges we also consider fractional probe charges, simulating for the first time the phenomenon of partial string breaking, with probe charges that get only partially screened. Finally in section \ref{conclusion} we present our conclusions. Technical details on our MPS simulations and on some perturbative weak-coupling calculations can be found in the appendices.

\section{Setup}\label{Hamiltonian}

\subsection{Hamiltonian and gauge symmetry}
\noindent The Schwinger model is (1+1)-dimensional QED with one fermion flavor. We start from the Lagrangian density in the continuum:
\be \mathcal{L} = \bar{\psi}\left(\gamma^\mu(i\partial_\mu+g A_\mu) - m\right) \psi - \frac{1}{4}F_{\mu\nu}F^{\mu\nu}\,.\label{Lagrangian} \ee
One then performs a Hamiltonian quantization in the timelike axial gauge ($A_0=0$), which can be turned into a lattice system by the Kogut-Susskind spatial discretization \cite{Kogut}. The two-component fermions are sited on a staggered lattice. These fermionic degrees of freedom can be converted to spin-1/2 degrees of freedom by a Jordan-Wigner transformation with the eigenvectors $\{\ket{s_n}_n: s_n \in \{-1,1\} \}$ of $\sigma_z(n)$ as basis of the local Hilbert space at site $n$. The compact gauge fields $\theta(n)=a g A_1(n)$, live on the links between the sites. Their conjugate momenta $E(n)$, with $[\theta(n),E(n')]=ig\delta_{n,n'}$ correspond to the electric field. The commutation relation determines the spectrum of $E(n)$ up to a constant: $E(n)/g = \alpha(n)+p$ , with $\alpha(n)\in \mathbb{R}$ corresponding to the background electric field at link $n$ and $p\in\mathbb{Z}$. 

In this formulation the gauged spin Hamiltonian derived from the Lagrangian density (\ref{Lagrangian}) reads (see \cite{Banks,Kogut} for more details):
\bea\label{equationH0} H&=& \frac{g}{2\sqrt{x}}\Biggl(\sum_{n \in \mathbb{Z}} \frac{1}{g^2} E(n)^2 + \frac{\sqrt{x}}{g} m \sum_{n \in \mathbb{Z}}(-1)^n\sigma_z(n) \nonumber
\\ &&+ x \sum_{n \in \mathbb{Z}}(\sigma^+ (n)e^{i\theta(n)}\sigma^-(n + 1) + h.c.)\biggl) \eea
where $\sigma^{\pm} = (1/2)(\sigma_x \pm i \sigma_y)$ are the ladder operators. Here we have introduced the parameter $x$ as the inverse lattice spacing in units of $g$: $x \equiv 1/(g^2a^2)$. The continuum limit will then correspond to $x\rightarrow \infty$.  
Note the different second (mass) term in the Hamiltonian for even and odd sites which originates from the staggered formulation of the fermions. In this formulation the odd sites are reserved for the charge $-g$ `quarks', where spin-up, $s = +1$, corresponds to an unoccupied site and spin-down, $s = -1$, corresponds to an occupied site. The even sites are reserved for the charge $+g$ `antiquarks' where now conversely spin-up corresponds to an occupied site and spin-down to an occupied site. 

In the time-like axial gauge the Hamiltonian is still invariant under the residual time-independent local gauge transformations generated by:
\begin{align} gG(n) = & E(n)-E(n-1)-\frac{g}{2}( \sigma_z(n) + (-1)^n )\,.\label{gauss} \end{align}
As a consequence, if we restrict ourselves to physical gauge-invariant operators $O$, with $[O,G(n)]=0$, the Hilbert space decomposes into dynamically disconnected superselection sectors, corresponding to the different eigenvalues of $G(n)$. In the absence of any background charge, the physical sector then corresponds to the $G(n)=0$ sector. Imposing this condition (for every $n$) on the physical states is also referred to as the Gauss law constraint, as this is indeed the discretized version of $\partial_z E - \rho=0$, where $\rho$ is the charge density of the dynamical fermions. 

The other superselection sectors correspond to states with background charges. Specifically, if we want to consider two probe charges, one with charge $-gQ$ at site $0$ and one with opposite charge $+gQ$ at site $k$, we have to restrict ourselves to the sector:
\begin{align} gG(n) =  gQ(\delta_{n,0} - \delta_{n,k})\label{spingauss0}\,.\end{align} Notice that we will consider both integer and noninteger (fractional) charges $Q$.  

As in the continuum case \cite{ColemanCS}, we can absorb the probe charges into a background electric field string that connects the two sites. This amounts to the substitution $E(n) = g[L(n) + \alpha(n)]$ where $\alpha(n)$ is only nonzero in between the sites: $\alpha(n) = -Q\Theta(0 \leq n < k)$; and $L(n)$ has an integer spectrum: $L(n)=p\in \mathbb{Z}$. In terms of $L(n)$ the Gauss constraint now reads:
 \be\label{spingauss} G(n) = L(n) - L(n-1) -\frac{ \sigma_z(n) + (-1)^n }{2} = 0\,,  \ee and we finally find the Hamiltonian \footnote{One could have started from two flavor QED, with integer charges $ag$ for the light fermions and $bg$ for the heavy fermions ($a,b\in \mathbb{N}$); and a discrete integer spectrum $L(n)=p\in \mathbb{Z}$ for the original electric field. If we impose $L(n)=0$ at the boundaries we get an effective spectrum from Gauss's law: $L(n)=ap+bq$ ($p,q \in \mathbb{Z}$). If we then further limit ourselves to states with only one heavy fermion located at site $0$ and one heavy anti-fermion located at site $k$, the resulting effective spectrum will read $L(n)=ap-b$ for the sites in between the heavy fermion pair and $L(n)=ap$ for the other sites. Upon the redefinitions $g\rightarrow g/a$ and $L(n)\rightarrow L(n)/a-b/a$ we then indeed recover the effective Hamiltonian (\ref{equationH}) for fractional background charge $Q=b/a$. }:
 \bea\label{equationH} H&=& \frac{g}{2\sqrt{x}}\Biggl(\sum_{n \in \mathbb{Z}} [L(n) + \alpha(n)]^2 + \frac{\sqrt{x}}{g} m \sum_{n \in \mathbb{Z}}(-1)^n\sigma_z(n) \nonumber
\\ &&+ x \sum_{n \in \mathbb{Z}}(\sigma^+ (n)e^{i\theta(n)}\sigma^-(n + 1) + h.c.)\biggl),\eea
in accordance with the continuum result of \cite{ColemanCS}. 

In the following sections we will obtain ground-state approximations of this Hamiltonian, for different values of $m/g$, different values of the probe charge $Q$ and different distances $L=k/\sqrt{x}$ (in physical units $g=1$) of the charge pair, all this for different lattice spacings $1/\sqrt{x}$, focusing on the continuum limit $x\rightarrow \infty$. 

An important point regarding the continuum limit is that the ground-state energy of the Schwinger model is UV divergent but that this UV divergence does not depend on the background field $\alpha(n)$. If we write $\mathcal{E}_0=2N\epsilon_0$ (with $N=|\mathbb{Z}|$) for the ground-state energy of (\ref{equationH}) with zero background field $\alpha(n)=0$, we have $\sqrt{x}\epsilon_0\rightarrow -x/\pi$ for the energy density in the $x\rightarrow \infty$ limit \cite{Hamer}. For the modified ground-state energy in the presence of the probe charge $gQ$ pair at distance $L$  we can then write $\mathcal{E}_Q(L)=V_Q(L)+\mathcal{E}_0$, where the potential $V_Q(L)$ is now UV-finite. Notice that $V_Q(L)$ will also be IR ($N\rightarrow \infty$) finite (for finite $L$). 

\subsection{Gauge-invariant MPS}\label{subsec:GIMPS}
\noindent Consider now the lattice spin-gauge system (\ref{equationH}) on $2N$ sites. On site $n$ the matter fields are represented by the spin operators with basis $\{ \ket{s_n}_n: s_n \in \{-1,1\}\}$. The gauge fields live on the links, and on link $n$ their Hilbert space is spanned by the eigenkets $\{\ket{p_n}_{n}: p_n \in \mathbb{Z}\}$ of the angular operator $L(n)$. But notice that for our numerical scheme, we retain only a finite range: $p_{min}(n+1) \leq p_n \leq p_{max}(n+1)$. We will address the issue of which values to take for $p_{min}(n+1)$ and $p_{max}(n+1)$ later in this subsection. Furthermore, it will be convenient to block site $n$ and link $n$ into one effective site with local Hilbert space spanned by $\{\ket{s_n,p_n}_n \}$. 
Writing $ \kappa_n = (s_n,p_n)$ we introduce the multi-index 
$$\bm{\kappa} = \bigl((s_1,p_1),(s_2,p_2),\ldots,(s_{2N},p_{2N})\bigl) = (\kappa_1,\ldots,\kappa_{2N}).$$ 
With these notations we have that the effective site $n$ is spanned by $\{\ket{\kappa_n}_n \}$. Therefore the Hilbert space of the full system of $2N$ sites and $2N$ links, which is the tensor product of the local Hilbert spaces, has basis $\{\ket{\bm{\kappa}}= \ket{\kappa_1}_1\ldots \ket{\kappa_{2N}}_{2N} \}$ and a general state $\ket{\Psi}$ is thus a linear combination of these $\ket{\bm{\kappa}}$: 
$$\ket{\Psi} = \sum_{\bm{\kappa}} C_{\kappa_1,\ldots,\kappa_{2N}}\ket{\bm{\kappa}}$$
 with basis coefficients $C_{\kappa_1,\ldots,\kappa_{2N}} \in \mathbb{C}$.\\
\\ 
A general MPS $\ket{\Psi(A)}$ now assumes a specific form for the basis coefficients \cite{Fannes}:  
\be\ket{\Psi(A)} = \sum_{\bm{\kappa}}\bm{v}_L^\dagger A_{\kappa_1}(1) A_{\kappa_2}(2)\ldots A_{\kappa_{2N}}(2N)\bm{v}_R \ket{\bm{\kappa}},\label{MPS}\ee
where $A_{\kappa_n}(n)$ is a complex $D(n)\times D(n+1)$ matrix with components $[A_{\kappa_n}(n)]_{\alpha \beta}$ and where $\bm{v}_L \in \mathbb{C}^{D(1)\times 1}, \bm{v}_R \in \mathbb{C}^{D(2N+1)\times 1}$ are boundary vectors. The MPS ansatz thus associates with each site $n$ and every local basis state $\ket{\kappa_n}_n =\ket{s_n,p_n}_n$ a matrix $A_{\kappa_n}(n)= A_{s_n,p_n}(n)$. The indices $\alpha$ and $\beta$ are referred to as virtual indices, and $D = \max_n [D(n)]$ is called the bond dimension. 

To better understand the role of the bond dimension in MPS simulations it is useful to consider the Schmidt decomposition (\ref{eq:SchmidtGeneral}) with respect to the bipartition of the lattice consisting of the two regions $\mathcal{A}_1(n) = \mathbb{Z}[1, \ldots, n]$ and $\mathcal{A}_2(n) = \mathbb{Z}[n+1,\ldots, 2N]$ \cite{Schollwoeck2011}:
\be \label{eq:MPSschmidt} \ket{\Psi(A)} = \sum_{\alpha = 1}^{D(n+1)} \sqrt{\lambda_\alpha(n)} \ket{\psi_\alpha^{\mathcal{A}_1(n)}}\ket{\psi_\alpha^{\mathcal{A}_2(n)}}. \ee
Here $\ket{\Psi_\alpha^{\mathcal{A}_1(n)}}$ (resp. $\ket{\Psi_\alpha^{\mathcal{A}_2(n)}}$) are orthonormal unit vectors living in the tensor product of the local Hilbert spaces belonging to the region $\mathcal{A}_1(n)$ (resp. $\mathcal{A}_2(n)$) and $\lambda_\alpha(n)$, called the Schmidt values, are non-negative numbers that sum to one. One can easily deduce that for a general MPS of the form (\ref{MPS}) at most $D(n+1)$ Schmidt values will be nonzero (for the cut at site $n$ (\ref{eq:MPSschmidt})). We refer to appendix \ref{appasymptotic} and \ref{appDMRG} for the computation of the Schmidt values for the specific case of our simulations and to \cite{Schollwoeck2011,HaegemanBMPS} for the general case. We thus see that taking a finite bond dimension for the MPS corresponds to a truncation in the Schmidt spectrum of a state. The success of MPS is then explained by the fact that ground states of local gapped Hamiltonians can indeed be approximated very efficiently in $D$ \cite{Hastings2007} and that the computation time for expectation values of local observables scales only with $D^3$, allowing for reliable simulations on an ordinary desktop.

Another advantage of MPS simulations is that one can work directly in the thermodynamic limit $N \rightarrow \infty$ \cite{HaegemanTDVP,HaegemanBMPS}, bypassing any possible finite-size artifacts. In the following, we work in this limit. In section \ref{section:String tension}, where the Hamiltonian is invariant under translations (over two sites), the tensors $A_{\kappa_n}(n)$ depend only on the parity of the site $n$, see eq. (\ref{uMPS}). While in section \ref{sec:potential} the MPS ansatz is not translational invariant in the bulk, see eq. (\ref{nonuMPS}). In that case the tensors will be fixed asymptotically ($n \gg 1$) to their ground-state value, anticipating that we approach the translational invariant ground state of the zero-background Hamiltonian. In both cases the MPS ansatz depends on a finite number of parameters. Finally, we note that, in the thermodynamic limit, the expectation values of local observables are independent of the boundary vectors $\bm{v}_L$ and $\bm{v}_R$. \\
\\
As explained in \cite{Buyens}, to parametrize gauge-invariant MPS, i.e. states that obey $G(n)\ket{\Psi(A)}=0$ for every $n$, it is convenient to give the virtual indices a multiple index structure $\alpha\rightarrow (q,\alpha_q); \beta \rightarrow (r,\beta_r)$, where $q$ resp. $r$ labels the eigenvalues of $L(n-1)$ resp. $L(n)$. One can verify that the condition $G(n)=0$ (\ref{spingauss}) then imposes the following form on the matrices:
\be
{[A_{s,p}(n)]}_{(q,\alpha_q),(r,\beta_r)} =  {[a_{s,p}(n)]}_{\alpha_q,\beta_r}\delta_{q+(s+(-1)^n)/2,r}\delta_{r,p}
\label{gaugeMPS},\ee
where $\alpha_q = 1\ldots D_q(n)$, $\beta_r = 1 \ldots D_r(n+1)$. The first Kronecker delta is Gauss' law (\ref{spingauss}) on the virtual level while the second Kronecker delta connects the virtual index $r$ with the physical eigenvalue $p$ of $L(n)$. Because the indices $q$ (resp. $r$) label the eigenvalues of $L(n-1)$ (resp. $L(n)$) and we only retain the eigenvalues of $L(n-1)$ in the interval $\mathbb{Z}[p_{min}(n),p_{max}(n)]$ (resp. of $L(n)$ in the interval $\mathbb{Z}[p_{min}(n+1),p_{max}(n+1)]$) we have that $D_q(n) = 0$ for $q > p_{max}(n)$ and $q < p_{min}(n)$. The formal total bond dimension of this MPS is $D(n) = \sum_{q = p_{min}(n)}^{p_{max}(n)} D_q(n)$, but notice that, as (\ref{gaugeMPS}) takes a very specific form, the true variational freedom lies within the matrices $a_{s,p}(n) \in \mathbb{C}^{D_q(n) \times D_r(n+1)}$. 

Gauge invariance is, of course, also reflected in the Schmidt decomposition (\ref{eq:MPSschmidt}): for states of the form (\ref{gaugeMPS}) the Schmidt values can be labeled with the same double index $\alpha \rightarrow (q,\alpha_q)$. More specifically, the Schmidt decomposition (\ref{eq:MPSschmidt}) now reads (see Appendix A and C):
\be \label{eq:MPSschmidtGauge} \ket{\Psi(A)} = \hspace{-1.5mm}\sum_{q = p_{min}(n+1)}^{p_{max}(n+1)}\hspace{-1.5mm} \sum_{\alpha_q=1}^{D_q(n+1)} \sqrt{\lambda_{q,\alpha_q}(n)} \ket{\psi_{q,\alpha_q}^{\mathcal{A}_1(n)}}\ket{\psi_{q,\alpha_q}^{\mathcal{A}_2(n)}}. \ee

As before, we observe that taking a finite bond dimension $D_q(n+1)$ corresponds to a truncation in the Schmidt spectrum, now of the charge sector $q$. The choice for the different bond dimensions $D_q(n+1)$ in the different simulations should then be such that the discarded Schmidt values for each charge sector are {\em sufficiently} small. For our simulations with zero background, $\alpha(n)=0$, in \cite{Buyens} we could take $D_q=0$ for $|q|>3$, i.e. $p_{min} = -3$ and $p_{max} = +3$. For the simulations with a nonzero background field we find that for the same accuracy it suffices to consider eight $q$-sectors. But -- not surprisingly given the first term in the Hamiltonian (\ref{equationH}) -- we find the relevant eigenvalues sectors of $L(n)$ to be centered around a dominant sector $p_0$ that can be shifted away from $p_0=0$ for some sites $n$. The largest Schmidt value in each $q$-sector decreases as we move farther away from $q = p_0$. When $\vert q-  p_0\vert \gtrsim 4$ all the Schmidt values $\lambda_{q,\alpha_q}(n)$ are sufficiently small and we can safely take $D_q = 0$, i.e. $p_{max} \gtrsim p_0 + 4$ and $p_{min} \lesssim p_0 - 4$.  We refer to appendices \ref{appasymptotic} and \ref{appDMRG} for more details on the weight of the different sectors for the different simulations, see in particular figs. \ref{fig:STBDa} and  \ref{fig:STBDb} and figs. \ref{fig:maxChargec} and \ref{fig:maxCharged} for some explicit examples.  

From the Schmidt spectrum (\ref{eq:MPSschmidtGauge}) one can extract different measures for the entanglement. In the following we will always use the von Neumann entropy \footnote{Notice that for gauge theories the full von Neumann entropy (\ref{EntropyI}) is not equivalent to the LOCC distillable entanglement \cite{Ghosh:2015iwa,VanAcoleyen:2015ccp}. Nevertheless it is the full entropy that is supposed to be calculated by e.g. the AdS/CFT method. We leave the study of the distillable entanglement for the Schwinger model for future work.}. For the half-chain cut at site $n$, to which we associate the position $z=(n+1/2)a$ in physical units, we then have: \be S(z) = -\sum_{q}\sum_{\alpha_q}\lambda_{q,\alpha_q}(n)\log[\lambda_{q,\alpha_q}(n)]\,.\label{EntropyI}\ee

\section{Asymptotic string tension} \label{section:String tension} 

\noindent We first study the large distance behavior of the potential as captured by the asymptotic string tension $ \sigma_Q = \lim_{L \rightarrow + \infty}V_Q(L)/L$. This is the quantity that indicates whether the probe charges are asymptotically confined ($\sigma_Q \neq 0$) or not ($\sigma_Q=0$). For the Schwinger model $\sigma_Q$ has been computed analytically in the strong-coupling expansion \cite{Iso,ColemanCS, Adam, Armoni}. At the numerical front the most successful computation to date used finite-lattice scaling methods in a Hamiltonian formulation \cite{Hamer}. An advantage of our MPS simulations is that in contrast to \cite{Hamer} we can directly work in the thermodynamic limit ($N\rightarrow \infty$), leaving only the $x\rightarrow \infty$ interpolation to extract the continuum results. The challenge of taking this continuum limit now lies in the diverging correlation length $\xi/a$ (in lattice units), as MPS simulations require larger bond dimensions for growing correlation length \cite{TNS2}.   

To find the asymptotic string tension we put a probe charge $-gQ$ at $-\infty$ and a probe charge $gQ$ at $+\infty$.  As we explain in the previous section, a probe charge pair translates to a background electric field $\alpha(n)$ in the Hamiltonian (\ref{equationH}). In this case the background electric field will be uniform: $\alpha(n)=-Q, \forall n$. The Hamiltonian is then invariant under $T^2$, a translation over two sites. In accordance with this symmetry the appropriate MPS variational ground-state ansatz takes the form
\begin{multline} \Ket{\Psi\bigl(A(1),A(2)\bigl)} \\ = \sum_{\bm{\kappa}} \bm{v}_L^\dagger \left(\prod_{n \in \mathbb{Z}}A_{\kappa_{2n-1}}(1)A_{\kappa_{2n}}(2n)\right) \bm{v}_R \ket{\bm{\kappa}}, \label{uMPS}\end{multline}
where 
$$\kappa_n = (s_n,p_n) \in \{ -1,1\}\times\mathbb{Z}[p_{min}(n+1),p_{max}(n+1)],$$ $\ket{\bm{\kappa}} = \ket{\{ \kappa_{n}\}_{n \in \mathbb{Z}}}$, $\bm{v}_L \in \mathbb{C}^{D(1)\times 1}, \bm{v}_R \in \mathbb{C}^{D(1) \times 1}$, and $A_{\kappa}(n) \in \mathbb{C}^{D(n) \times D(n+1)}$ takes the form (\ref{gaugeMPS}) ($n = 1,2$). This corresponds to a general MPS (\ref{MPS}) in the thermodynamic limit ($N \rightarrow + \infty$) where the tensors $A_{\kappa_n}(n)$ depend only on the parity of the site $n$: $A_{\kappa_{2n-1}}(2n-1) = A_{\kappa_{2n-1}}(1)$ and $A_{\kappa_{2n}}(2n) = A_{\kappa_{2n}}(2), \forall n$. As a consequence $D_q(n),p_{min}(n)$ and $p_{max}(n)$ also depend on the parity of $n$. 

As we explain in appendix \ref{appasymptotic} we were able to accurately approximate the ground state and its finite energy per site $\epsilon_Q = \mathcal{E}_Q/2N$, with $\mathcal{E}_Q$ the total infrared divergent energy, within the class of states (\ref{uMPS}). Therefore, we perform imaginary time evolution ($d\tau = i dt$) of the Schr\"{o}dinger equation, $i\partial_t \Ket{\Psi\bigl(A(1),A(2)\bigl)}  = H\Ket{\Psi\bigl(A(1),A(2)\bigl)}$, with the time-dependent variational principle (TDVP) \cite{HaegemanTDVP,HaegemanBMPS}. In appendix \ref{appasymptotic} we also explain how we chose the virtual dimensions $\{D_q(1), D_q(2)\}$ and $\{p_{min/max}(1),p_{min/max}(2)\}$ by investigating the Schmidt spectrum. In \cite{Buyens}, we found the energy of the vacuum $\mathcal{E}_0 = 2N \epsilon_0$ for the zero-background field $\alpha(n)=0$. In the same fashion, we now compute the string tension $\sigma_Q$ as the extra energy density, induced by the uniform background electric field: $\sigma_Q = (\mathcal{E}_Q - \mathcal{E}_0)/L$ where $L$ is the length of our lattice. In units $g = 1$ we have $L = 2N / \sqrt{x}$, and therefore $\sigma_Q(x) = \sqrt{x}(\epsilon_Q - \epsilon_0)$.

From the numerical point of view it is important to take the convergence of this UV-finite quantity $\sigma_Q(x)$ as criterion for halting the imaginary TDVP time evolution.  As we explain in more detail in appendix \ref{appasymptotic} we computed values for $\sigma_Q(x)$  in this way, for $x = 100,200,300,400,600,800$ and perform a polynomial extrapolation in $1/\sqrt{x}$ similar to \cite{Hamer}. This indeed allows us to recover a finite value for $\lim_{x\rightarrow\infty}\sigma_Q(x)$, thereby explicitly verifying that the UV divergencies in the energy densities $\sqrt{x}\epsilon_Q$ and $\sqrt{x}\epsilon_0$ cancel out.

\begin{figure}
\begin{subfigure}[b]{.24\textwidth}
\includegraphics[width=\textwidth]{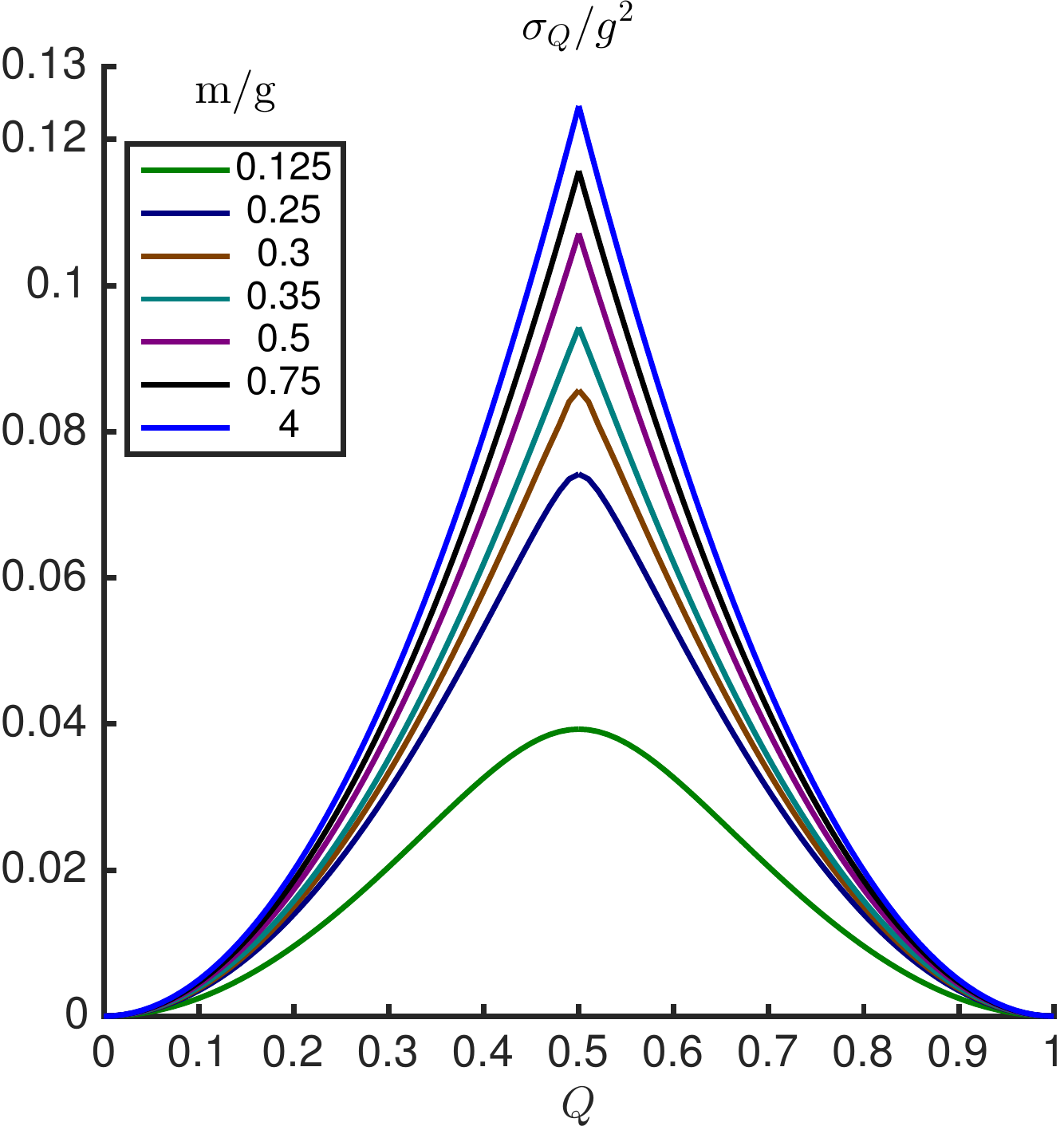}
\caption{\label{fig:StringTensiona}}
\end{subfigure}\hfill
\begin{subfigure}[b]{.24\textwidth}
\includegraphics[width=\textwidth]{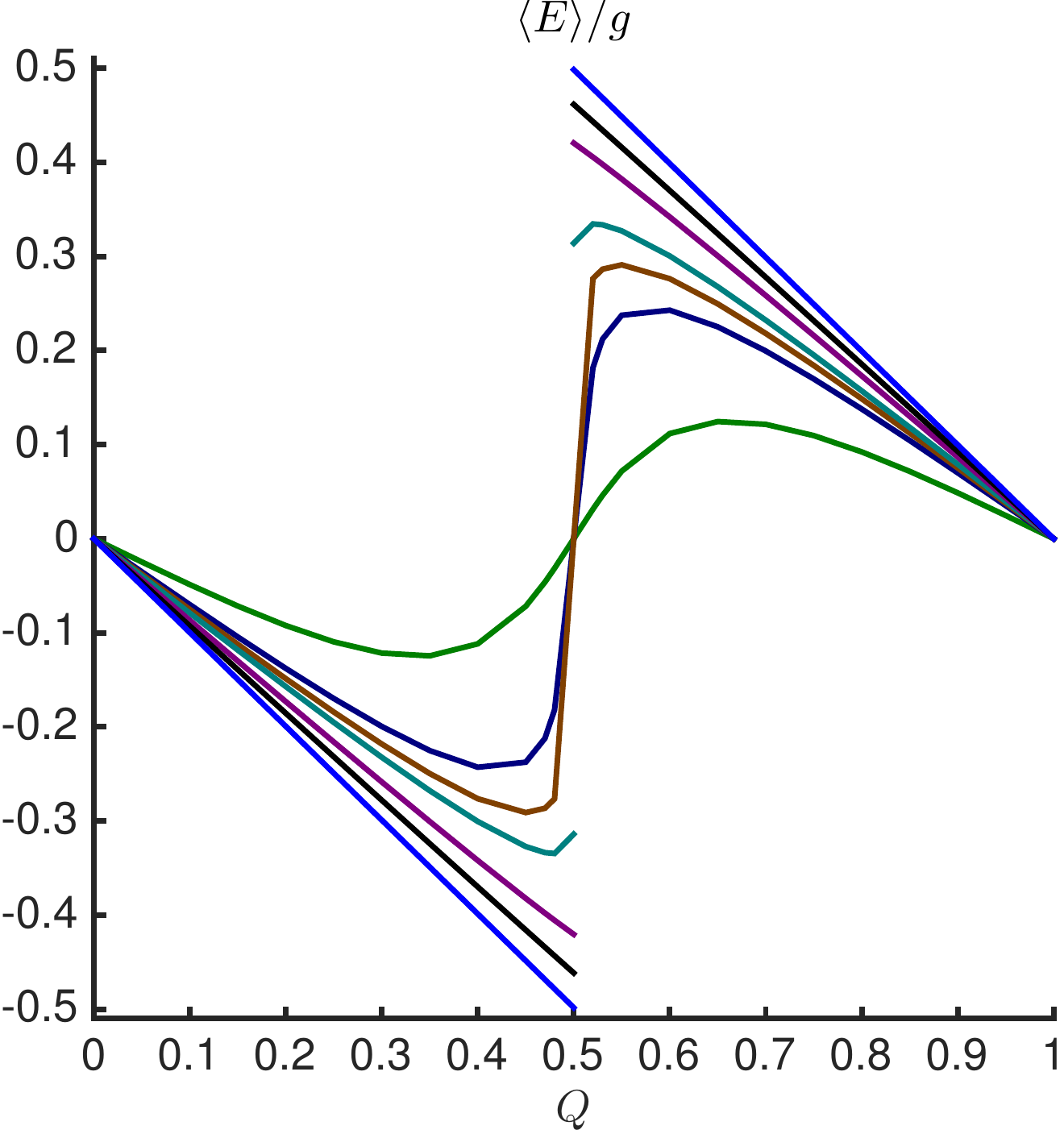}
\caption{\label{fig:StringTensionb}}
\end{subfigure}\hfill
\vskip\baselineskip
\begin{subfigure}[b]{.24\textwidth}
\includegraphics[width=\textwidth]{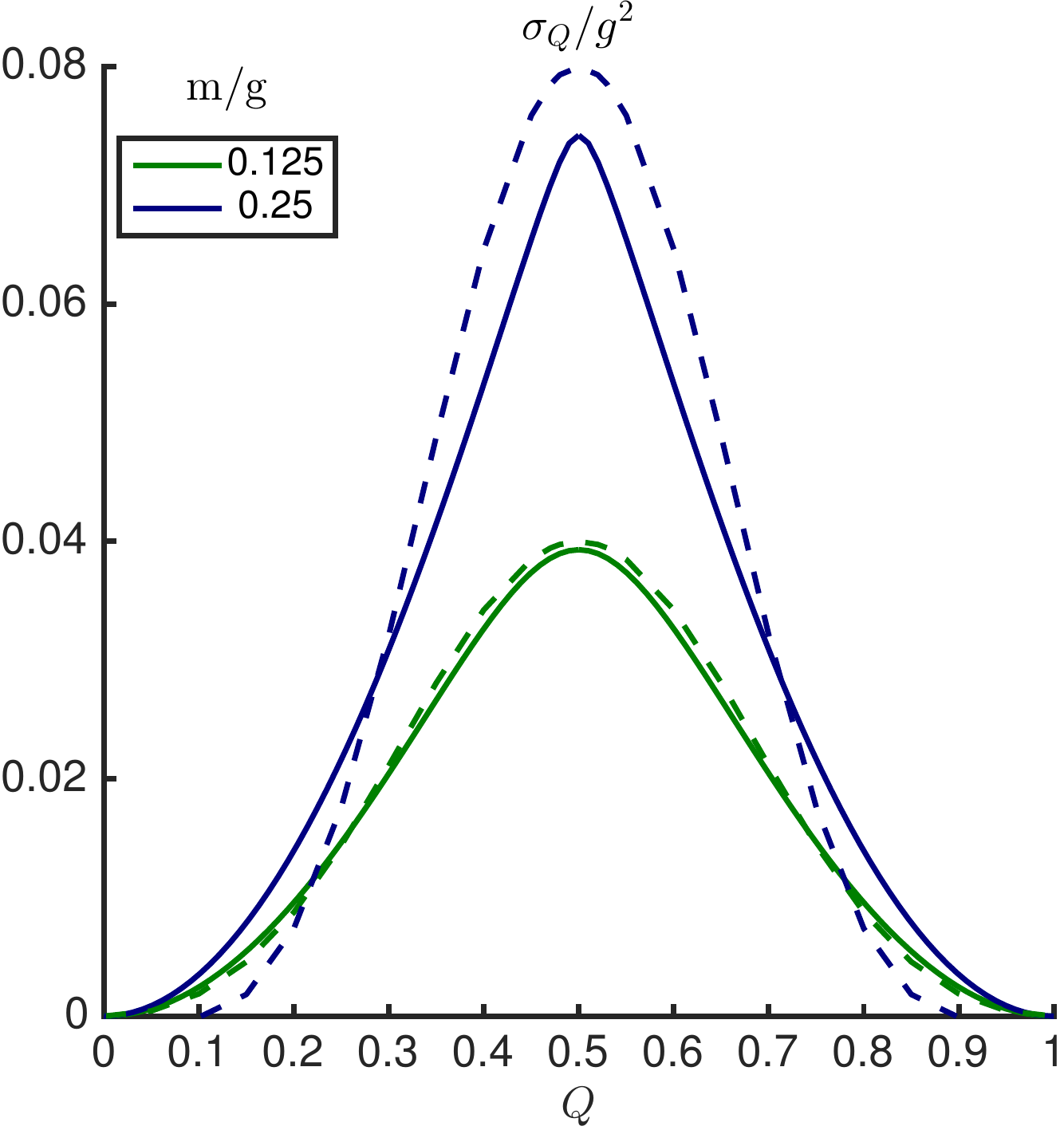}
\caption{\label{fig:StringTensionc}}
\end{subfigure}\hfill
\begin{subfigure}[b]{.24\textwidth}
\includegraphics[width=\textwidth]{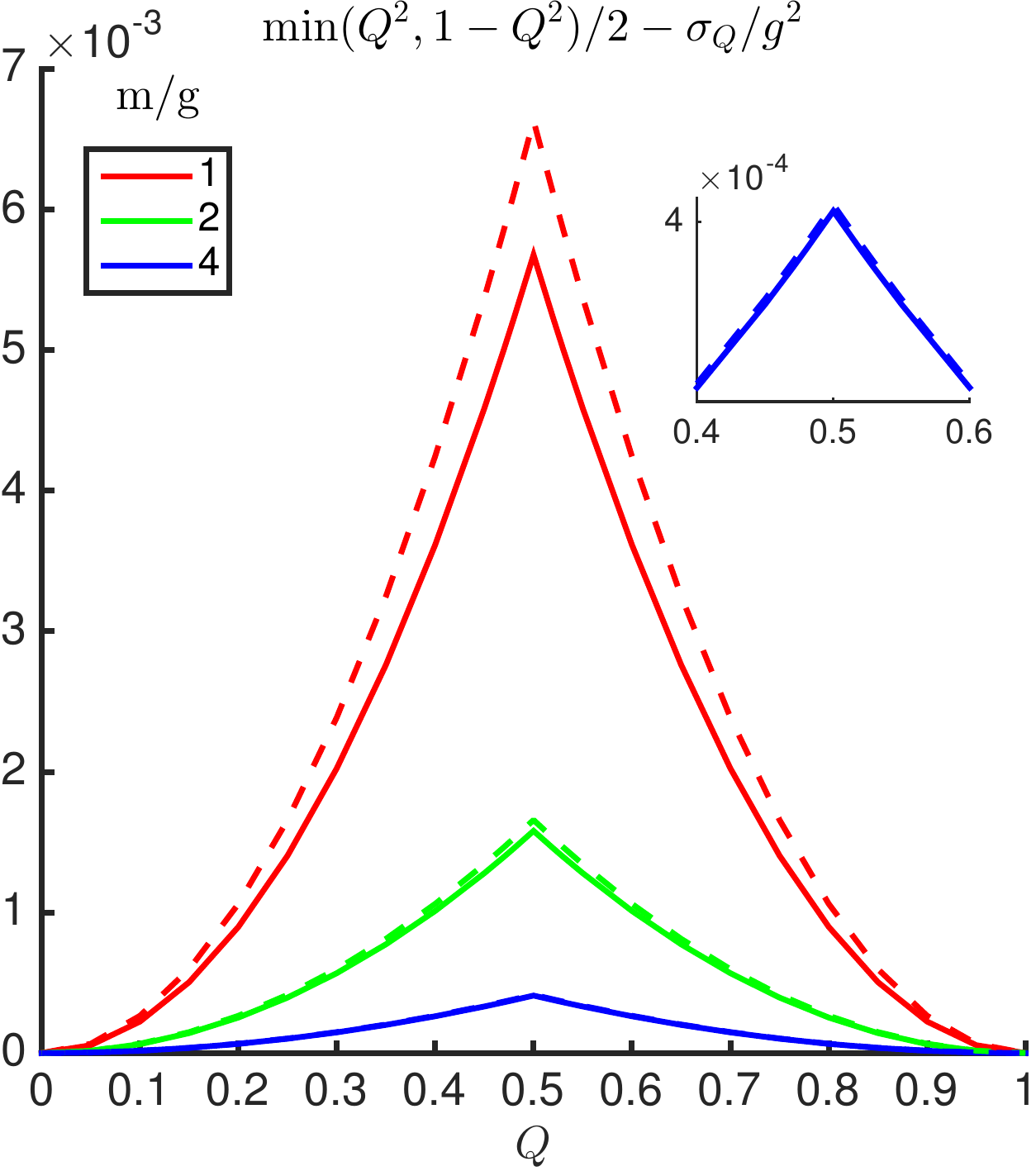}
\caption{\label{fig:StringTensiond}}
\end{subfigure}
\vskip\baselineskip
\captionsetup{justification=raggedright}
\caption{\label{fig:StringTension}(a): string tension $\sigma_Q$. (b): electric field per site. (c): comparison with the strong-coupling result (\ref{tensionsc}) (dashed line) for $m/g = 0.125$ and $m/g = 0.25$.  (d): comparison with the weak-coupling result (\ref{tensionwc}) (dashed line) for $m/g=1,2,4$. Inset: zooming in on the $m/g=4$ curve.  }
\end{figure}

In fig. \ref{fig:StringTensiona} we plot our result for the continuum string tension $\sigma_Q$ computed for different values of the mass $m/g$ as a function of the charge $g Q$ of the external quark-antiquark pair. Note that we only consider $Q$-values $\in [0,1[$ as the string tension is periodic in $Q$: $Q\rightarrow Q-p$ upon $L(n)\rightarrow L(n)+p$ for $p\in\mathbb{Z}$ in the Hamiltonian (\ref{equationH}). Note also that one can combine this transformation for $p=1$ with a $CT$-transformation ($C=$ charge conjugation): \bea L(n) \rightarrow 1 - L(n+1) &\quad \theta(n) \rightarrow - \theta(n+1) \nonumber\\
 \sigma_z(n) \rightarrow - \sigma_z(n+1) &\quad \sigma^{\pm}(n) \rightarrow \sigma^{\mp}(n+1) \,.\label{CT}\eea This transformation gives $Q\rightarrow 1-Q$ in the Hamiltonian (\ref{equationH}) and therefore $\sigma_Q =\sigma_{1-Q}$. So for our calculations we can restrict ourselves to values $Q\in[0,1/2]$. In practice we consider the explicit values: $Q = 0.05,0.10,0.15,...,0.45, 0.47, 0.48, 0.5$ and perform an interpolating fit. 
 
Our considered values for $m/g$ interpolate between the strong- and weak-coupling regime. In the strong-coupling regime $m/g \ll 1$ the string tension is computed in mass perturbation theory from the bosonized field theory up to order $\mathcal{O}((m/g)^3)$ \cite{Adam}
\be\label{tensionsc}\frac{ \sigma_Q}{g^2} \approx \frac{m}{g}\Sigma(1 - \cos(2\pi Q)) + \frac{m^2\Sigma^2E_+\pi}{4g^2}(1-\cos(4\pi Q)) \ee
where $\Sigma = 0.15993, E_{+} = -8.9139$. As one can observe in fig. \ref{fig:StringTensionc} for $m/g\rightarrow 0$ our results indeed converge to this analytic result that is plotted with a dashed line for $m/g=0.125$ and for $m/g=0.25$. 

In the weak-coupling regime $g/m\ll 1$ we can easily compute the string tension in standard perturbation theory from the continuum Lagrangian (\ref{Lagrangian})  (see appendix \ref{pertsigma}). Up to order $(g/m)^4$  we find the string tension for $Q\leq 1/2$: \be \frac{\sigma_Q}{g^2} \approx \frac{Q^2}{2}\left(1-\frac{g^2}{m^2}\frac{1}{6\pi}\right)\,,\label{tensionwc}\ee with the value for $Q>1/2$ following from the identification $\sigma_Q=\sigma_{1-Q}$ for the compact formulation of QED$_2$ that we are considering. In fig. \ref{fig:StringTensiond} one can again observe the convergence of our numerical results to this analytic result, now for $g/m\rightarrow 0$. Notice here that we subtract the leading order term of (\ref{tensionwc}).

\begin{figure}
\begin{subfigure}[b]{.24\textwidth}
\includegraphics[width=\textwidth]{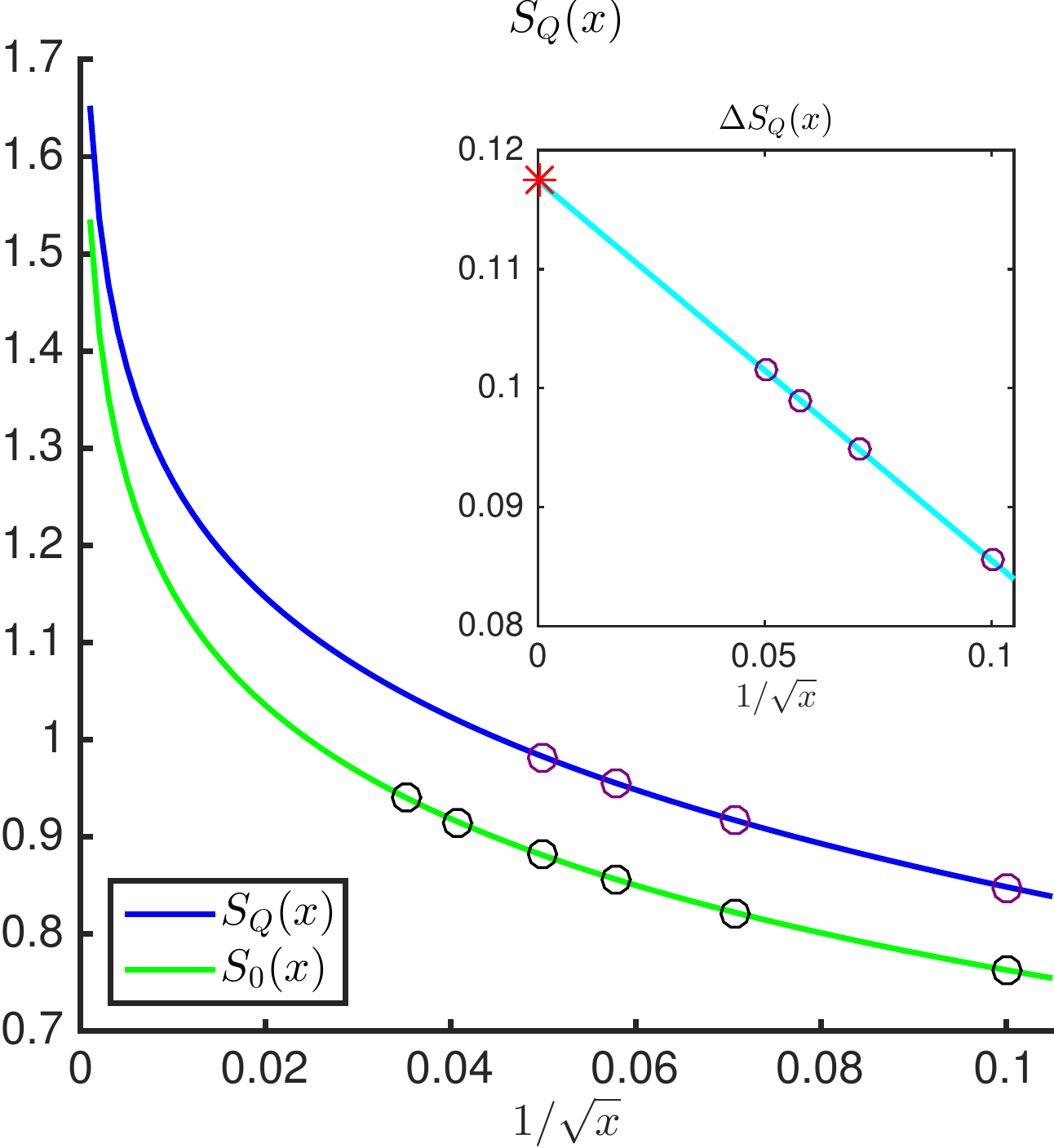}
\caption{\label{fig:UEntropya}}
\end{subfigure}\hfill
\begin{subfigure}[b]{.24\textwidth}
\includegraphics[width=\textwidth]{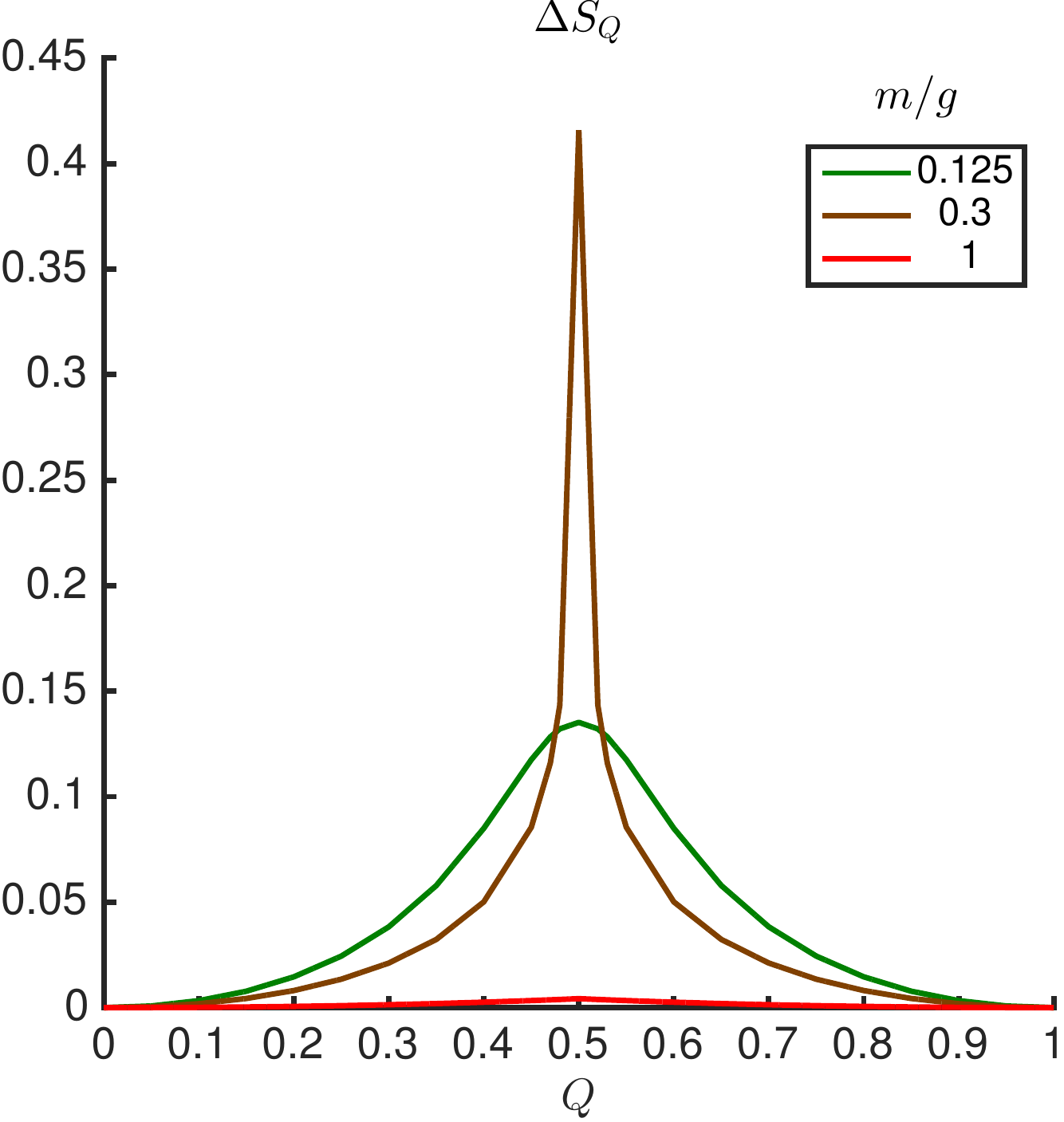}
\caption{\label{fig:UEntropyb}}
\end{subfigure}\hfill
\vskip\baselineskip
\captionsetup{justification=raggedright}
\caption{\label{fig:UEntropy}(a): m/g = 0.25, Q = 0.45. Fit of the form $(-1/6)\log(1/\sqrt{x}) + A + C/\sqrt{x}$ to $S_Q(x)$ and $S_0(x)$. Inset: linear fit to $\Delta S_Q(x)$. (b): $\Delta S_Q$ for different values of $m/g$.  }
\end{figure}

Comparing the strong- and weak-coupling regime we observe an important difference: in the strong-coupling limit $\sigma_Q$ is differentiable at $Q = 1/2$ whereas in the weak-coupling limit this is not the case. Therefore there exists a critical mass $(m/g)_c$ with the property that $\sigma_Q$ is differentiable at $Q = 1/2$ for $(m/g) < (m/g)_c$ and not differentiable at $Q = 1/2$ for $(m/g) > (m/g)_c$. This point $(m/g)_c$ corresponds to the first-order phase transition for the Hamiltonian $H_Q$ (\ref{equationH}) at $Q=1/2$ \cite{Hamer}. $H_{Q=1/2}$ is symmetric under the $CT$ transformation (\ref{CT}) and the point $(m/g)_c$ separates the unbroken phase $m/g<(m/g)_c$ from the spontaneously broken phase $m/g>(m/g)_c$ that was originally predicted by Coleman \cite{Coleman}. This relationship of the breaking of $CT$-symmetry with the nondifferentiability of $\sigma_Q$ can be made more concrete by noting that
\be\label{orderParameter} \frac{d \sigma_Q}{d Q} = -\frac{1}{2}\left\langle \sum_{n=1,2}  (L(n) - Q)\right\rangle_Q\equiv -\frac{1}{2g}E_Q\ee
where $\langle\ldots \rangle_Q$ denotes the expectation values with respect to the ground state of $H_Q$. We now have the relation $E_Q=-E_{1-Q}$ from the $CT$-transformation (\ref{CT}), which indeed makes it a good order parameter for the $CT$ breaking at $Q=1/2$. 

We perform an independent computation of $E_Q$, again for $Q = 0.05,0.10,0.15,...,0.45, 0.47, 0.48, 0.5$, and now using values $x = 100,200,300,400$ for our continuum extrapolation (see appendix \ref{appasymptotic}). Our results are displayed in fig. \ref{fig:StringTensionb}. At $Q \rightarrow 1/2$ we find for $m/g = 0.3$, $E_Q/g=0$ up to a numerical error of $4 \times 10^{-3}$ while for $m/g = 0.35$ we find $E_Q/g=0.314(2)$, consistent with the value $(m/g)_c\approx 0.33$ that was obtained in \cite{Byrnes} and also consistent with the behavior of $\sigma_Q$ in fig. \ref{fig:StringTensiona}. 

Finally we also compute the half-chain entropy $S$ (\ref{EntropyI}) for different values of $Q$ and $m/g$, which in this translational-invariant case does not depend on the position of the cut. As such the entropy is a UV divergent quantity, but one expects the divergence to come from the fermion kinetic term in the Hamiltonian (\ref{equationH}) and therefore be $Q$-independent. Specifically, the general results of Cardy and Calabrese \cite{Calabrese2004} predict for two fermionic degrees of freedom a UV divergence (with correlation length $\xi$ in physical units):
\bea \label{eq:CardyCalabrese} S_Q(x) &\sim& \frac{1}{6}\log\left(\frac{\xi}{a}\right)\\
&=& -\frac{1}{6}\log (1/\sqrt{x}) + (\mbox{finite terms as } x \rightarrow +\infty)\nonumber\eea 
with $a = 1/g\sqrt{x}$ the lattice spacing. This is precisely what we find in our simulations. As an illustration, in fig. \ref{fig:UEntropya} we show a fit of the form $(-1/6)\log(1/\sqrt{x}) + A + C/\sqrt{x}$ through our data of $S_Q(x)$ for $Q = 0$, $Q = 0.45$ and $m/g = 0.25$. $A$ and $B$ are obtained by a linear fit through $S_Q(x) + (1/6)\log(1/\sqrt{x})$, see subsection \ref{subsec:continuumExtraEntropy} of appendix \ref{appasymptotic}. There, we also explicitly extract the coefficient $-1/6$ of the logarithmic term by a logarithmic fit to the data. The errors are of the order $10^{-3}$ for $m/g \gtrsim 0.5$ and only of order $10^{-4}$ for $m/g \lesssim 0.5$, see table \ref{fig:tableEntropyQ0} in appendix \ref{subsec:continuumExtraEntropy}. 

The universality of the logarithmic UV divergence then allows us to define a UV-finite renormalized entropy $\Delta S_Q\equiv S_Q-S_0$, with a finite continuum value that can be obtained by a polynomial extrapolation in $1/\sqrt{x}$, see inset fig. \ref{fig:UEntropya}. Contrary to the string tension and the electric field, we found sometimes that the results at $x = 100$ and the continuum results differ by a factor of order one or have different sign. We refer the reader to subsection \ref{subsec:continuumExtraEntropy} in appendix \ref{appasymptotic} and, in particular, to fig. \ref{fig:ExtrapolationEntr} for the details about the continuum extrapolation. In fig. \ref{fig:UEntropyb} we show this renormalized entropy $\Delta S_Q$ as a function of $Q$ for different values of $m/g$. Most notably we observe an (almost) divergent behavior for $m/g=0.3$ at $Q\rightarrow 1/2$ close to the critical point $Q=1/2, (m/g)_c \approx 0.33$. From (\ref{eq:CardyCalabrese}) we indeed expect a growing entropy for growing correlation length. By the same argument one can understand the behavior at small $Q$-values: there the correlation length (inverse mass gap) increases with growing $g/m$ \cite{Buyens3}, which is indeed paralleled by the behavior of $\Delta S_Q$.

\section{From small to large distances}\label{sec:potential} 
\noindent Now, we consider the situation where the external quark and antiquark pair are separated over a finite length $L$.  On a lattice with spacing $a$ and interquark distance $L = ka$, the pair introduces a nonuniform background electric field $\alpha(n)= -Q\Theta(0 \leq n < k)$ in the Hamiltonian (\ref{equationH}). As ansatz for our MPS trial state  $\ket{\Phi(\bm{B})}$ for the ground state we now write:
\begin{multline} \ket{\Phi(\bm{B})} =  \sum_{\bm{\kappa}} \bm{v}_L^\dagger \left(\prod_{n < r_L}A_{\kappa_{n}}(n)\right) \\  \left(\prod_{n = r_L}^{r_R-1}B_{\kappa_{n}}(n)\right)   \left(\prod_{n \geq r_R}A_{\kappa_{n}}(n)\right) \bm{v}_R  \ket{\bm{\kappa}}, \label{nonuMPS}\end{multline}
where $r_L \ll 0 \leq k \ll r_R$ and $A_\kappa(n) = A_\kappa(n \mbox{ mod 2})$ corresponds to the MPS approximation (\ref{uMPS}) of the ground state of the zero-background Hamiltonian ($\alpha(n) = 0$) and depends only on the parity of $n$. This is a MPS (\ref{MPS}) in the thermodynamic limit ($N \rightarrow + \infty$) where we take $A(n) = B(n)$ for $r_l \leq n \leq r_R-1$ and take the $A(n)$ corresponding to the ground state (\ref{uMPS}) for $\alpha(n) = 0$ to the left and to the right of the $B(n)$'s ($n < r_L$ and $n \geq r_R$). 

The idea behind this ansatz is that the nonuniform background electric field changes the vacuum and breaks translation invariance (all $B(n)$ are different) but that asymptotically ($\vert n \vert \gg 1$) it does not affect the vacuum. Again, gauge invariance (\ref{spingauss}) is imposed if $B(n)$ takes the form (\ref{gaugeMPS}) with general matrices $b_{s,p}(n) \in \mathbb{C}^{D_q(n) \times D_r(n+1)}$ ($q \in \mathbb{Z}[p_{min}(n),p_{max}(n)];$ $p, r \in \mathbb{Z}[p_{min}(n+1),p_{max}(n+1)])$. Note that we allow different bond dimensions on different sites. Also $a_{s,p}(n)$ is of the form (\ref{gaugeMPS}) as we impose this to determine the ground state of the zero-background electric field Hamiltonian.

Because (\ref{nonuMPS}) is linear in each of the $B_n$ we can use the DMRG-method \cite{White} to obtain the best approximation for the ground state within the manifold of gauge-invariant states, by optimizing on the UV- and IR-finite quantity $V_Q(L)$. By looking at the Schmidt spectrum we are able to fix the values of the virtual dimension $D_q(n)$ and the minimum and maximum eigenvalues $p_{min}(n+1)$ and $p_{max}(n+1)$ of $L(n)$ we retain in our numerical scheme to obtain an accurate approximation of the ground state. The choices for $r_L$ and $r_R$, which vary between $-k/2-250 \leq r_L \leq -k/2-150$ and $k/2 +150 \leq r_R \leq k/2 +250$ are checked a posteriori by verifying the convergence of local observables at large distances to their zero-background value. We refer the reader to Appendix \ref{appDMRG} for the details. 

\begin{figure}
\begin{subfigure}[b]{.24\textwidth}
\includegraphics[width=\textwidth]{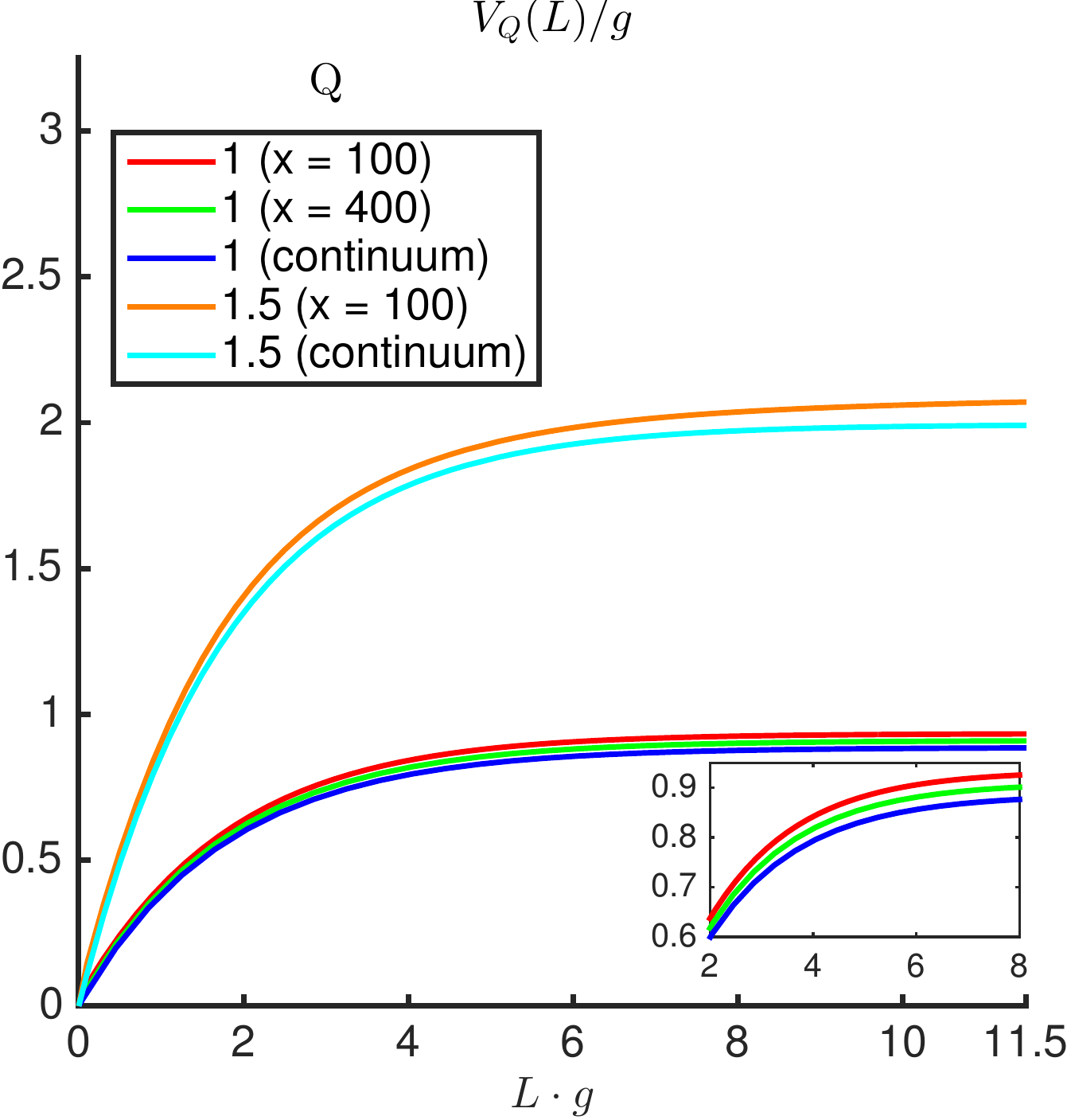}
\caption{\label{fig:mdivgzeroa}}
\end{subfigure}\hfill
\begin{subfigure}[b]{.24\textwidth}
\includegraphics[width=\textwidth]{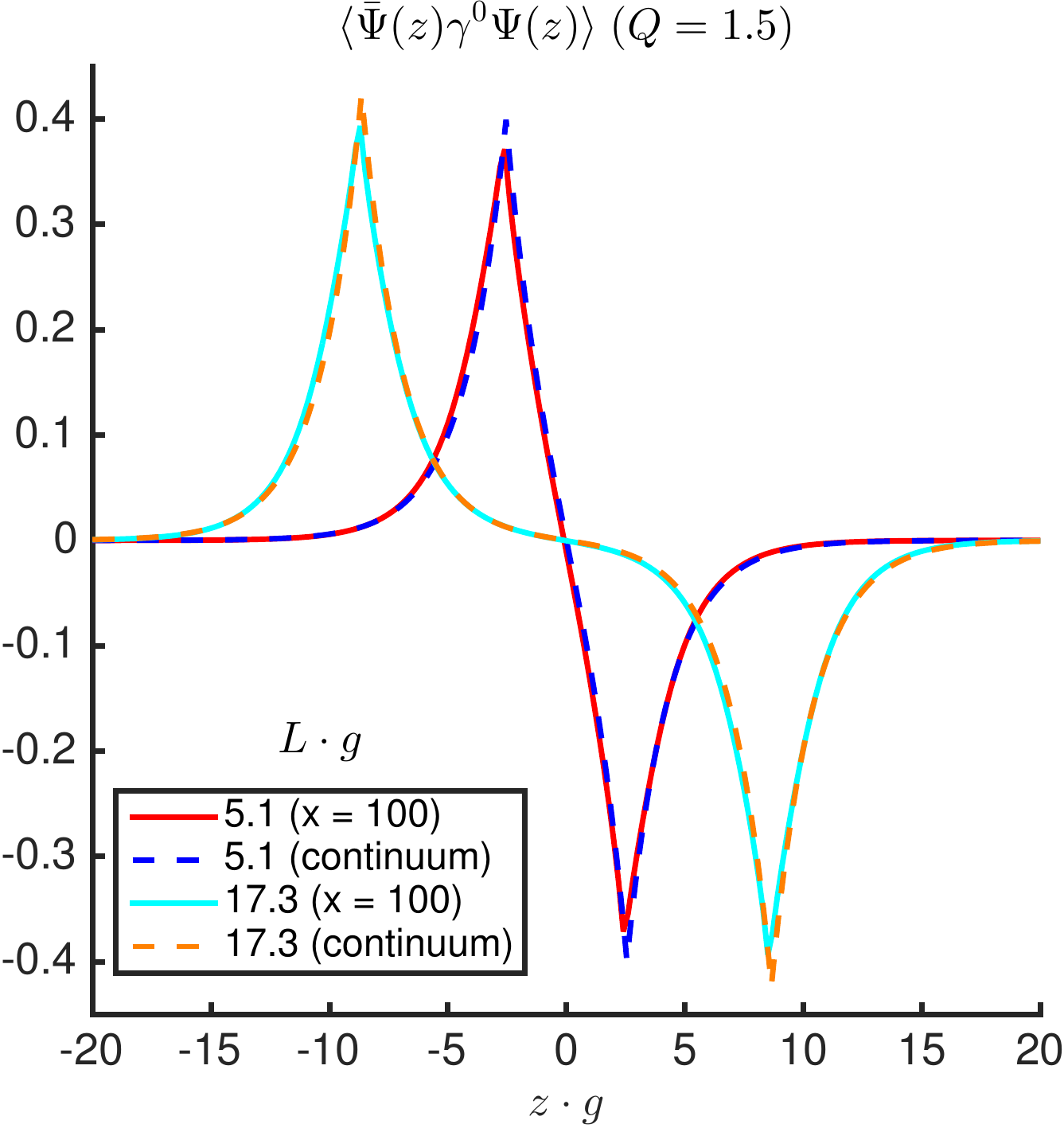}
\caption{\label{fig:mdivgzerob}}
\end{subfigure}\vskip\baselineskip
\captionsetup{justification=raggedright}
\caption{\label{fig:mdivgzero} $m/g = 0$. (a): Potential for $Q = 1$ and $Q = 1.5$ compared with exact result in the continuum (\ref{strongCouplingPotential}). Inset: convergence for $x \rightarrow + \infty$ to (\ref{strongCouplingPotential}) for $Q = 1$. (b): Distribution of fermion charge for $Q = 1.5$ for different separation lengths of the quark and antiquark for $x = 100$. The results are compared with the exact result (\ref{cdSC}). }
\end{figure}

\subsection{The case m/g=0: screening $\grave{\textrm{a}}$ la Higgs   }\label{mg0}
\noindent We first discuss our results for the $m/g=0$ case. This is a special case, as the asymptotic string tension $\sigma_Q$ vanishes for all values (integer or fractional) of the charge. Physically, this is interpreted as a manifestation of a Higgs mechanism \cite{ColemanCS}, suppressing the long-range Coulomb force and replacing it with a short-range Yukawa force thereby effectively screening all charges. Another reason that makes the $m/g=0$ case special is that it can be solved analytically \cite{Iso}, which allows for benchmarking of numerical results. Previous numerical calculations for this case were performed with Monte Carlo simulations on the bosonized version of the theory \cite{Bender}.      

In fig. \ref{fig:mdivgzeroa} we plot our results for the potential for $m/g = 0$ for $Q = 1$ and $Q = 1.5$. This can be compared with the exact continuum result  \cite{Iso}:
\be\label{strongCouplingPotential} V_{Q}(L) =  \frac{\sqrt{\pi}gQ^2}{2}\left(1 - e^{- Lg/\sqrt{\pi}}\right), \ee which is indeed of the Yukawa type. We find very good agreement already for $x=100$, for both $Q=1$ and $Q=1.5$. For $Q=1$ we also perform a computation for $x=400$,  in the inset of fig. \ref{fig:mdivgzeroa} one can observe the rate of convergence towards the continuum $x\rightarrow \infty$ in this case. 

The charge density $\langle \bar{\psi}(z)\gamma^0\psi(z) \rangle$ of the light quarks is, of course, also an interesting quantity to compute, as it explicitly shows the screening of the external probe charges. The analytical result for the probe charge pair put at $\pm L/2$ reads \cite{Iso}:
\be\label{cdSC} \langle \bar{\psi}(z)\gamma^0\psi(z) \rangle = \frac{gQ}{2\sqrt{\pi}}\left( e^{-g\vert z + L/2 \vert/\sqrt{\pi} } - e^{-g\vert z- L/2 \vert / \sqrt{\pi}}\right).\ee This indeed corresponds to a charge distribution with two `clouds' of oppositely charged light (in this case massless) quarks, around the external quark and antiquark, that for large distance $L$ have exactly the same total charge $\pm Q$ as the external pair. On the lattice the charge density at $z = (2n - 1/2)a$ is computed as $\sqrt{x}\langle \sigma_z(2n-1) + \sigma_z(2n) \rangle /2$. In fig. \ref{fig:mdivgzerob}, we plot this density for $Q = 1.5$ where the charges are separated at distances $Lg = 5.1$ and $Lg = 17.3$. Here, too, our results for $x = 100$ are already very close to the continuum result. 

\begin{figure}
\begin{subfigure}[b]{.24\textwidth}
\includegraphics[width=\textwidth]{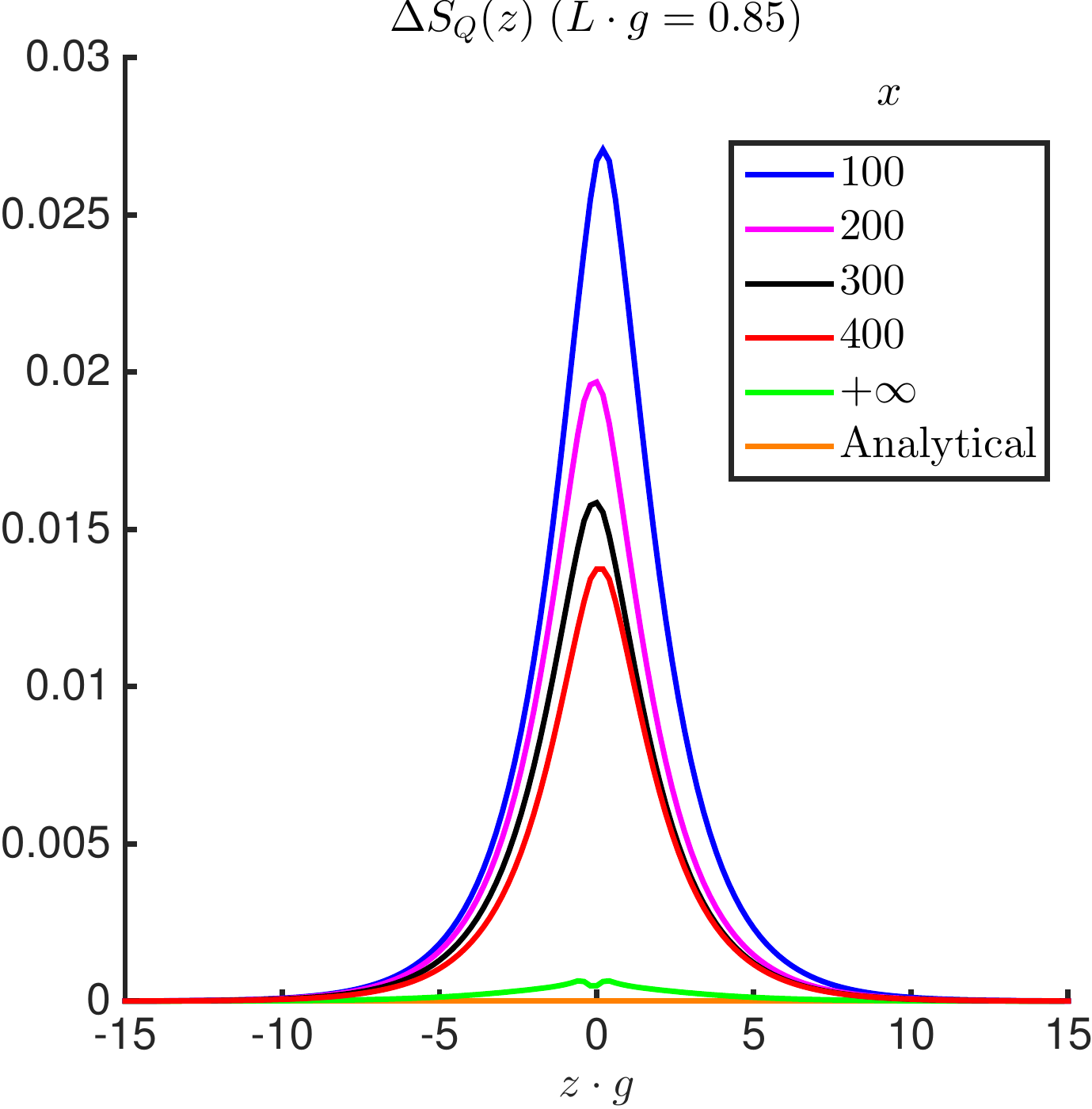}
\caption{\label{fig:mdivg0Entropya}}
\end{subfigure}\hfill
\begin{subfigure}[b]{.24\textwidth}
\includegraphics[width=\textwidth]{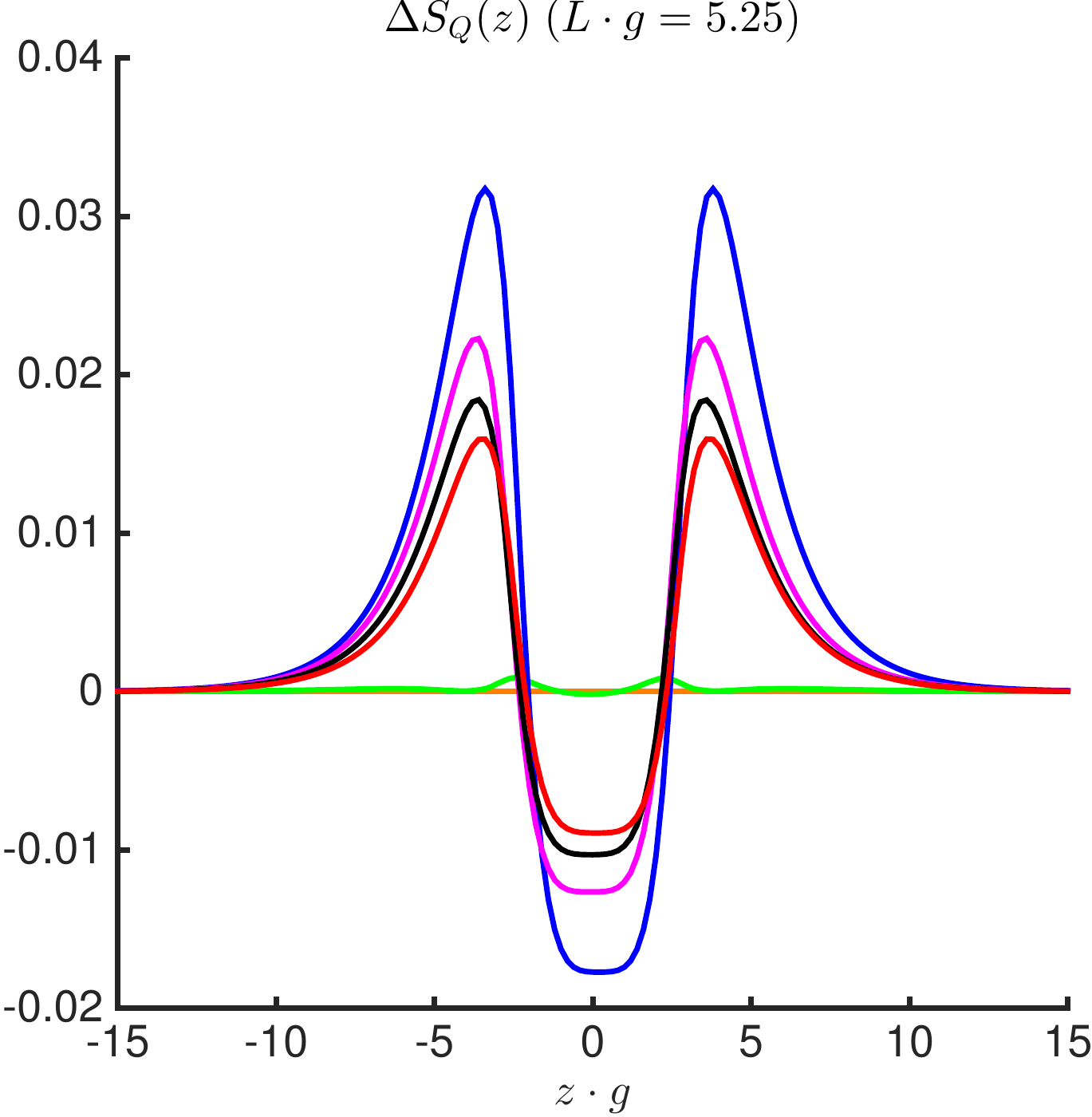}
\caption{\label{fig:mdivg0Entropyb}}
\end{subfigure}
\captionsetup{justification=raggedright}
\vskip\baselineskip
\caption{\label{fig:mdivg0Entropy} $m/g = 0$, $Q = 1$. Spatial profile of $\Delta S_Q$ for different values of $L$ and scaling to the continuum limit ($x \rightarrow + \infty$). (a) $Lg = 0.85$.  (b):  $Lg = 15.65$.}
\end{figure}

In fig. \ref{fig:mdivg0Entropy} we show the spatial profile of the renormalized half-chain von Neumann entropy $\Delta S_{Q}(z) = S_{Q}(z) - S_{0}(z)$ for different values of $Lg$. We compute this quantity for $z = (n+1/2) a$ with $n$ even and perform an interpolating fit. When the heavy quarks are close to each other, $\Delta S_Q(z)$ shows a peak in the middle between the charges and falls off very fast with $\vert z g\vert$, see fig. \ref{fig:mdivg0Entropya}. For larger values of $Lg$ a cloud of light quarks forms around each of the heavy charges which clearly leaves its imprints on the spatial profile of the von Neumann entropy, see fig. \ref{fig:mdivg0Entropyb}. $\Delta S_Q(z)$ is nonzero around each of the heavy charges and is zero around $z g \approx 0$. 

The observed spatial profiles of the von Neumann entropy, however, are lattice artifacts and vanish in the continuum limit ($x \rightarrow + \infty$). Indeed, from the bosonized Hamiltonian for $m/g = 0$, it can be observed that any position-dependent electric background field can be transformed away \cite{Abdalla0,ColemanCS} up to a position-dependent constant. Therefore the von Neumann entropy of the ground state with $\alpha(n) \neq 0$ and $\alpha(n)= 0$ are the same, hence $\Delta S_Q (z) = 0$ for any value of $Lg$. By investigating the scaling towards $x \rightarrow + \infty$ in figs. \ref{fig:mdivg0Entropya} and \ref{fig:mdivg0Entropyb} we indeed observe that $\Delta S_Q(z)$ tends towards a very small value for $x \rightarrow + \infty$. Note that we here need to perform an interpolation because we can only take $Lg$ to be an integer multiple of $1/\sqrt{x}$. Specifically, we perform simulations for $x = 400$ and $Lg = 0.85,5.25,15.65$. For $x = 100, 200, 300$ we first do simulations for $L_1g < 0.85, 5.25, 15.65$ and $L_2g > 0.85, 5.25, 15.65$. Afterwards we do a simple linear interpolation between $L_1g$ and $L_2g$ to obtain the curve for $L g = 0.85, 5.25, 15.65$. By performing a linear extrapolation in $1/\sqrt{x}$ through $x = 100, 200, 300, 400$ and $x = 200,300,400$ we find that in fig. \ref{fig:mdivg0Entropya} the continuum extrapolation of the maxima for $zg \approx 0$ yields the estimate $\Delta S _Q(0) \approx 5 (7) \times 10^{-4}$, while extrapolating the maxima in fig. \ref{fig:mdivg0Entropyb} around $z g = \pm 10$ gives $\Delta S _Q(\pm10) \approx 2 (5) \times 10^{-4}$, consistent with $\Delta S_Q(z) = 0$. The interpolation to obtain the curves for $Lg = 0.85,5.25,15.65$ for all values of $x$ leads to relative large errors such that a continuum extrapolation of $\Delta S_Q(z)$ for all values of $zg$ is less reliable. Therefore, for $Lg \leq 0$, we perform a linear extrapolation in $1/\sqrt{x}$ through our values for $x = 100,200,300,400$ and obtained the results for $Lg \geq 0$ by a reflection of the results for $Lg \leq 0$ (green line). This can be compared with the analytical result $\Delta S_Q(z) = 0$ (orange).

\subsection{The case $Q=1$: string breaking}
\noindent For $m/g\neq 0$, the asymptotic string tension $\sigma_Q$ vanishes only for integer charges $Q$. This is taken to be an indication for a screening $\grave{\textrm{a}}$ la QCD \cite{ColemanCS}, where the potential exhibits a string tension ($V_Q(L) \propto L$) at short distances, but flattens out completely at large distances, at least for integer charges $Q$. At these large distances it becomes energetically favorable to materialize light (yet massive) (anti)fermions out of the vacuum that bind to the external quark and antiquark, resulting in two charge neutral mesons. 

Historically, for QCD, lattice Monte-Carlo simulations succeeded first to calculate numerically the short distance confining behavior of the potential -- both in the quenched and unquenched approximation -- via the expectation value of the Wilson \cite{Bali,Greensite}. The detection of string breaking has posed a larger challenge. A main problem with the use of the standard Wilson loop is the poor overlap with the broken-string two-meson state. This problem was finally overcome by including light quark propagators in the Wilson loop and analyzing its mixing with the standard Wilson loop \cite{Philipsen,Bali2}. 

\begin{figure}
\begin{subfigure}[b]{.24\textwidth}
\includegraphics[width=\textwidth]{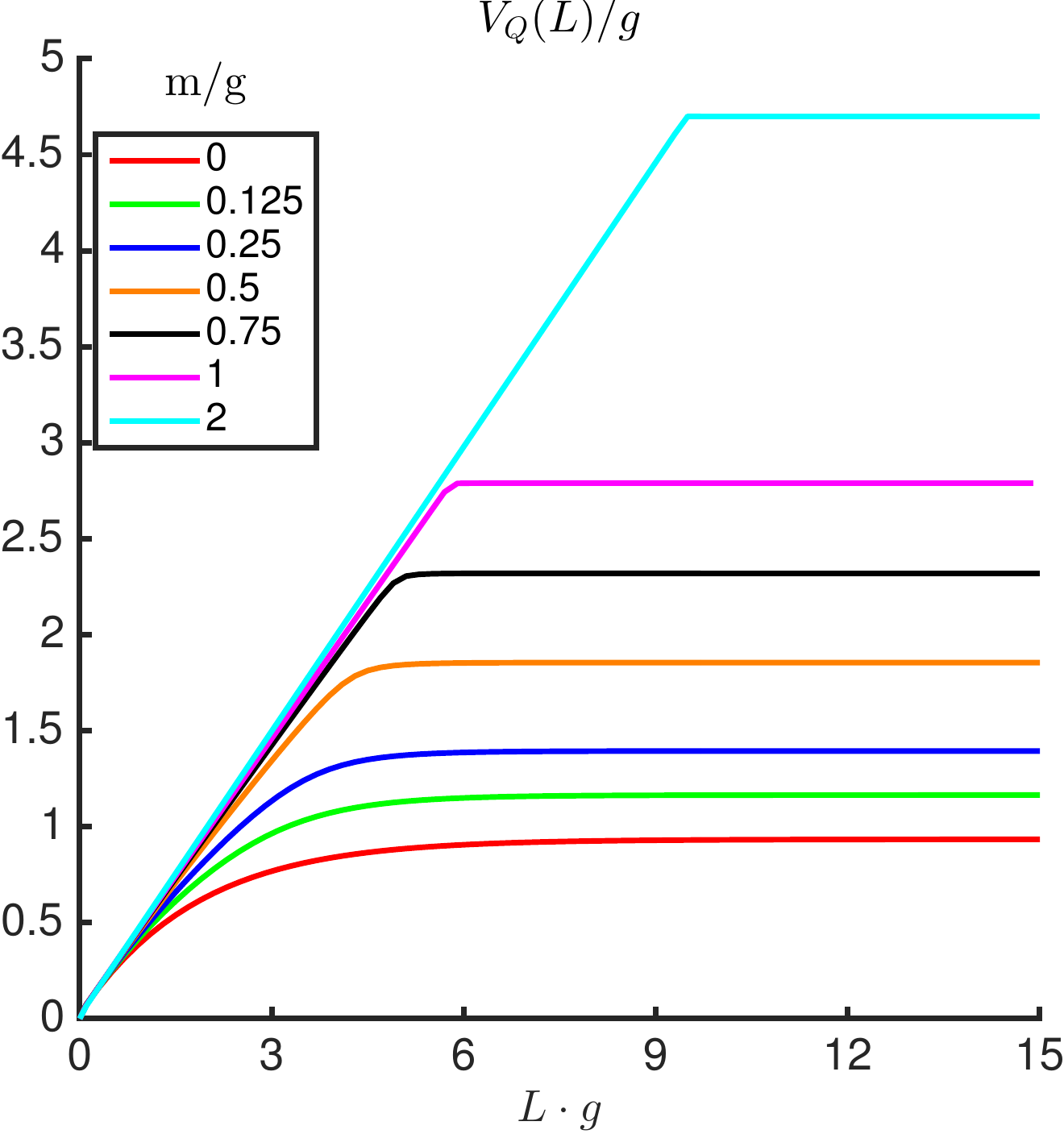}
\caption{\label{fig:Q1a}}
\end{subfigure}\hfill
\begin{subfigure}[b]{.24\textwidth}
\includegraphics[width=\textwidth]{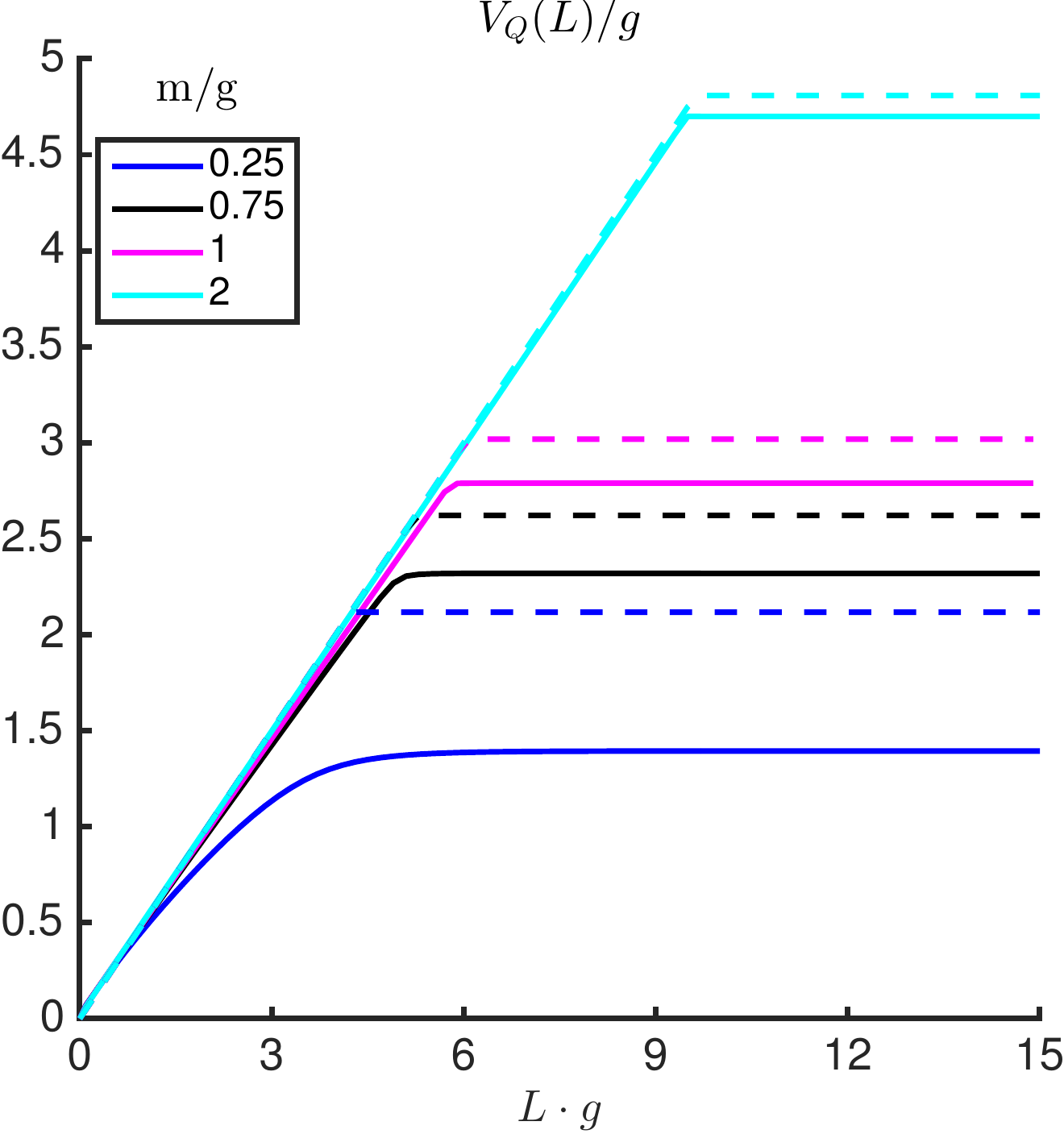}
\caption{\label{fig:Q1b}}
\end{subfigure}\hfill
\vskip\baselineskip
\begin{subfigure}[b]{.24\textwidth}
\includegraphics[width=\textwidth]{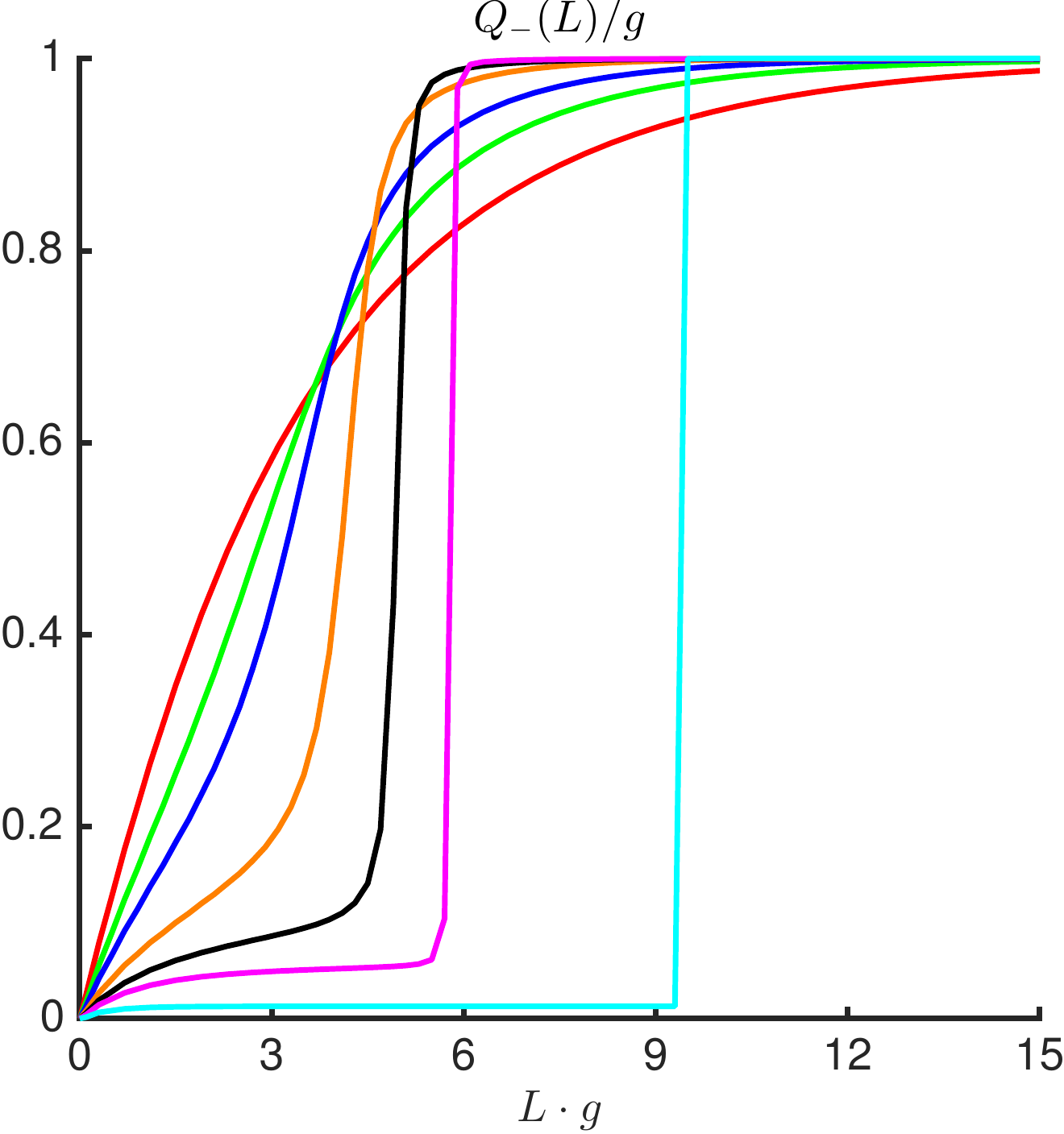}
\caption{\label{fig:Q1c}}
\end{subfigure}\hfill
\begin{subfigure}[b]{.24\textwidth}
\includegraphics[width=\textwidth]{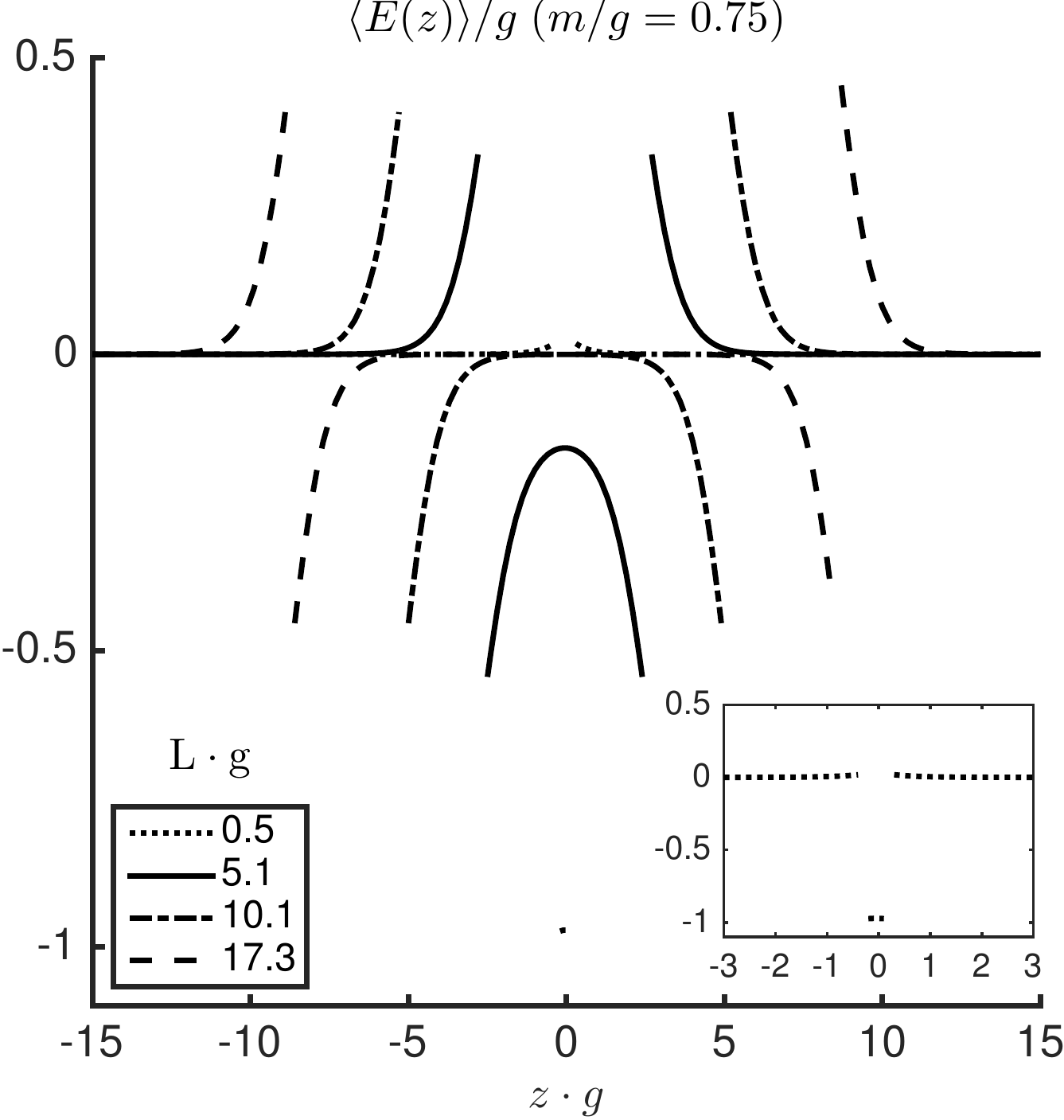}
\caption{\label{fig:Q1d}}
\end{subfigure}
\vskip\baselineskip
\begin{subfigure}[b]{.24\textwidth}
\includegraphics[width=\textwidth]{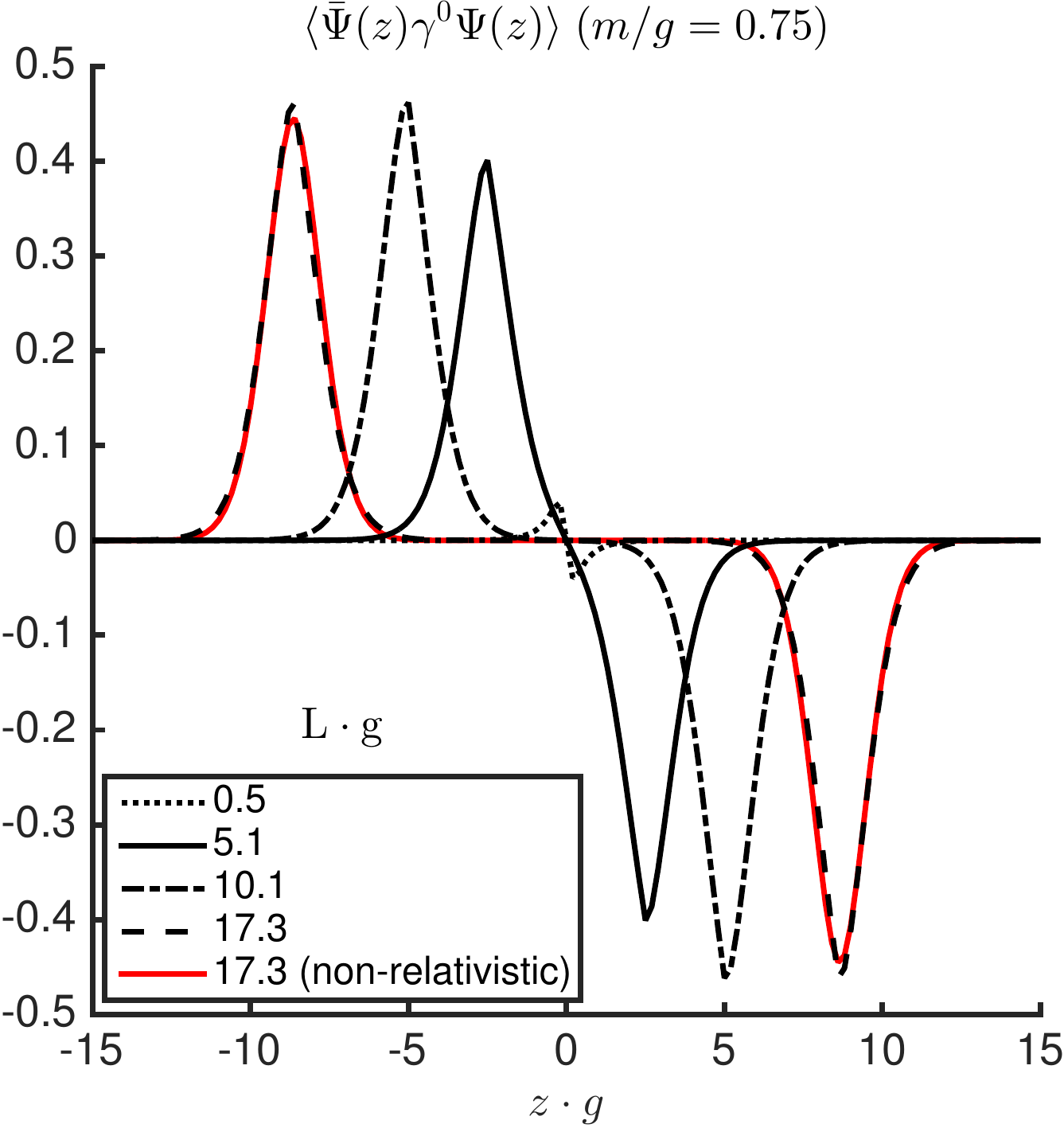}
\caption{\label{fig:Q1e}}
\end{subfigure}\hfill
\begin{subfigure}[b]{.24\textwidth}
\includegraphics[width=\textwidth]{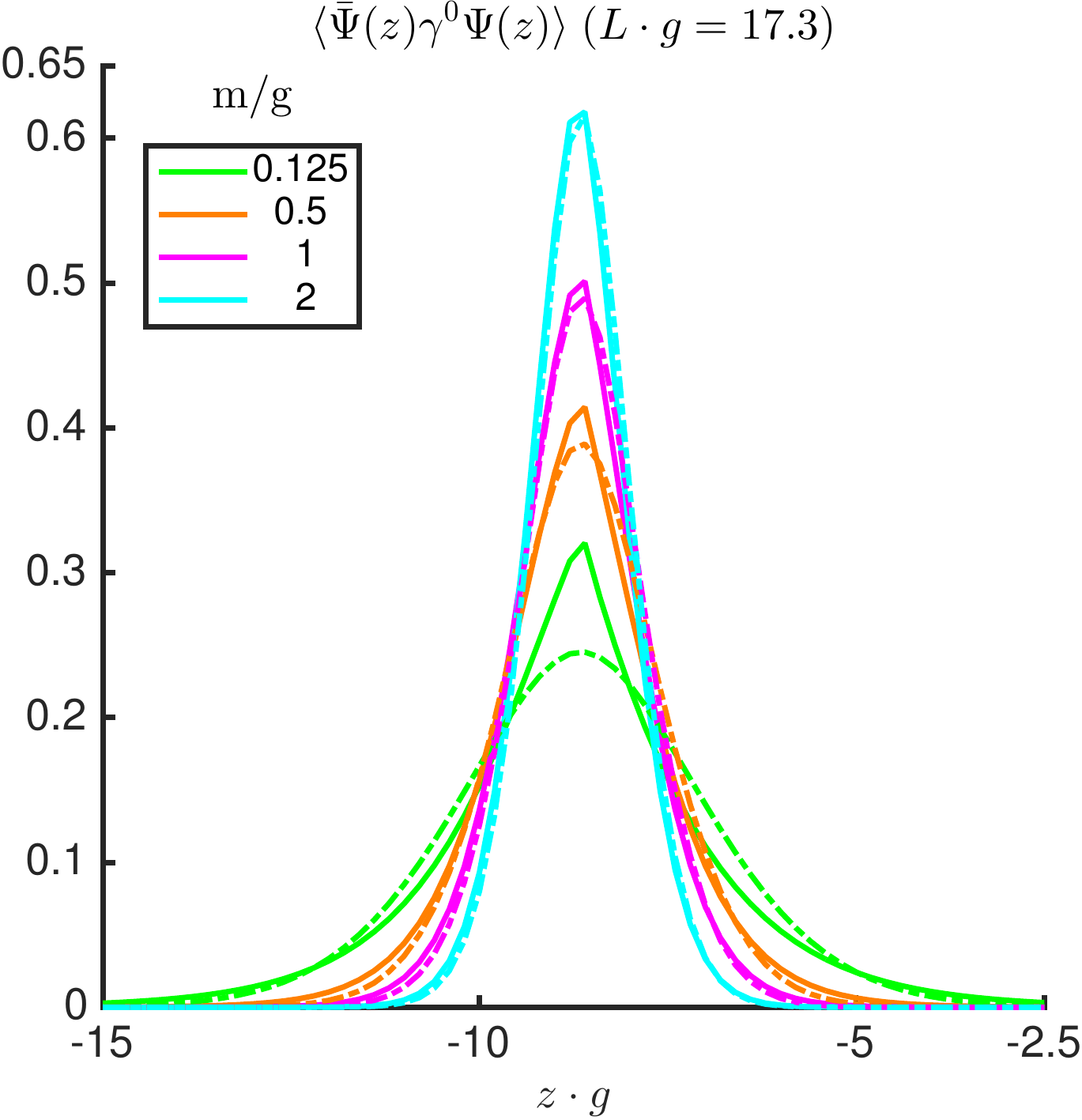}
\caption{\label{fig:Q1f}}
\end{subfigure}
\captionsetup{justification=raggedright}
\vskip\baselineskip
\caption{\label{fig:Q1} $Q = 1, x = 100$. (a): Quark-antiquark potential for different values of $m/g$. (b): Comparison of potential with nonrelativistic limit result (dashed line) for m/g = 0.25, 0.75, 1, 2. (c): The total charge of the light fermions on the negative axis $Q_-(L)$ for different values of $m/g$. (d): Electric field for $m/g = 0.75$. (e): Charge distribution for $m/g = 0.75$. For $Lg = 17.3$ we compare with the charge distribution of the nonrelativistic meson state (full red line). (f): Comparison of the charge density of the left cloud (full line) with that of the nonrelativistic meson state, eq (\ref{eqnonRelMeson}), (dashed line) for $Lg = 17.3$, now for $m/g=0.125, 0.5, 1, 2.$ }
\end{figure}

For the Schwinger model the string-breaking phenomenon has been confirmed in mass perturbation theory \cite{Armoni} and in a semiclassical approximation of the bosonized version of the theory \cite{Abdalla, Rothe}. At the numerical level, for $Q=1$, lattice Monte Carlo simulations have detected both the confining and string-breaking behavior of the potential \cite{Bender, Korcyl}. In \cite{Bender} the problem with the Wilson loop was avoided by computing instead the expectation value of the bosonized Hamiltonian, while \cite{Korcyl} turned to very high statistics thereby explicitly showing the poor overlap of the Wilson loop with the broken-string ground state.

For the local quantities (charge density =$\bar{\psi}(z)\gamma^0\psi(z)$, electric field=$E(z)$)  and the potential, we restrict ourselves from now on to lattice spacing $x=100(=1/g^2a^2)$; from the previous subsection we can expect these results already to be quite close to the continuum. In fig. \ref{fig:Q1a} we display our results for the potential, and this for different values of $m/g$. We compute explicitly the ground-state energy at $Lg = 0.1, 0.3, \ldots 15.3$ and perform an interpolating fit. We clearly find a transition from the confining behavior, associated with the string state,  towards the constant behavior associated with the broken-string two-meson state. This transition happens more suddenly for larger values of $m/g$, which is in qualitative agreement with the semiclassical results from the bosonized theory \cite{Abdalla, Rothe}. This is also what one would expect from the nonrelativistic weak-coupling regime, where the transition can be understood as a level crossing between the zero-particle string state and the two-particle broken-string meson state (see appendix \ref{apppert}). The dashed lines in fig. \ref{fig:Q1b} corresponding to this nonrelativistic result for $\mathcal{E}_{string}= L g^2/2$ and $\mathcal{E}_{2meson}=2m+1.0188 \frac{g^{4/3}}{m^{1/3}}$ were plotted for comparison. We can indeed observe the convergence towards this result for increasing values of $m/g$.

We further illustrate this behavior in fig. \ref{fig:Q1c}, where we plot the total charge $Q_-$ of the light fermions on the negative $z-$axis: 
\be Q_{-} = g\int_{-\infty}^{0} dz\; \langle\bar{\psi}(z)\gamma^0\psi(z)\rangle\,. \ee
One can observe indeed that the interpolation between $Q_{-}=0$ (string state) for small $L$ and $Q_{-}=1$ (meson state) for large $L$ becomes more and more discontinuous for growing $m/g$ in accordance with the nonrelativistic level-crossing picture.

In figs. \ref{fig:Q1d} and fig. \ref{fig:Q1e} we investigate the interpolation from the string state to the string-broken state in more detail for $m/g=0.75$ by plotting the charge density and electric field. For $L/g=0.5$ there is only a very small charge cloud around the external quark and antiquark, notice also the very short electric field string displayed at the bottom of fig. \ref{fig:Q1d}. At $L/g=5.1$ the clouds start to build up, lowering the electric field value at the center. At $L/g=10.1$ the string is completely broken, the electric field at the center has vanished, and we have two clouds of total charge $\pm 1$ around the external quark and antiquark.  At $L/g=17.3$ the two isolated mesons are simply separated over a larger distance, with a quasi-identical charge distribution around the external quarks as for $L/g=10.1$. 

The full red line in fig. \ref{fig:Q1e} is the charge distribution $\pm|\phi(z)|^2$ for the nonrelativistic meson state for $Lg = 17.3$, with 
\be\label{eqnonRelMeson} \phi(z)=\mathcal{N} Ai\left((g^2m)^{1/3}|z\pm L/2|-1.0188\right)\ee 
the ground state of the one-particle problem in a linear potential (appendix \ref{apppert}), where $Ai$ is the Airy function \cite{Airy} and $\mathcal{N}$ the normalization factor. As one can observe, the charge distribution from this nonrelativistic picture matches very well our exact (numerical) result. In fig. \ref{fig:Q1f} we compare the charge cloud at the negative $z-$axis with the nonrelativistic result for other values of $m/g$. One can again observe the convergence to the nonrelativistic result for growing $m/g$, notice that already for $m/g=0.5$ the match is quite good.
\begin{figure}
\begin{subfigure}[b]{.24\textwidth}
\includegraphics[width=\textwidth]{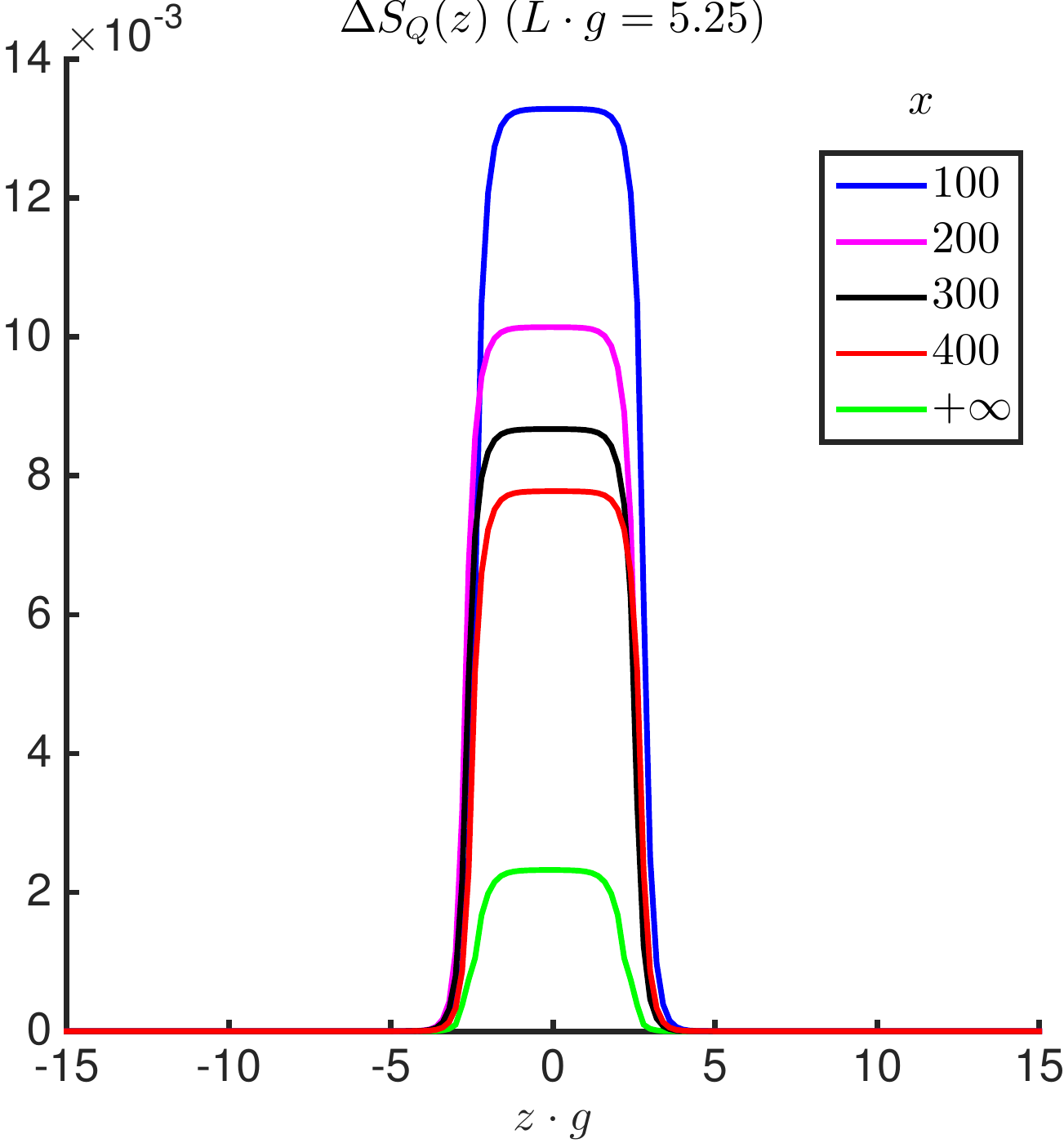}
\caption{\label{fig:mdivg2Entropya}}
\end{subfigure}\hfill
\begin{subfigure}[b]{.24\textwidth}
\includegraphics[width=\textwidth]{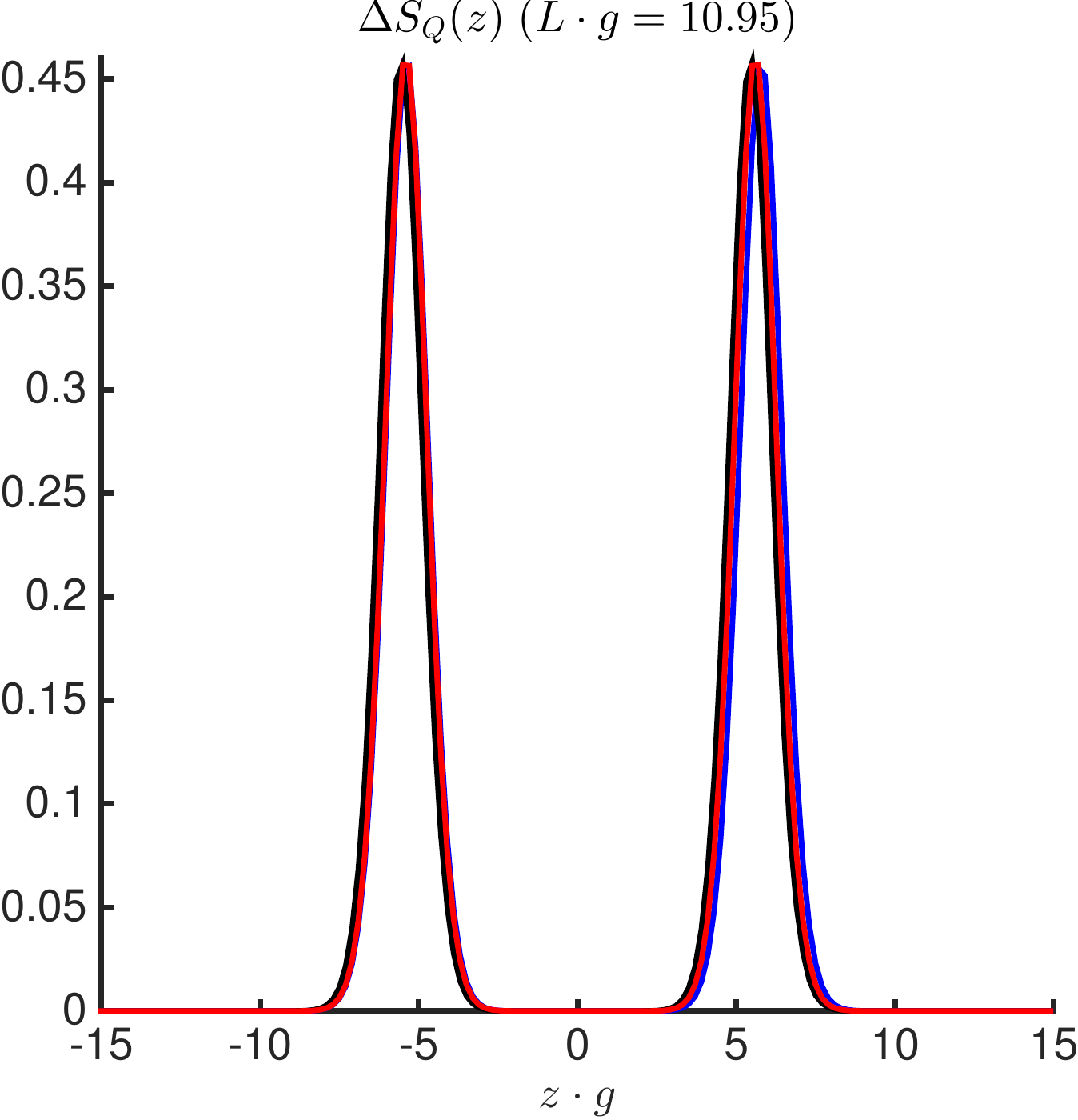}
\caption{\label{fig:mdivg2Entropyb}}
\end{subfigure}\hfill
\captionsetup{justification=raggedright}
\vskip\baselineskip
\caption{\label{fig:mdivg2Entropy} $m/g = 2$, $Q = 1$. $\Delta S_Q(z)$ for different values of $L$ and scaling to the continuum limit. (a) $Lg = 5.25$. (b) $Lg = 10.95$.}
\end{figure}

For the renormalized von Neumann entropy $\Delta S_Q (z)$, we also find a characteristic picture, both for the string state and the string-broken state, see fig. \ref{fig:mdivg2Entropy} for the case $m/g = 2$. For the string state, $L g \lesssim 9.5$, the entropy shows a constant surplus in between the probe charges, similar to the electric field. But notice that this effect becomes very small in the continuum limit (green line), we find an extrapolated value: $\Delta S_g (z) \approx 2.0(5) \times 10^{-3}$ for $zg \in [-2.5,2.5]$. For the string-broken case $L g \gtrsim10$, see fig. \ref{fig:mdivg2Entropyb}, we find that the entropy now shows two clouds around the heavy quark and the heavy antiquark, similar to the charge density. But notice that in contrast to the string state the entropy now survives the continuum limit, with the $x=100$ value already close to the continuum extrapolation. 

\begin{figure}
\begin{subfigure}[b]{.24\textwidth}
\includegraphics[width=\textwidth]{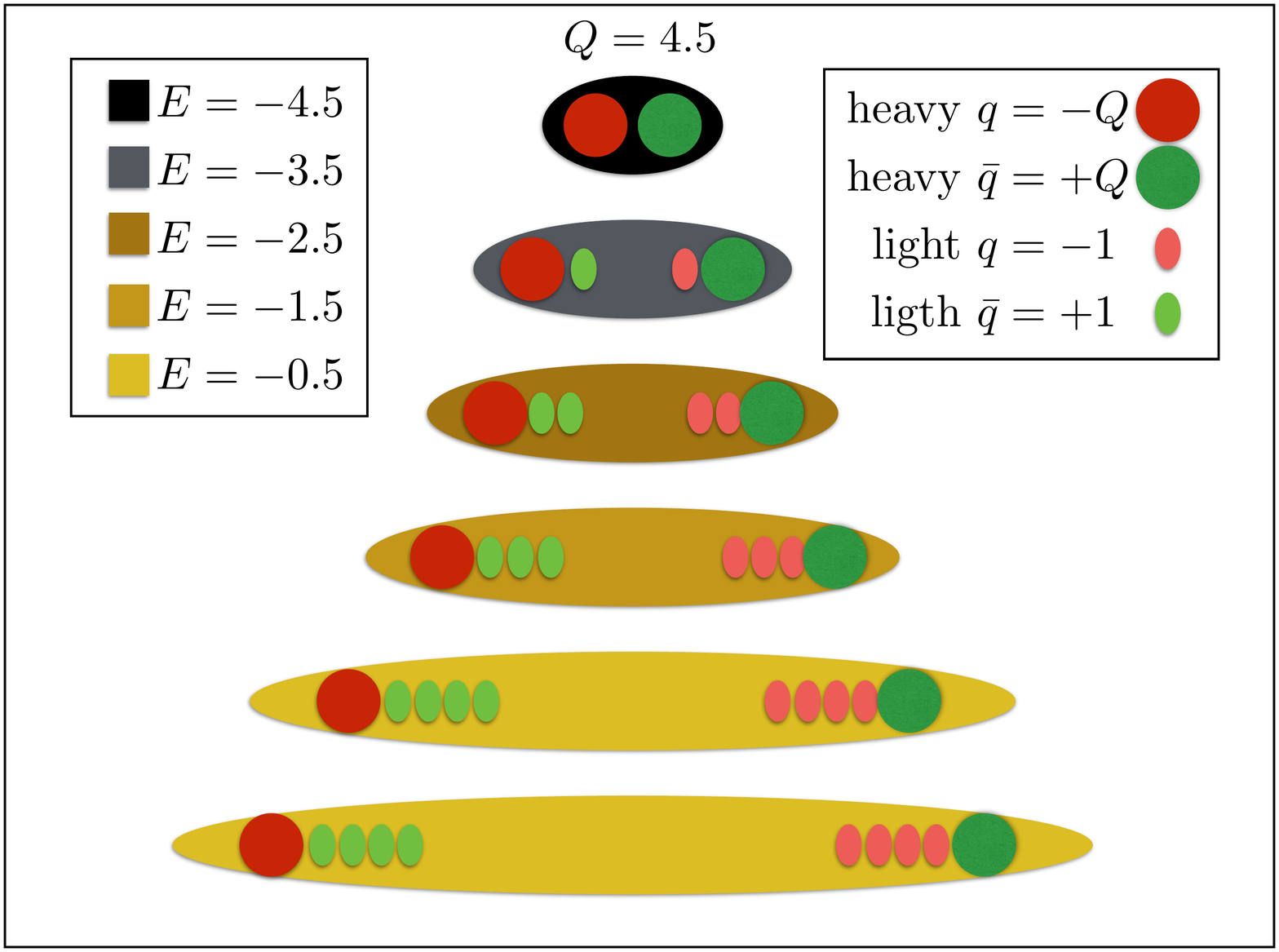}
\caption{\label{fig:cartoon1}}
\end{subfigure}\hfill
\begin{subfigure}[b]{.24\textwidth}
\includegraphics[width=\textwidth]{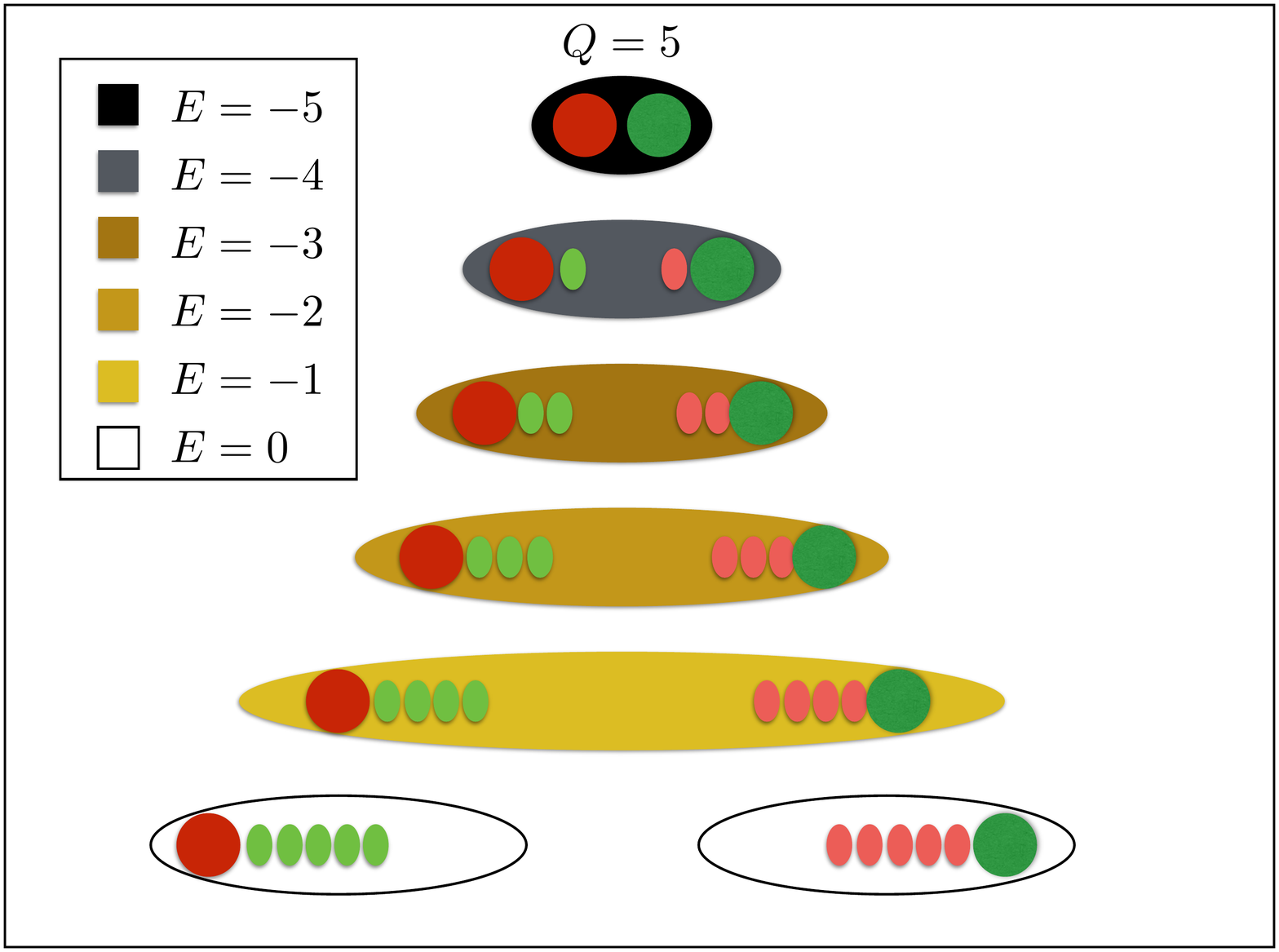}
\caption{\label{fig:cartoon2}}
\end{subfigure}
\captionsetup{justification=raggedright}
\vskip\baselineskip
\caption{\label{fig:cartoon} Cartoon picture of string breaking in the nonrelativistic limit $m/g\rightarrow \infty$. The electric field gets successively screened by light quarks (antiquarks) that bind to the external charges with charge +Q (-Q). This leads to the formation of two mesons: one meson existing of the heavy quark and light antiquarks with charge +1 and one meson existing of the heavy antiquark and the light quarks with charge -1. (a) $Q = 4.5$: the remaining electric field in between with net charge $-0.5$ confines the two meson configurations asymptotically. (b) $Q = 5$: the electric field is entirely screened and the meson configurations are deconfined.}
\end{figure}

\subsection{General $Q$: partial string breaking}
\noindent We now finally turn our attention to the general case $Q\neq 1$. In this case we should have the interesting phenomenon of partial string breaking. Indeed, in the nonrelativistic limit $m/g\rightarrow \infty$ of string breaking due to meson formation, probe charges $Q$ can only be screened by an integer number: $Q\rightarrow \tilde{Q}=Q-n$, where $n$ is the number of light (anti-)quarks that bind to the external charges. For nonzero $\tilde{Q}$, i.e. when $Q$ is noninteger, this still leaves a string between the two separated meson configurations. A visualization of this process in the nonrelativistic limit is shown in fig. \ref{fig:cartoon}.

\begin{figure}
\begin{subfigure}[b]{.24\textwidth}
\includegraphics[width=\textwidth]{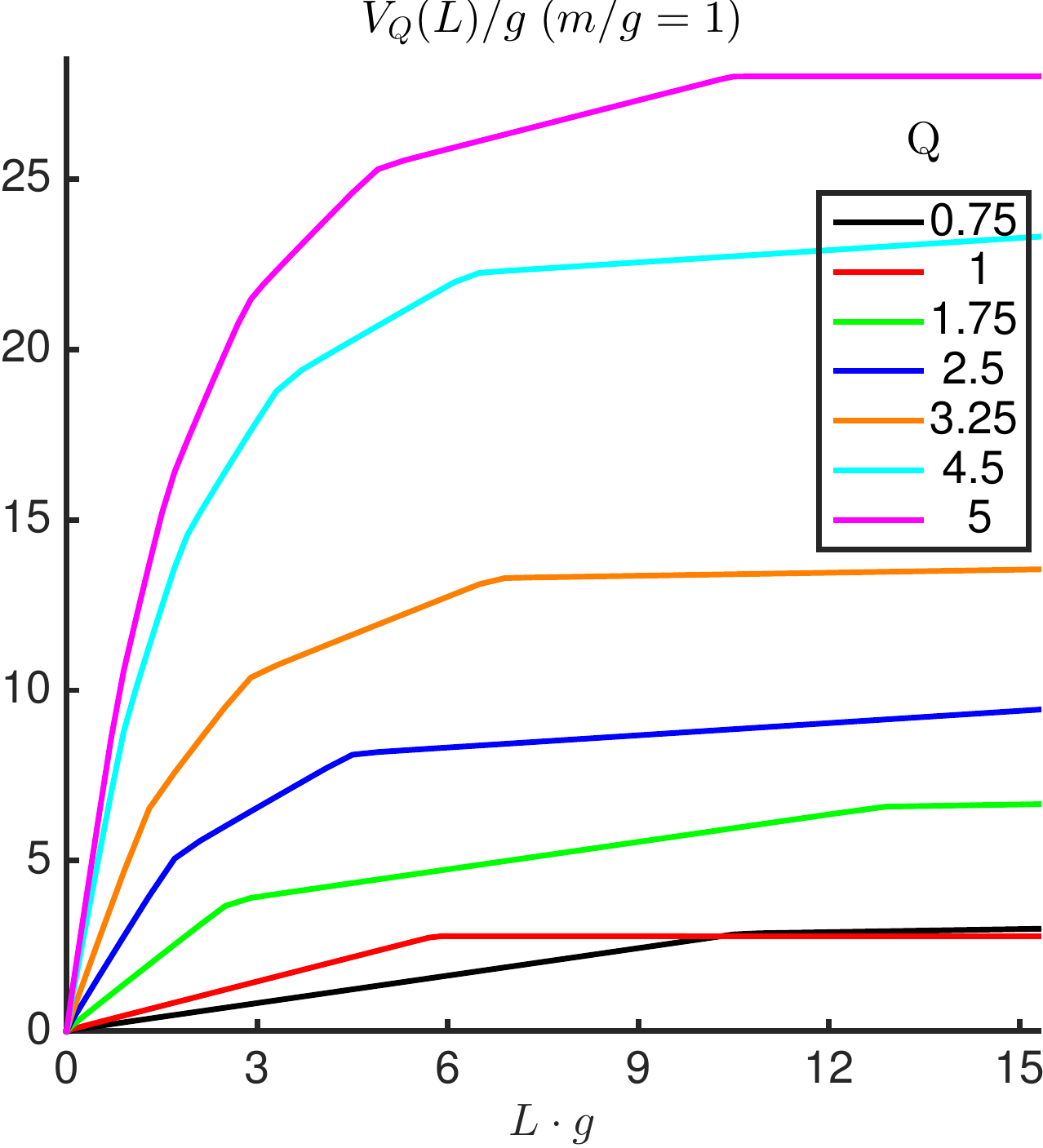}
\caption{\label{fig:StringBreakinga}}
\end{subfigure}\hfill
\begin{subfigure}[b]{.24\textwidth}
\includegraphics[width=\textwidth]{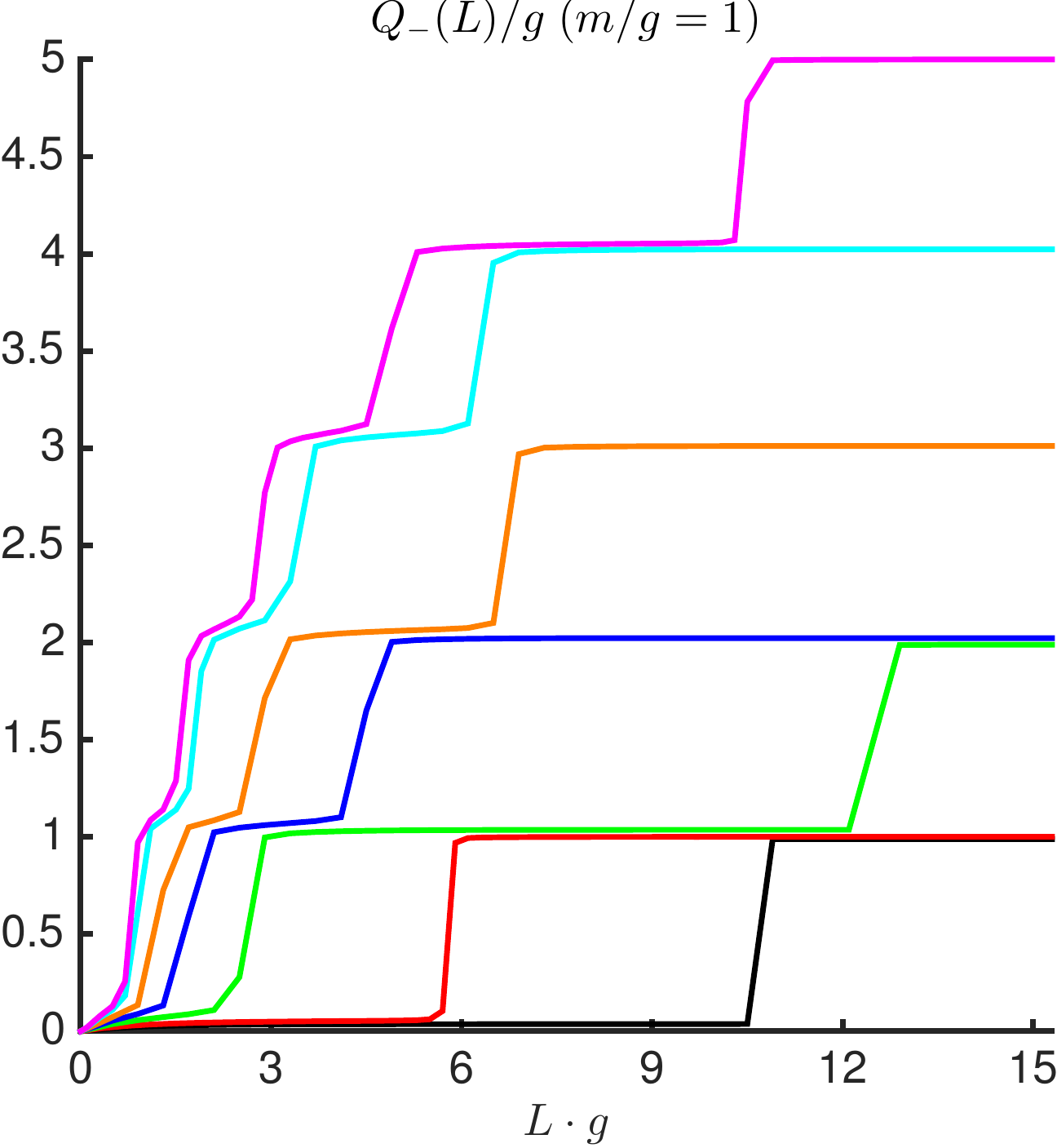}
\caption{\label{fig:StringBreakingb}}
\end{subfigure}\hfill
\vskip\baselineskip
\begin{subfigure}[b]{.24\textwidth}
\includegraphics[width=\textwidth]{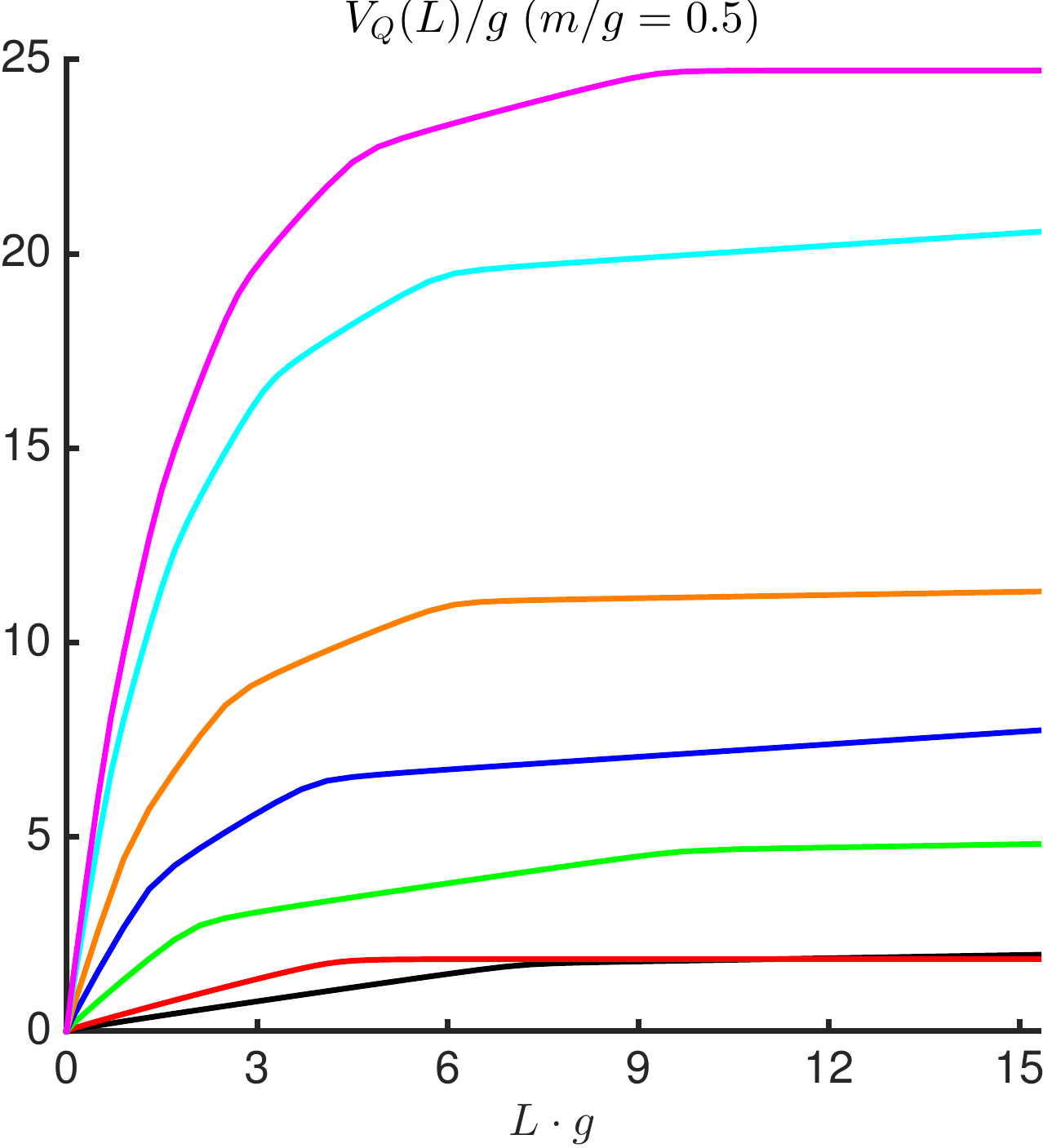}
\caption{\label{fig:StringBreakingc}}
\end{subfigure}\hfill
\begin{subfigure}[b]{.24\textwidth}
\includegraphics[width=\textwidth]{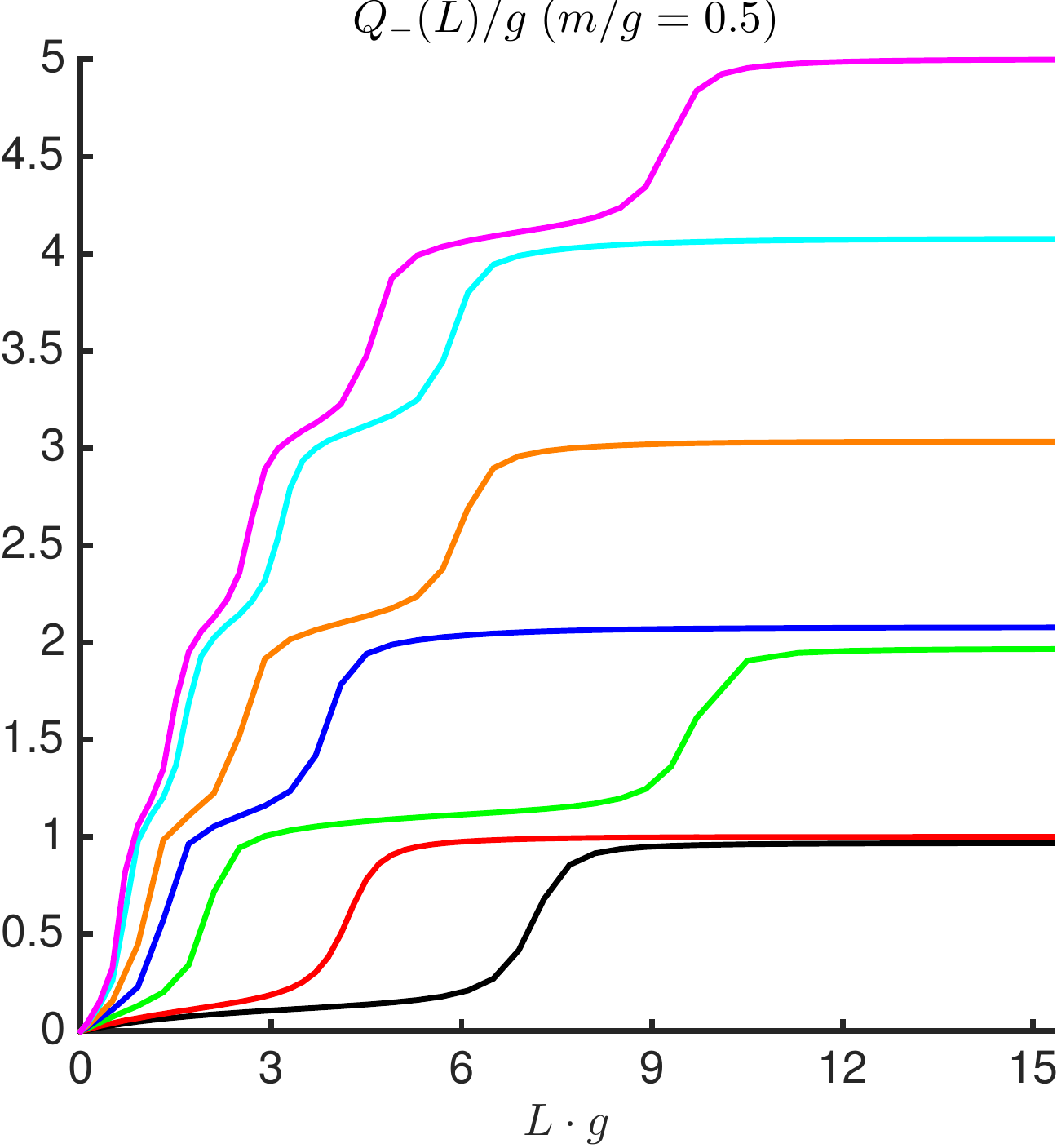}
\caption{\label{fig:StringBreakingd}}
\end{subfigure}
\vskip\baselineskip
\begin{subfigure}[b]{.24\textwidth}
\includegraphics[width=\textwidth]{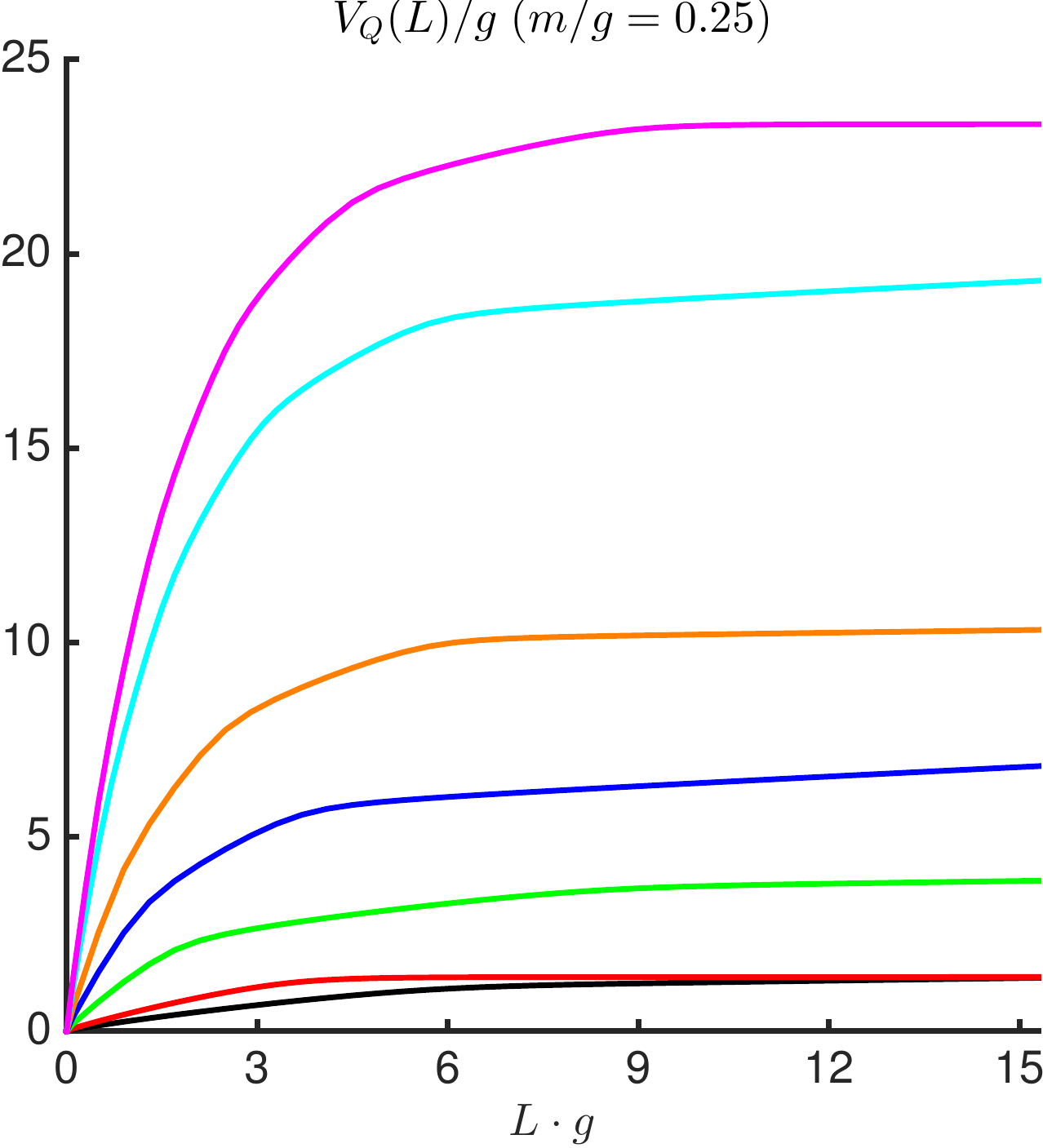}
\caption{\label{fig:StringBreakinge}}
\end{subfigure}\hfill
\begin{subfigure}[b]{.24\textwidth}
\includegraphics[width=\textwidth]{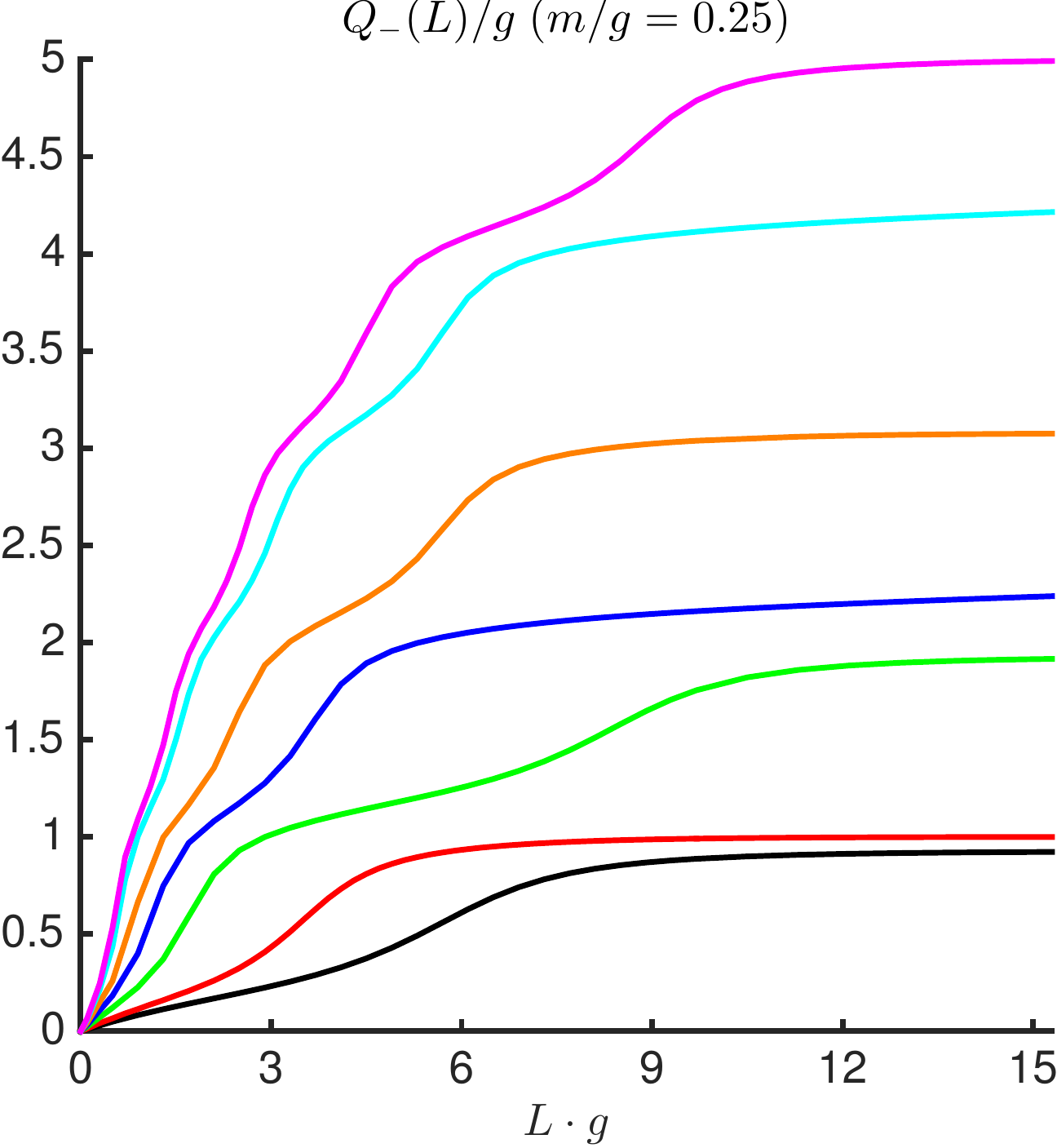}
\caption{\label{fig:StringBreakingf}}
\end{subfigure}
\captionsetup{justification=raggedright}
\vskip\baselineskip
\caption{\label{fig:StringBreaking}$x = 100.$ (a): Quark-antiquark potential for $m/g = 1$ for different values of $Q$. (b): $Q_{-}(L)$ for $m/g = 1$ for different values of $Q$. (c) and (d): the same quantities for $m/g=0.5$. (e) and (f): the same quantities for $m/g=0.25$  }
\end{figure}

Our simulations allow us to verify to what extent this picture is realized for finite $m/g$. In fig. \ref{fig:StringBreaking} we plot our results for different values of $Q$, both fractional and integer. We do indeed recover partial string breaking, largely following the nonrelativistic picture. To our knowledge this is the first successful simulation of partial string breaking in the Schwinger model, a previous Monte Carlo simulation \cite{Korcyl} failed to detect the phenomenon. 

In fig. \ref{fig:StringBreakingb} we plot, as in the previous section, the evolution of the total dynamical charge $Q_-$ at the negative $z$-axis, for $m/g=1$. For all values of $Q$ this charge $Q_-$ indeed makes quasidiscrete jumps of $\Delta Q_-\approx +1$ which should correspond to (partial) string breakings. As we see in fig. \ref{fig:StringBreakinga} these jumps indeed correlate with jumps in the string tensions in the different regions of the potentials. For $m/g=0.5$ we still find jumps of $Q_-$ but they are smoothened out, as can be seen in figs. \ref{fig:StringBreakingc} and \ref{fig:StringBreakingd}. For $m/g=0.25$ the jumps are even more smoothened out as can be seen in figs.  \ref{fig:StringBreakinge} and \ref{fig:StringBreakingf}. This smoothened behavior, similar to what we obtain in the $Q=1$ case, is expected as we go further from the nonrelativistic large $m/g$ regime. But still note the contrast with the behavior in the massless limit $m/g=0$ of subsection \ref{mg0}, where the charge $Q_-$ grows continuously to the external value $Q$, assuring a complete screening.

\begin{table}
\begin{tabular}{| c| | c | c | c | }
        \hline
        & \multicolumn{3}{c | }{$m/g$}\\
        \hline
     $Q$ &   0.25 &    0.5 & 1\\
     \hline
     \hline
   0.75& $2.5 (6)\times 10^{-5}$  & $5 (2) \times 10^{-7}$ & $-7 (2) \times 10^{-10}$\\
   1&$8 (2) \times 10^{-7}$   & $5 (2) \times 10^{-9}$ & $-4 (4) \times 10^{-11}$ \\
   1.75&$2.6 (7)  \times 10^{-4}$ & $1.7 (7) \times 10^{-5}$  &$8 (8) \times 10^{-8}$\\
   2.5&$3.0 (1)\times 10^{-3}$  &$2.5 (5) \times 10^{-6}$ & $-1 (1) \times 10^{-9}$ \\
   3.25&$2.2 (1) \times 10^{-5}$  &$2.1(8)  \times 10^{-7}$  &$-1 (1) \times 10^{-9}$ \\
   4.5&$4.0 (2)  \times 10^{-3}$  & $1.0 (2) \times 10^{-5}$ &$-1 (1)  \times 10^{-9}$ \\
   5&$2.1 (6) \times 10^{-4}$  & $1.0 (5) \times 10^{-5}$ &  $-2  (2) \times 10^{-8}$ \\
    \hline
\end{tabular}
\captionsetup{justification=raggedright}
\caption{\label{fig:tableVm} $x = 100$. Values for the difference $(\Delta V_Q/\Delta L - \sigma_Q)/g^2$ where  $\Delta V_Q/\Delta L$ is the mean of the backward differences at $Lg = 15.3$ with $\Delta L\; g =$ $0.4$,$0.8$,$1.2$,$1.6$.}
\end{table}

\begin{table}
\begin{tabular}{| c| | c | c | c | }
        \hline
        & \multicolumn{3}{c | }{$m/g$}\\
        \hline
     $Q$ &   0.25 &    0.5 & 1\\
     \hline
     \hline
   0.75    &      0.9225         &  0.9675&  0.9891\\
   1&   0.9995&  1.0000&   1.0000\\
   1.75&   1.9157&   1.9665 &  1.9891\\
   2.5&   2.2384&      2.0778 &  2.0223\\
   3.25&   3.0748&      3.0332&  3.0111\\
   4.5&   4.2150&   4.0770&  4.0230\\
      5 &   4.9922&   4.9990&   5.0000\\
    \hline
\end{tabular}
\captionsetup{justification=raggedright}
\caption{\label{fig:tableQm} $x = 100$. Values for $Q_{-}$ at $Lg = 15.3$ for $m/g = 0.25$, $m/g = 0.5$ and $m/g = 1$ .}
\end{table}

For $L$ going from 0 to $\infty$, different partial string breakings should lead to the asymptotic behavior of the potential that we examine in section \ref{section:String tension}. In table \ref{fig:tableVm} we show the difference of the slope of the potential around $Lg = 15.3$ with the asymptotic string tension at $x = 100$ that we calculate in the previous section. The former is estimated as the mean of  the backward differences 
\be\frac{1}{g^2} \frac{\Delta V_Q}{\Delta L} = \frac{V_Q(15.3 g) - V_Q(Lg)}{Lg} \left(\approx \frac{1}{g^2}\frac{dV_Q}{dL} \right)\ee
for $Lg =  13.7, 14.1, 14.5, 14.9$. The error is computed as the standard deviation of these backward differences. One observes that for $m/g = 1$ the string tension has already converged to the asymptotic result, almost up to the numerical precision,  while for $m/g=0.5$ we are already very close to the asymptotic result and for $m/g = 0.25$ there is a slightly larger (but still very small) difference.

For integer values of $Q$, the asymptotic string tension vanishes, so asymptotically we expect $Q_{-}\rightarrow Q$, corresponding to a complete screening.  For the values $Q=1$ and $Q=5$ that we consider, this is already almost satisfied at $Lg=15.3$, as can be seen in table \ref{fig:tableQm}. In the nonrelativistic limit for general $Q$, the total dynamical charge $Q_{-}$ that is produced asymptotically, will be the integer number that minimizes $|Q-Q_{-}|$.  For finite $m/g$ we expect corrections to the nonrelativistic limit, but as one can see in the table these corrections are still very small for $m/g=1$ and $m/g=0.5$. Notice also that for the half-integer values $Q=2.5$ and $4.5$, for which we have spontaneous symmetry breaking in the asymptotic limit (see section \ref{section:String tension}), we find $Q_{-}$ approaching the smallest of the two possible nonrelativistic values $Q_{-}\approx Q - 1/2$. 

\begin{figure}
\begin{subfigure}[b]{.24\textwidth}
\includegraphics[width=\textwidth]{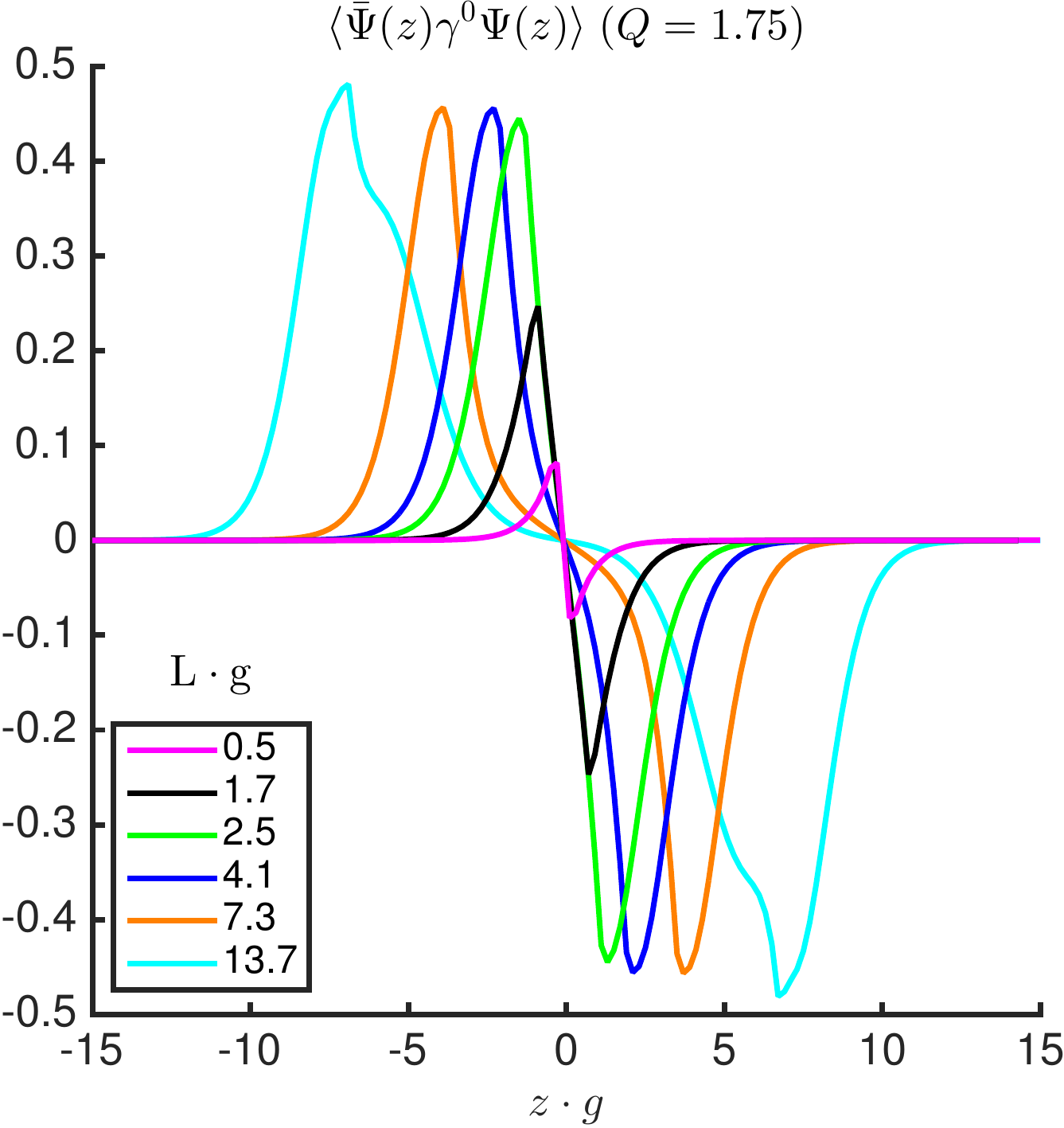}
\caption{\label{fig:ElFielda}}
\end{subfigure}\hfill
\begin{subfigure}[b]{.24\textwidth}
\includegraphics[width=\textwidth]{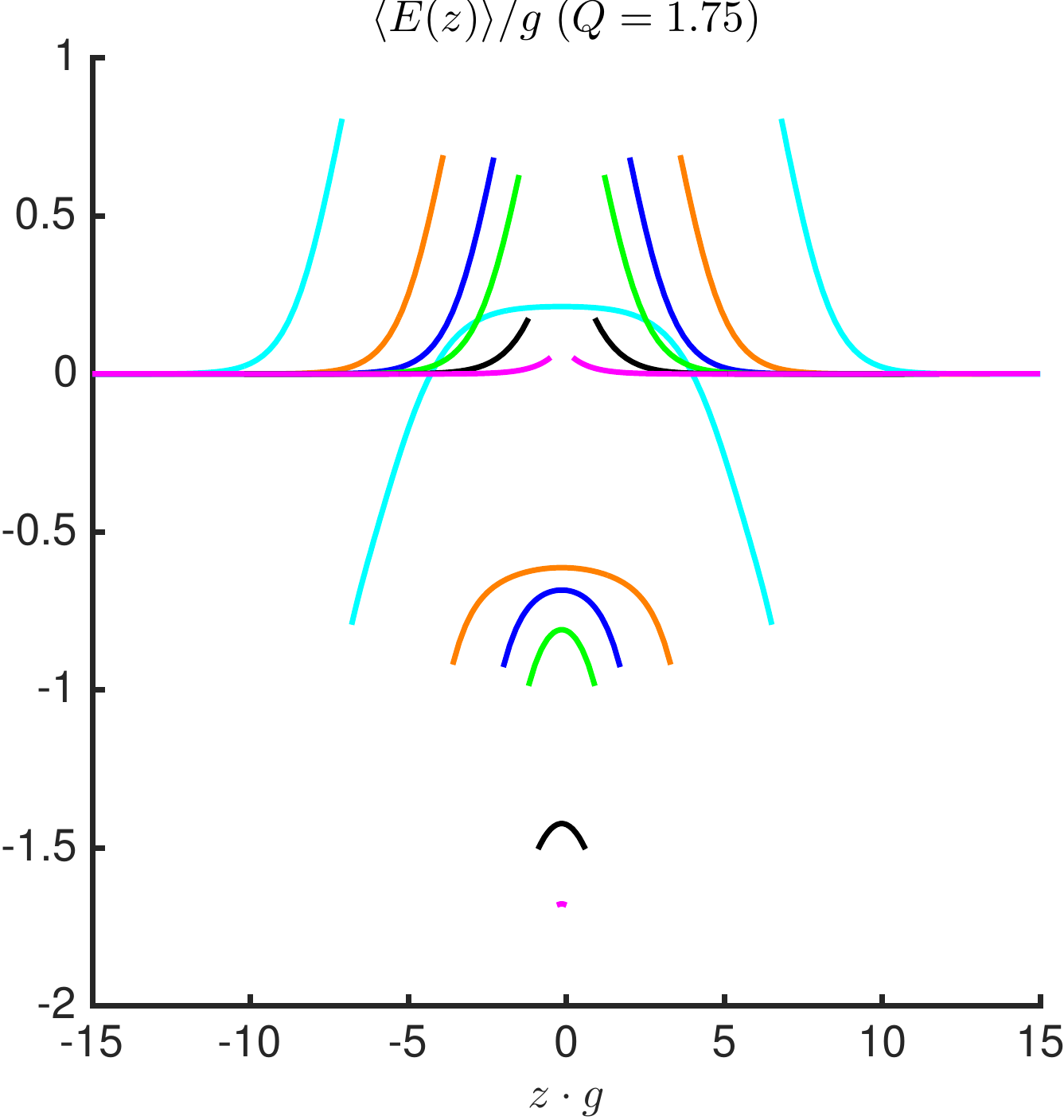}
\caption{\label{fig:ElFieldb}}
\end{subfigure}\hfill
\vskip\baselineskip
\begin{subfigure}[b]{.24\textwidth}
\includegraphics[width=\textwidth]{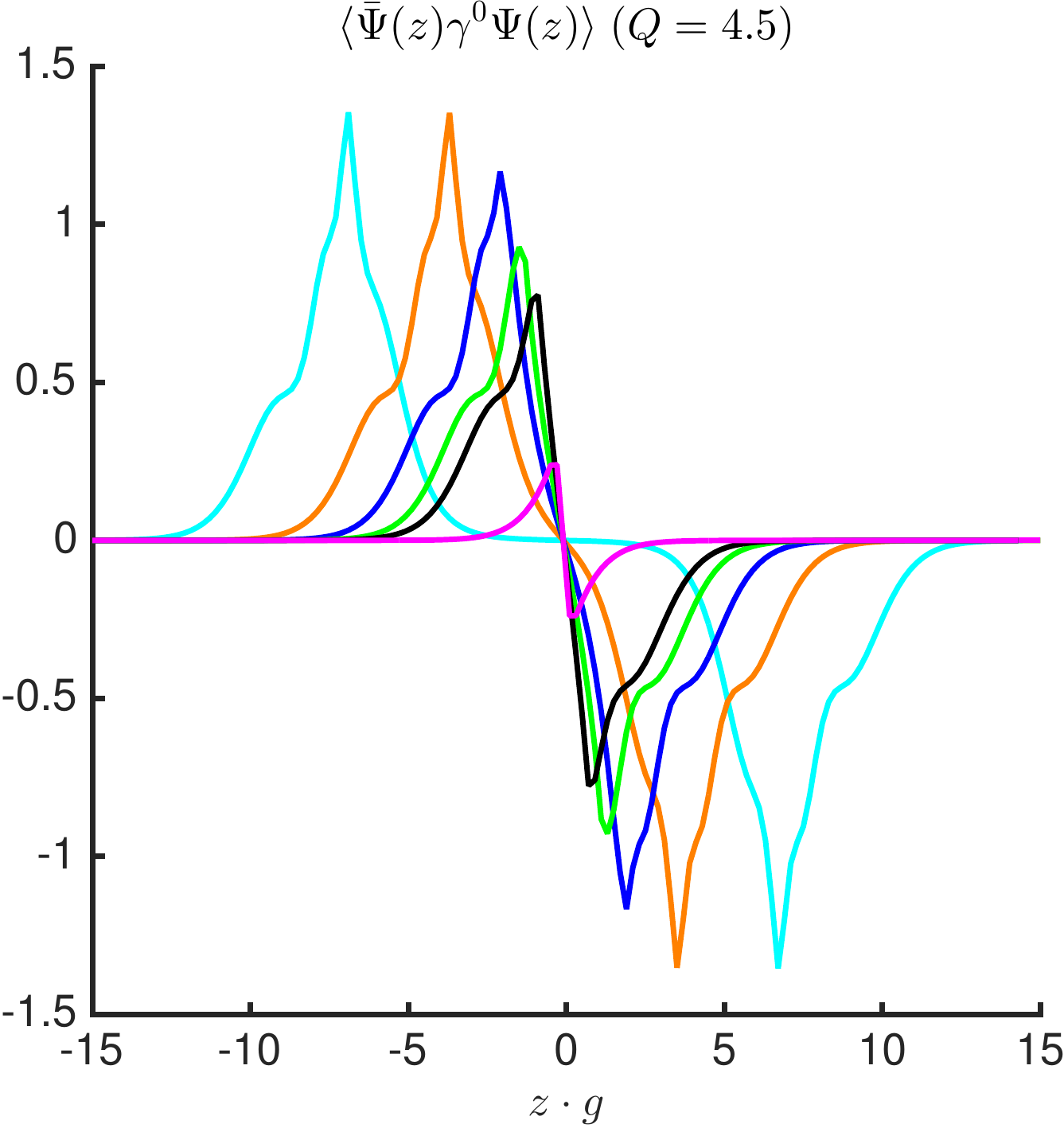}
\caption{\label{fig:ElFieldc}}
\end{subfigure}\hfill
\begin{subfigure}[b]{.24\textwidth}
\includegraphics[width=\textwidth]{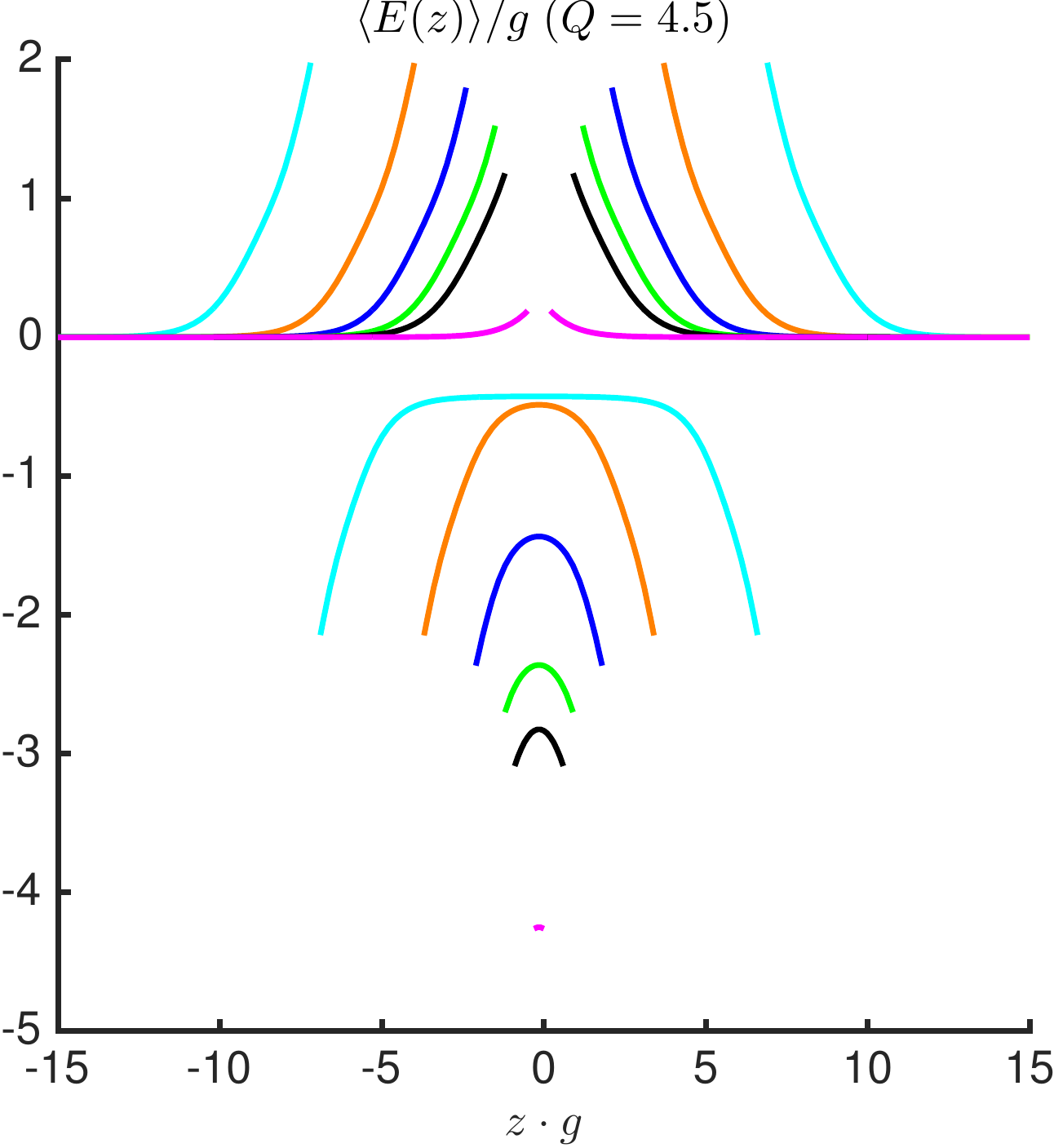}
\caption{\label{fig:ElFieldd}}
\end{subfigure}
\vskip\baselineskip
\begin{subfigure}[b]{.24\textwidth}
\includegraphics[width=\textwidth]{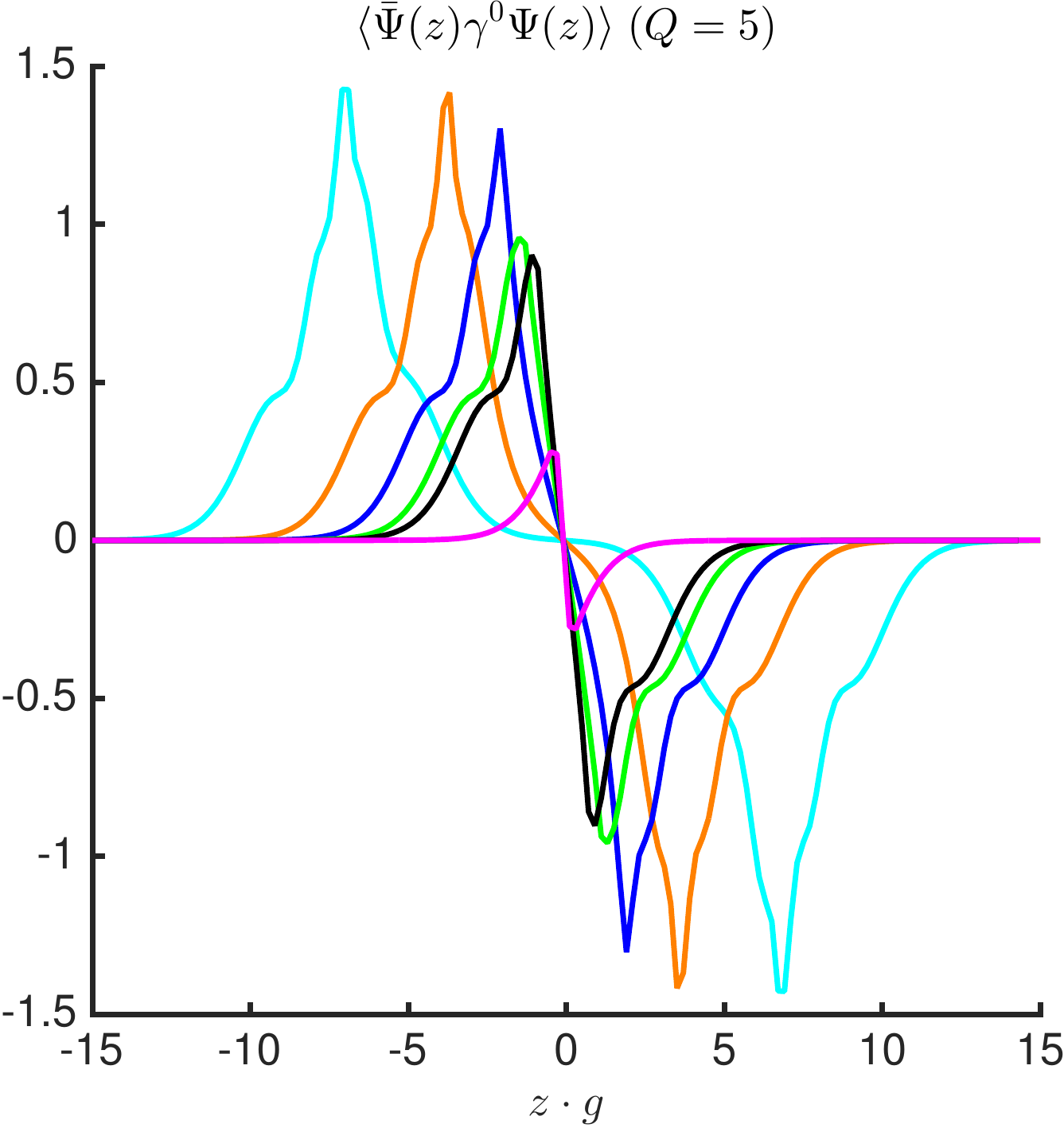}
\caption{\label{fig:ElFielde}}
\end{subfigure}\hfill
\begin{subfigure}[b]{.24\textwidth}
\includegraphics[width=\textwidth]{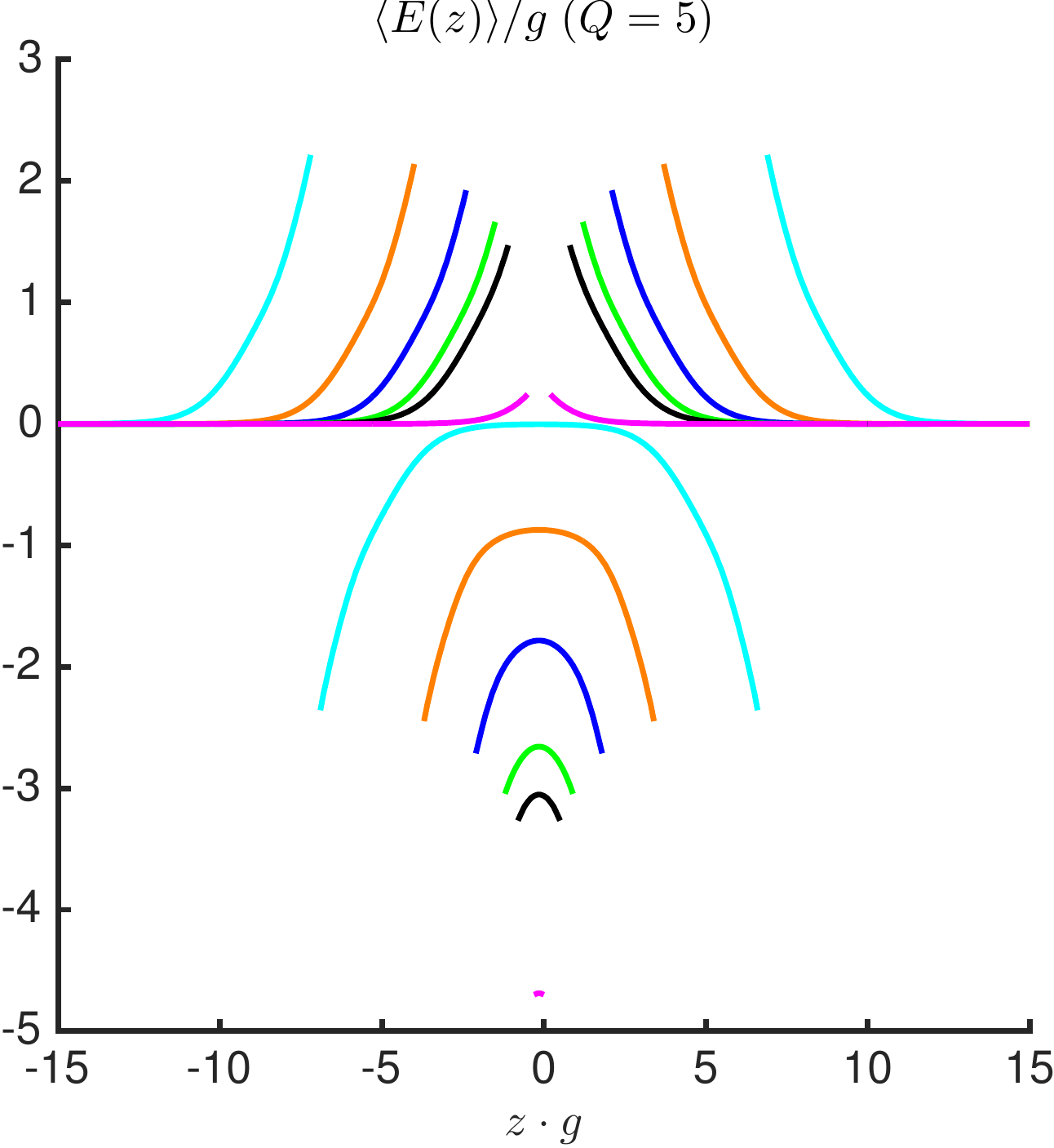}
\caption{\label{fig:ElFieldf}}
\end{subfigure}
\captionsetup{justification=raggedright}
\vskip\baselineskip
\caption{\label{fig:ElField}$m/g = 0.5, x = 100$. Left: Charge distribution for $Q = 1.75$ (a), $Q = 4.5$ (c) and $Q = 5$ (e) for different values of the separation length $L$. Right: Electric field for $Q = 1.75$ (b), $Q = 4.5$ (d) and $Q = 5$ (f) for different values of the separation length $L$.    }
\end{figure}

In fig. \ref{fig:ElField} we show the spatial charge distribution and electric field for different distances of the probe quarks. For $Q=1.75$ we have two partial string breakings. The first one, around $Lg\approx 1.7$ (see fig. \ref{fig:StringBreaking}) brings the electric field string at the center from $E/g\approx-1.7$ to $E/g\approx-0.7$. After the second partial string breaking, around $Lg\approx 9$, the probe charge is `overscreened', $Q_{-}\approx 2$, leading to a final electric field string with opposite sign $E/g \approx+ 0.2$. Notice that in contrast to the $Q=1$ case, the charge clouds at large separation of the probe quarks are not symmetric around the position of the probes. This is expected, as the remaining confining force between the two (charged) `mesons' distorts the charge distribution. For $Q=4.5$ we have a similar picture, but now, after the final partial string breaking, the probe charge is `underscreened', $Q_{-}\approx 4$, resulting in a final negative electric field string $E/g\approx -0.4$. While for $Q=5$ the final string breaking is complete: the probe charge is screened entirely $Q_{-}\approx 5$, leading to a complete neutralization of the electric field string $E/g\approx 0$ at the center. In this case for large enough $Lg$ we expect the charge distributions to become fully symmetric around the probe charge positions. 
\begin{figure}
\begin{subfigure}[b]{.24\textwidth}
\includegraphics[width=\textwidth]{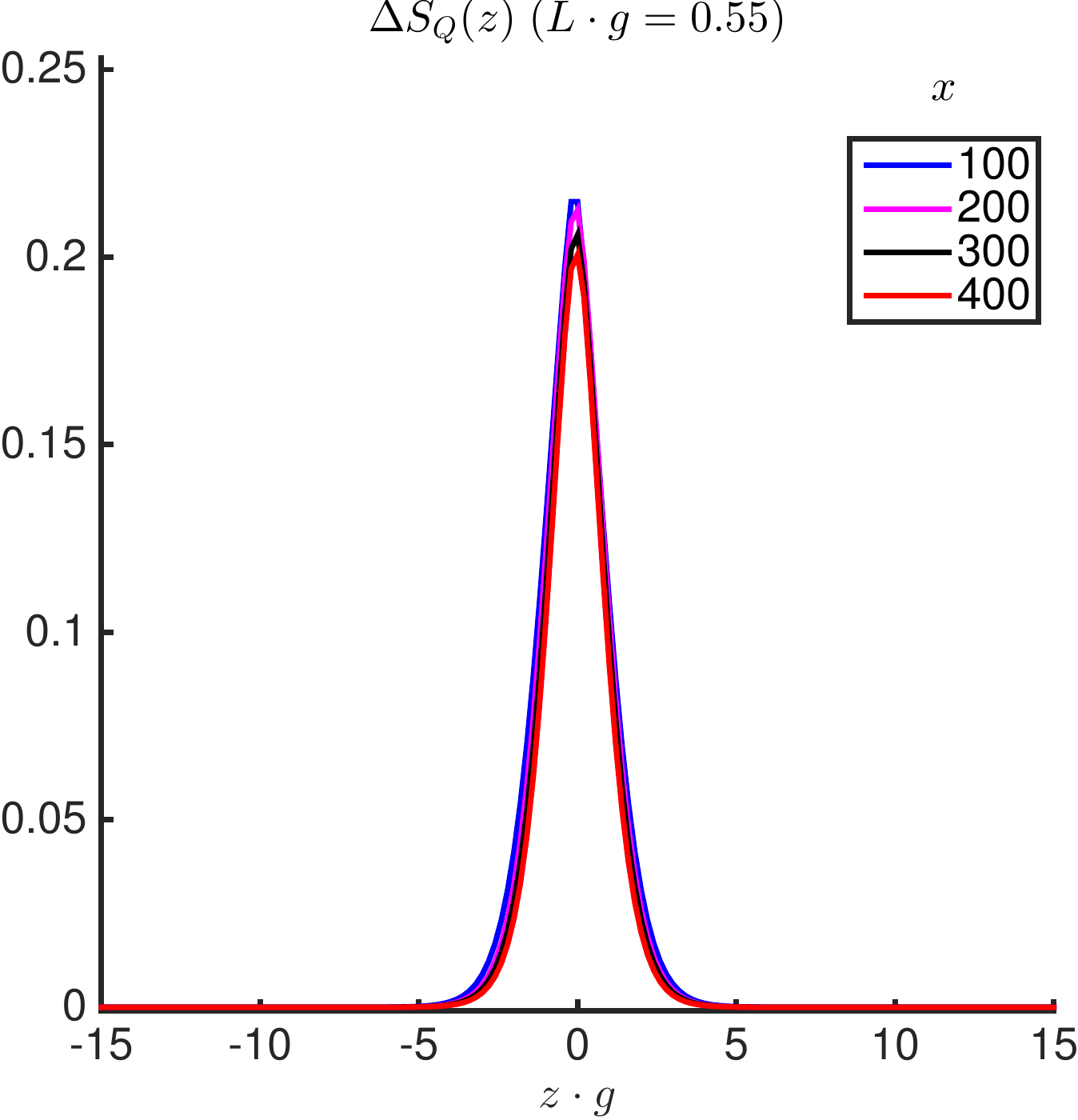}
\caption{\label{fig:mdivg50Entropya}}
\end{subfigure}\hfill
\begin{subfigure}[b]{.24\textwidth}
\includegraphics[width=\textwidth]{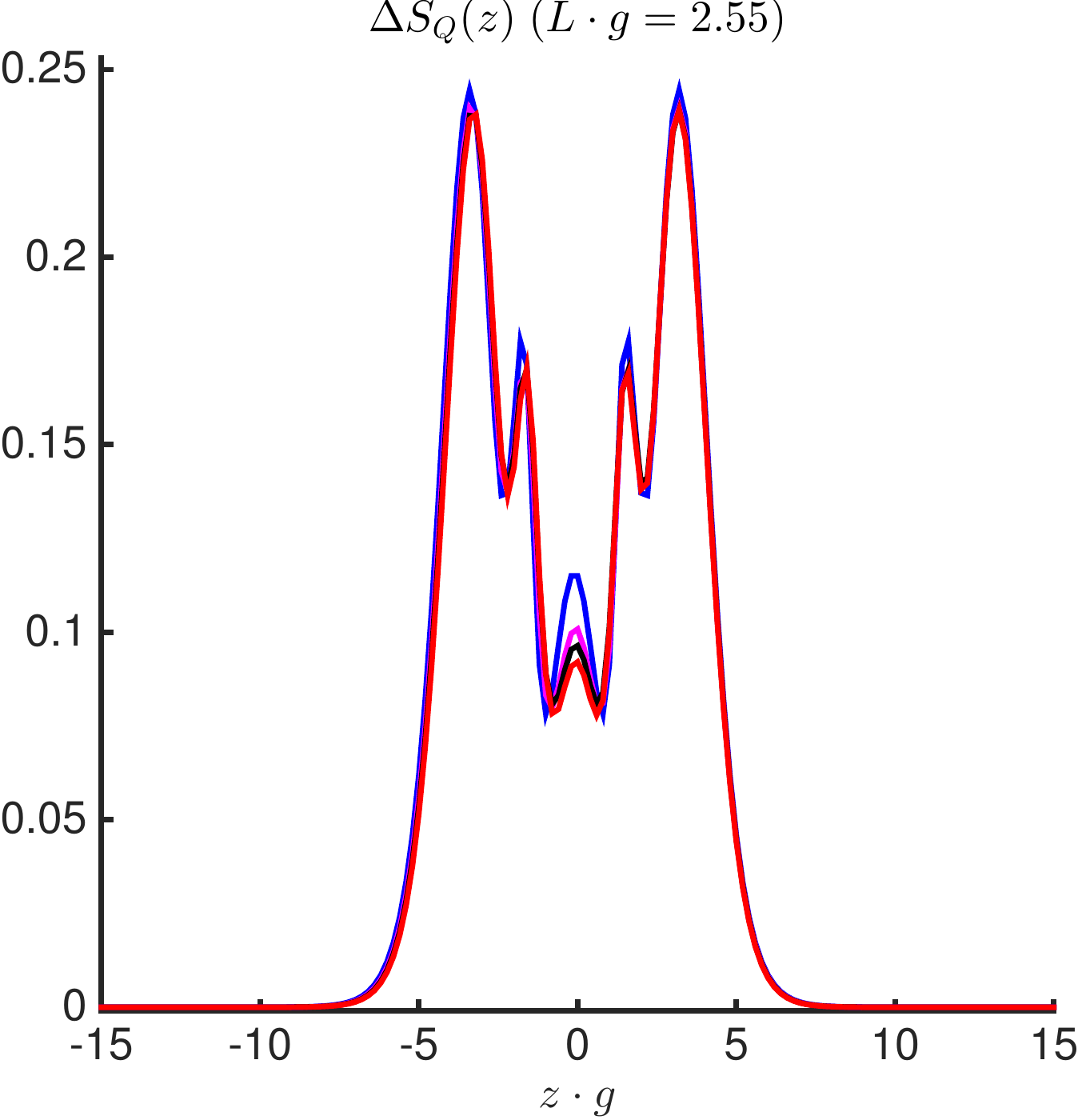}
\caption{\label{fig:mdivg50Entropyb}}
\end{subfigure}
\vskip\baselineskip
\begin{subfigure}[b]{.24\textwidth}
\includegraphics[width=\textwidth]{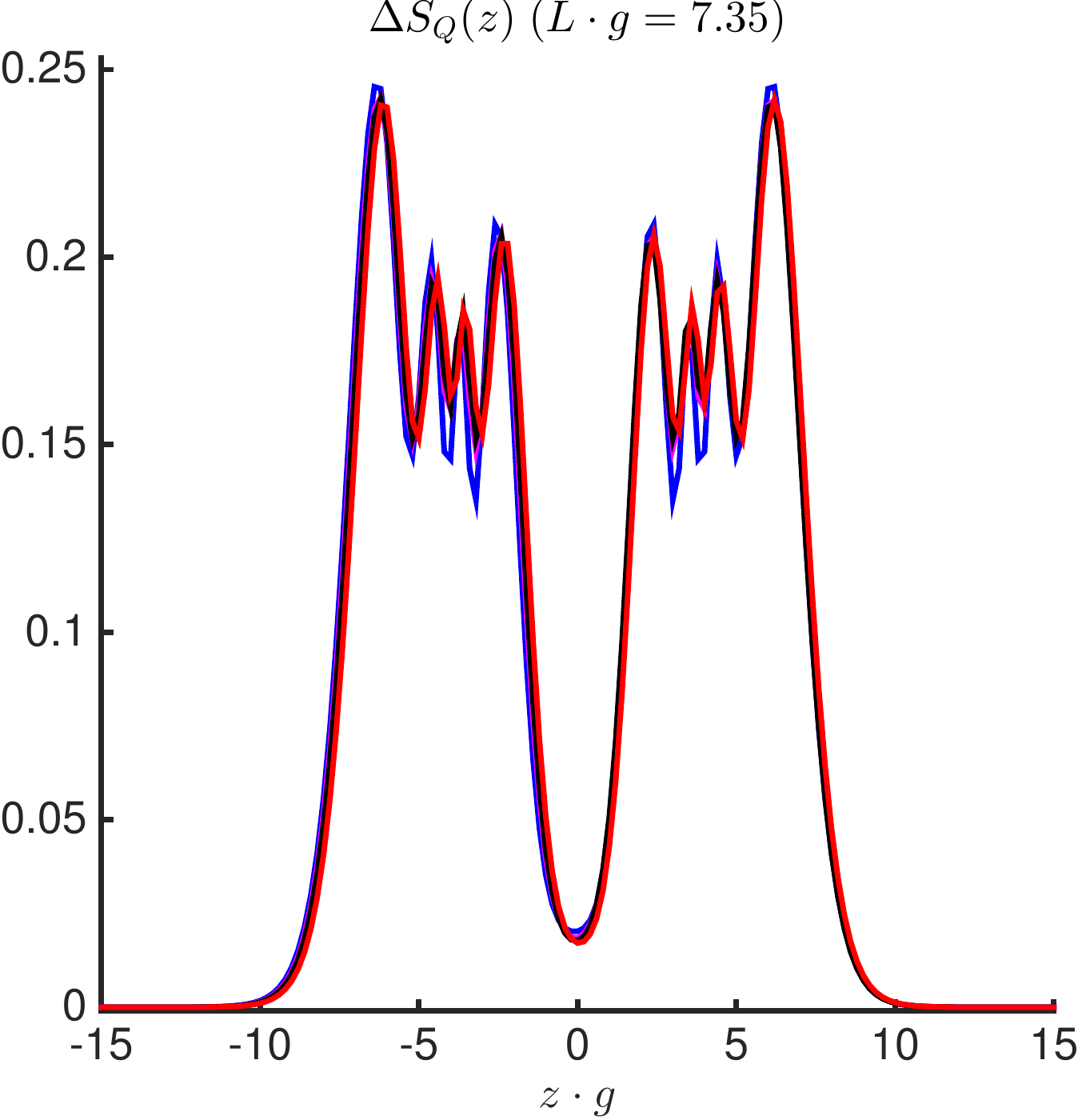}
\caption{\label{fig:mdivg50Entropyc}}
\end{subfigure}\hfill
\begin{subfigure}[b]{.24\textwidth}
\includegraphics[width=\textwidth]{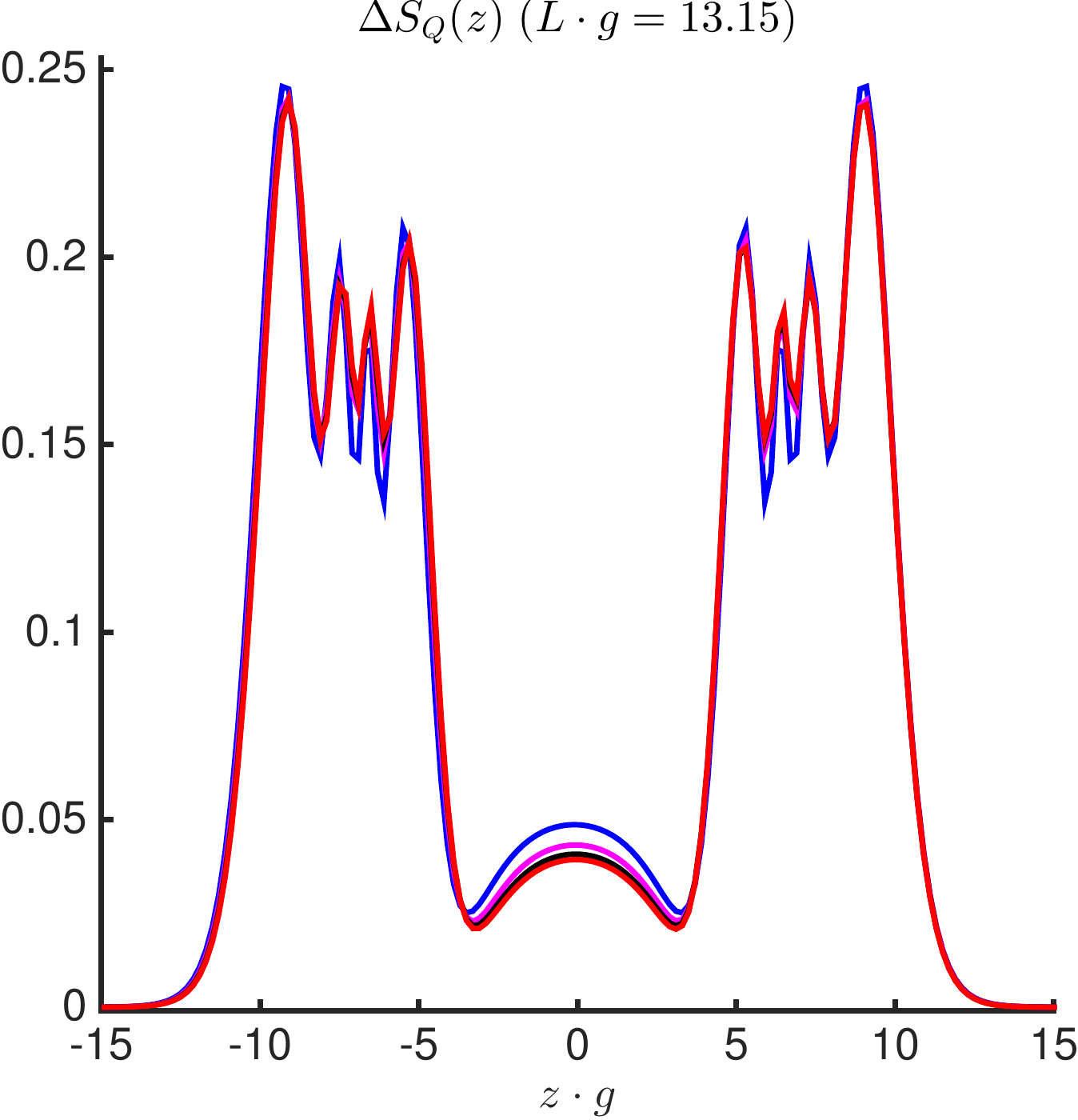}
\caption{\label{fig:mdivg50Entropyd}}
\end{subfigure}
\captionsetup{justification=raggedright}
\vskip\baselineskip
\caption{\label{fig:mdivg50Entropy} $m/g = 0.5$, $Q = 4.5$. $\Delta S_Q(z)$ for different values of $L$. We also show the scaling to $x \rightarrow + \infty$. (a) $Lg = 0.55$. (b) $Lg = 2.55$. (c) $Lg = 7.35$ (d) $Lg = 13.25$.}
\end{figure}

In fig. \ref{fig:mdivg50Entropy} we show the effect of different partial string breakings on the entropy profile $\Delta S_Q(z)$, for $m/g = 0.5$ and $Q = 4.5$. For the smallest interquark distance $Lg=0.55$, the entropy peaks at the center. At $Lg \gtrsim 2.55$ (after two string breakings, see fig. \ref{fig:StringBreaking}), we observe a profile with two peaks around the positions of the probe charges. At $Lg \gtrsim 7.35$ and $Lg \gtrsim 13.15$, after four string breakings, the profile now shows four peaks around the probe quark positions. In addition, we find an entropy surplus in the center, which now seems to be stable under the continuum extrapolation.

\begin{figure}
\begin{subfigure}[b]{.24\textwidth}
\includegraphics[width=\textwidth]{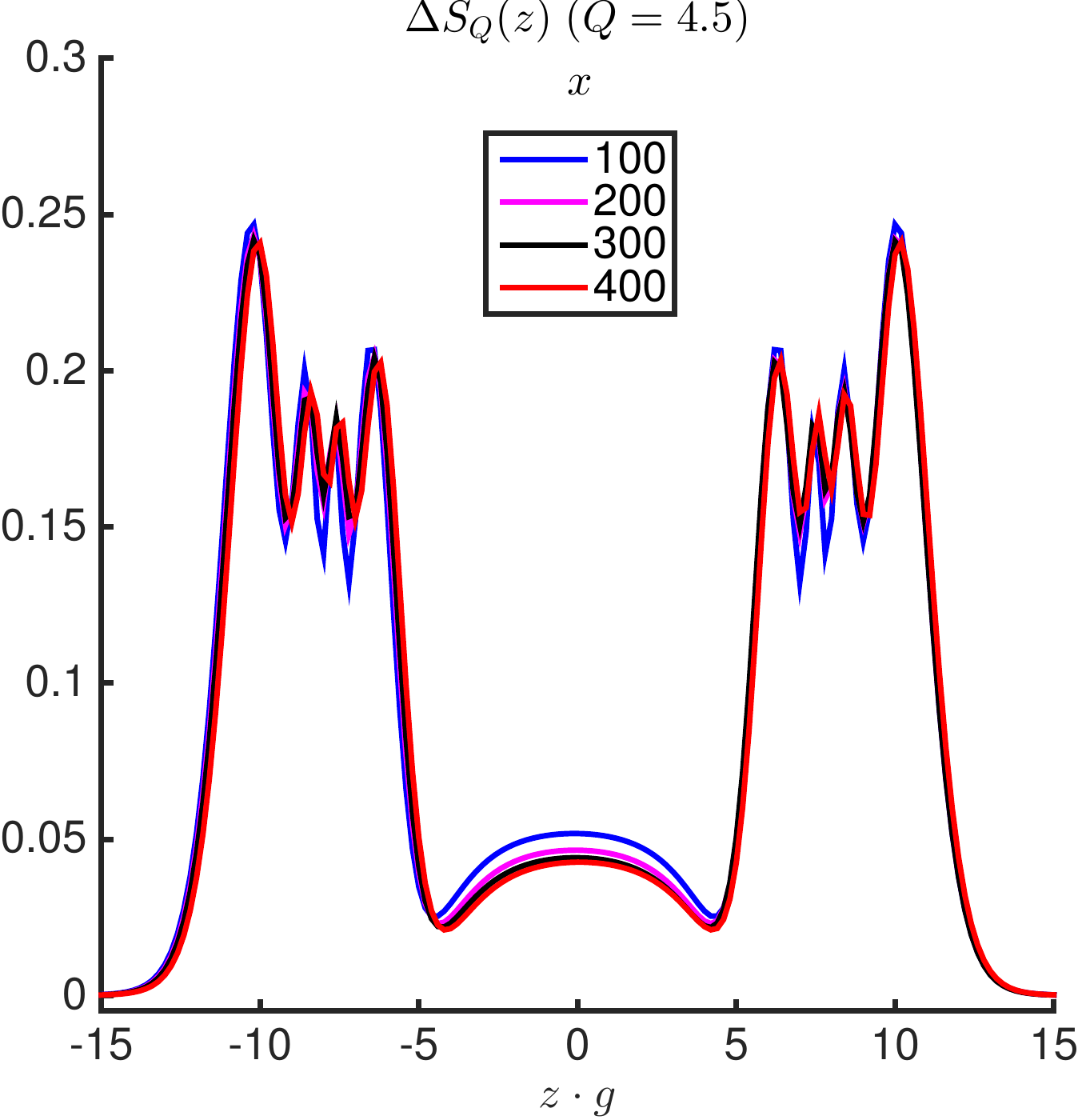}
\caption{\label{fig:mdivg50Entropya}}
\end{subfigure}\hfill
\begin{subfigure}[b]{.24\textwidth}
\includegraphics[width=\textwidth]{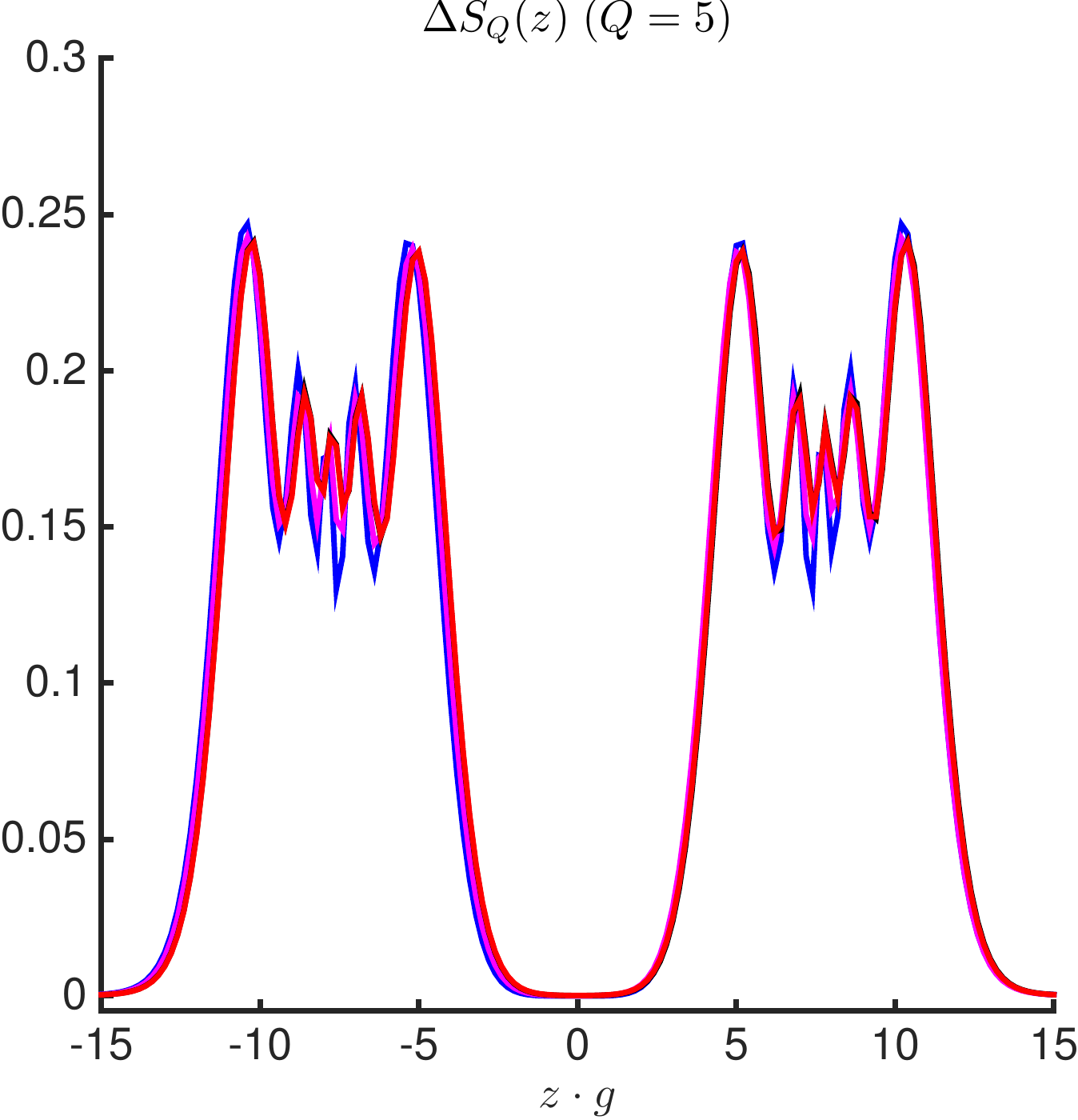}
\caption{\label{fig:mdivg50Entropyb}}
\end{subfigure}
\vskip\baselineskip
\begin{subfigure}[b]{.24\textwidth}
\includegraphics[width=\textwidth]{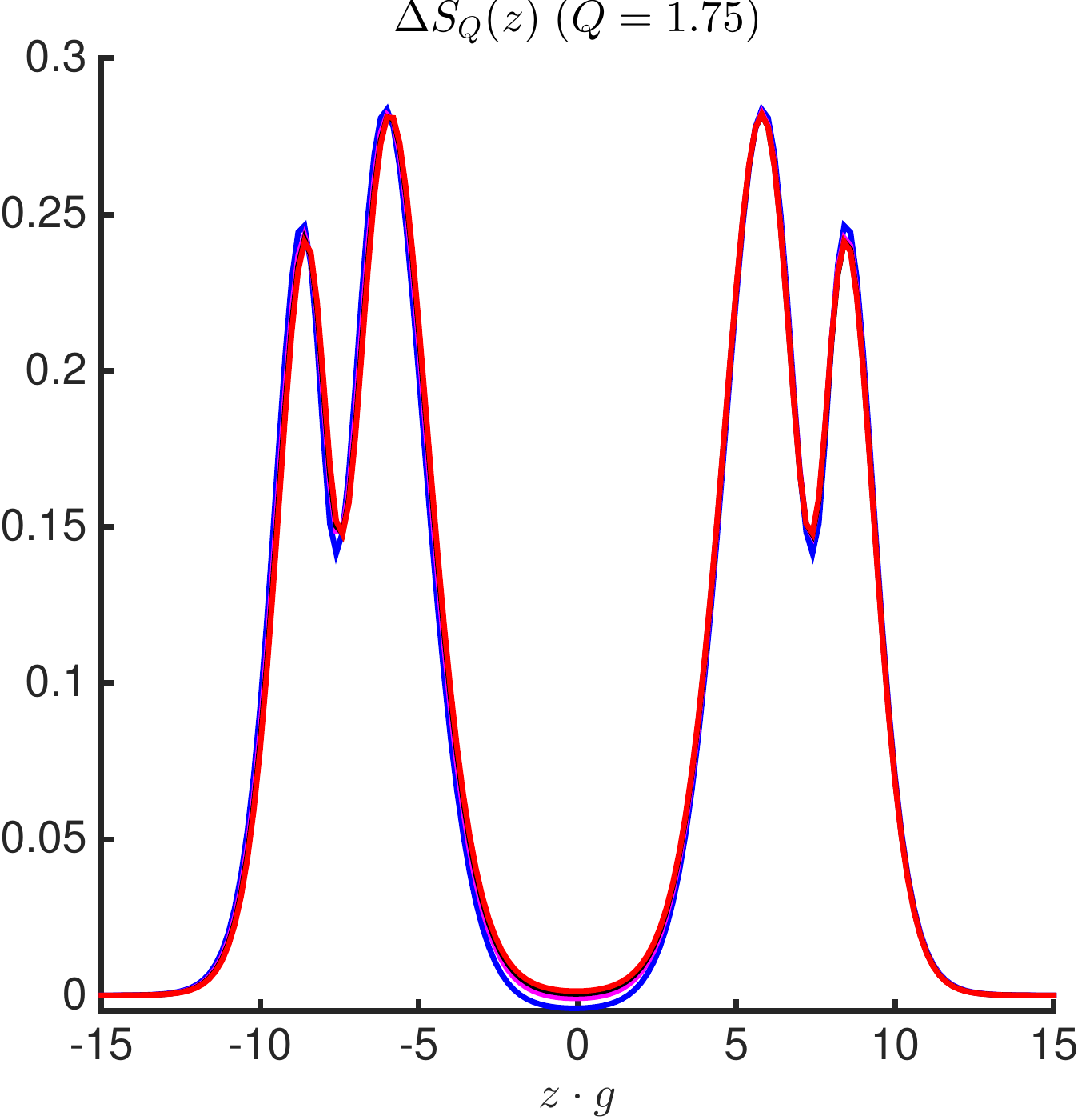}
\caption{\label{fig:mdivg50Entropyc}}
\end{subfigure}\hfill
\begin{subfigure}[b]{.24\textwidth}
\includegraphics[width=\textwidth]{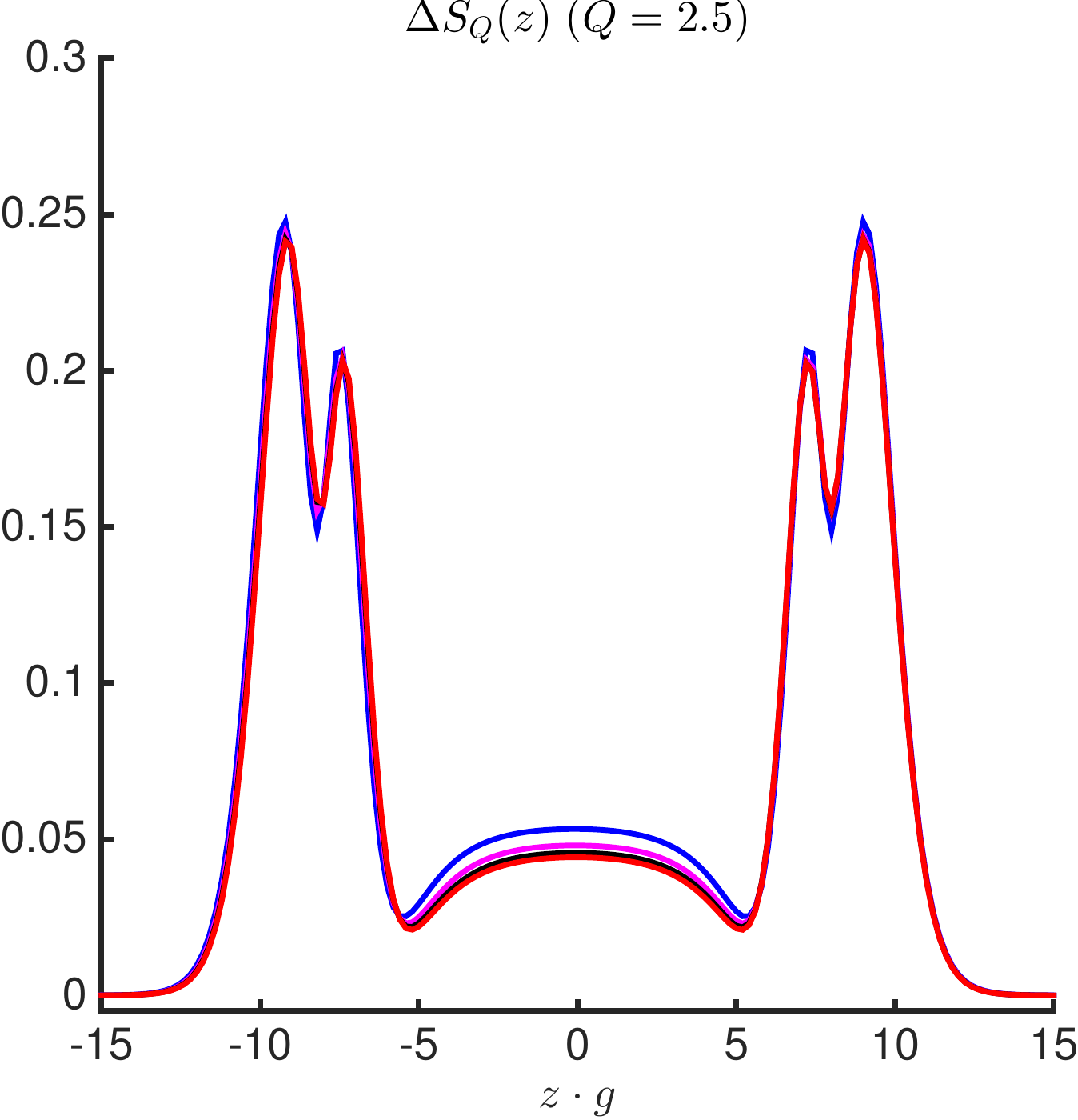}
\caption{\label{fig:mdivg50Entropyd}}
\end{subfigure}\hfill
\captionsetup{justification=raggedright}
\vskip\baselineskip
\caption{\label{fig:mdivg50EntropyDiffQ} $m/g = 0.5$, $L g = 15.25$. $\Delta S_Q(z)$ for different values of $Q$ and scaling to $x \rightarrow + \infty$. (a) $Q = 4.5$. (b) $Q = 5$. (c) $Q = 1.75$. (d) $Q = 2.5$. }
\end{figure}

In fig. \ref{fig:mdivg50EntropyDiffQ} we show that this characteristic imprint on the entropy is generic. We plot $\Delta S_Q(z)$ for $Lg = 15.25$ and different values of $Q$. For $Lg = 15.25$ all the partial string breakings have occurred and the final meson configurations around the external charge positions are formed. By counting the peaks one can again deduce the number of light elementary quarks (corresponding to the number of partial string breakings) in the meson states. For instance, for $Q = 4.5$, fig. \ref{fig:mdivg50Entropya}, we observe that there are four partial string breakings and for $Q = 5$, fig. \ref{fig:mdivg50Entropyb}, we observe that there are five partial string breakings. The spatial profiles do in fact differ by only one additional peak in each of the clouds for $Q = 5$ around $zg = \pm 5$.  Notice also the difference in the spatial profile for $Q = 1.75$ and $Q = 2.5$, see figs. \ref{fig:mdivg50Entropyc} and \ref{fig:mdivg50Entropyd}. In both cases two partial string breakings lead to the asymptotic meson state, but in the former case the final electric field is `overscreened' while in the latter case the final electric field is `underscreened'. Finally, notice that we can trust these results to be close to their continuum value, as the variation for the different $x$-values is very small. 

\section{Conclusions}\label{conclusion}
\noindent In this paper we employed the MPS formalism for a detailed numerical study of the confining mechanism in the static limit for the massive Schwinger model. Our Hamiltonian setup gives us direct access to the modified vacuum state in presence of two probe charges. This allowed us, not only to compute the interquark potential, but also the spatial profile of the electric field between the probe charges and the charge concentration of the light fermions. Even for relatively small $m/g$ the picture that emerged can be understood as a smoothened version of the nonrelativistic limit, with a level crossing between the electric string state that is the ground state at short distances and the broken-string two meson state that is the ground state at large distances. Here the two isolated mesons each consist of a light (anti-)quark cloud around the heavy probe charge, that is well described by the solution to the Schr\"{o}dinger equation of the appropriate one-particle problem. 

In the case of fractional probe charges, we clearly observed the expected partial string breaking. Again in accordance with the nonrelativistic picture we found the screening of the probe charges to happen in jumps $\Delta Q\approx 1$ of the light fermion charge; with these jumps becoming more and more discrete for growing $m/g$. 

Our tensor network simulations also give us direct access to the full Schmidt spectrum for the different bipartitions on the state. The numerical simulations show that the UV divergence in the corresponding von Neumann entropy is universal, allowing us to define a UV-finite renormalized entropy by subtracting the vacuum value. We have examined the imprint of both the string formation and string breaking on the profile of this renormalized entropy. Most notably we found that string breaking leaves a very distinct imprint on this entropy profile.   

We have checked our results not only against the predictions from the one-particle Schr\"{o}dinger equation (\ref{Schrodinger}), but also against the weak-coupling results from the original Lagrangian (\ref{Lagrangian}) and against the strong-coupling results from the bosonized field theory \cite{Coleman}. In the appropriate regimes we found nearly perfect agreement with these continuum analytic results. This not only demonstrates the potential of MPS simulations close to the continuum critical point of a lattice theory.  But it also serves as a nice, if not unexpected, cross-check of the consistency of all different descriptions of the Schwinger model.   

We have restricted ourselves to the study of the static limit of the confinement mechanism. An obvious future extension of our work is to consider the dynamical problem, simulating the real-time hadronization that takes place in a realistic scattering process. MPS real-time simulations of this type of problem were considered recently for $U(1)$ and $SU(2)$ quantum link lattice models in \cite{Montangero} and \cite{BanulsSB}. One could also approach this problem by first calculating the scattering eigenstates \cite{Vanderstraeten2014,Vanderstraeten2015a}. See also \cite{Hebenstreit2013a,Hebenstreit2013b} for an approach in the semiclassical limit.   

Of course it will also be very interesting to bring this type of analysis to higher dimensions. Specifically, present techniques should already allow one to simulate e.g., the static confinement problem for some simple 2+1 dimensional lattice models  (see \cite{gaugingStates} for homogeneous ground-state simulations of a $Z_2$ model and \cite{2015arXiv150708837Z} for simulations of a $U(1)$ lattice model with dynamical fermions). However, the current algorithms for PEPS simulations scale unfavorably with the bond dimension \cite{Lubasch2014} and we therefore expect that the successful simulation of specific microscopic gauge field Hamiltonians in the continuum limit will require new techniques. Still, given the continuous progress of PEPS methods \cite{Corboz2009,Jordan2008,Vanderstraeten2015b,Phien2015} we are hopeful on that front. In any case, in light of the potential for real-time and finite fermion density simulations, and of the new insight that might come from understanding the entanglement structure, it should certainly be worthwhile to further explore this direction, hopefully succeeding one day in the full simulation of the microscopic Hamiltonian of (3+1)-dimensional QCD.

\section*{Acknowledgements}
\noindent We acknowledge very interesting discussions with David Dudal. This work is supported by an Odysseus grant from the FWO, a PhD-grant from the FWO (B.B), a post-doc grant from the FWO (J.H.), the FWF grants FoQuS and Vicom, the ERC grant QUERG and the EU grant SIQS.

\bibliography{paperSB}

\newpage
\onecolumngrid
\appendix
\numberwithin{equation}{section}
\renewcommand\theequation{\Alph{section}.\arabic{equation}}

\section{Computation of the asymptotic string tension and electric field}\label{appasymptotic}
\noindent In this appendix we discuss the details of the computation of the asymptotic string tension $\sigma_Q$ and the electric field $E$. In subsection \ref{subsec:MPSansatz} we discuss how to obtain the electric field $E(x)$ and the energy density $\epsilon(x)$ for a fixed lattice spacing $ga = 1/\sqrt{x}$ and in subsection \ref{subsec:continuumExtra} we discuss how we to extrapolate these quantities to the continuum limit ($x \rightarrow + \infty$).

\subsection{MPS ansatz} \label{subsec:MPSansatz} 
\noindent To find the electric field $E(x)$ and the energy density $\epsilon(x)$ for a fixed lattice spacing $ga = 1/\sqrt{x}$ we need to find the ground state of the Schwinger Hamiltonian 
\be\label{Ham02} H_Q =  \frac{g}{2\sqrt{x}}\Biggl( \sum_{n \in \mathbb{Z}} [{L}(n) - Q]^2 + \frac{m}{g}\sqrt{x} \sum_{n \in \mathbb{Z}}(-1)^n(\sigma_z(n) + (-1)^{n}) + x \sum_{n \in \mathbb{Z}}(\sigma^+ (n)e^{i\theta(n)}\sigma^-(n + 1) + h.c.)\biggl) \ee
in an electric background field $Q \in [0,1[$. Taking into account the translation symmetry over two sites, we propose the following MPS ansatz in the thermodynamic limit, see eq. (\ref{uMPS}),
\be \Ket{\Psi\bigl(A(1),A(2)\bigl)} = \sum_{\bm{\kappa}} \bm{v}_L^\dagger \left(\prod_{n \in \mathbb{Z}}A_{\kappa_{2n-1}}(1)A_{\kappa_{2n}}(2n)\right) \bm{v}_R \ket{\bm{\kappa}},\ee 
with
\be \kappa_n = (s_n,p_n), s_n \in \{ -1,1\}, p_n \in \mathbb{Z}[p_{min}(n+1),p_{max}(n+1)], \ket{\bm{\kappa}} = \ket{\{\kappa_n\}_{n \in \mathbb{Z}}}. \ee
Gauss' law, $G(n)= L(n) - L(n-1) - (\sigma_z(n) + (-1)^n)/2$, imposes the following form for $A(1)$ and $A(2)$, see eq. (\ref{gaugeMPS}),
\be
{[A_{s,p}(n)]}_{(q,\alpha_q),(r,\beta_r)} =  {[a_{s,p}(n)]}_{\alpha_q,\beta_r}\delta_{q+(s+(-1)^n)/2,r}\delta_{r,p}, n = 1,2,
\ee
$q \in \mathbb{Z}[p_{min}(n),p_{max}(n)]; p,r \in \mathbb{Z}[p_{min}(n+1),p_{max}(n+1)]$,$s = \pm 1$, $\alpha_q = 1\ldots D_q(n)$, $\beta_r = 1\ldots D_r(n+1)$. The optimal approximation for the ground state within the class of MPS with fixed bond dimension is obtained by performing imaginary time evolution of the Schr\"{o}dinger equation using the time-dependent variational principle (TDVP) \cite{HaegemanTDVP}. When applying the TDVP, the sites $2n-1$ and $2n$ are blocked into one effective site $n$:
\be \bigl(A_{s_1,p_1}(1),A_{s_2,p_2}(2)\bigl)\rightarrow A_{s_1,p_1,s_2,p_2} = A_{s_1,p_1}(1)A_{s_2,p_2}(2)\ee
where
\be[A_{s_1,p_1,s_2,p_2}]_{(q,\alpha_q),(r,\beta_r)} = [a_{s_1,p_1,s_2}]_{\alpha_q,\beta_r}\delta_{p_2,p_1 + (s_2+1)/2}\delta_{p_2,r}\delta_{p_1,q + (s_1-1)/2}\label{BlockAGI} \ee
with $a_{s_1,p_1,s_2} \in \mathbb{C}^{D_q \times D_r}$, $D_q  = D_q(1)$. Note that $D_q = 0$ for $q < p_{min}$ and $q > p_{max}$ where $p_{min} = p_{min}(1)$ and $p_{max} = p_{max}(1)$. Gauss' law implies that $p_{min}(2) \in \{ p_{min}-1, p_{min}\}$ and  $p_{max}(2) \in \{p_{max} - 1,p_{max}\}$.
\\
\\
To check whether the obtained MPS is a good approximation we should look at the Schmidt values $\lambda_{q,\alpha_q}(n)$ associated with the bipartition $\{\mathcal{A}_1(2n) = (\mathbb{Z}[-\infty,2n], \mathcal{A}_2(2n) =  \mathbb{Z}[2n+1,+\infty]\}$ of the lattice. Translation symmetry implies that the Schmidt values are independent of the effective site $n$: $\lambda_{q,\alpha_q}^{2n} = \lambda_{q,\alpha_q}, \forall n$. Similar to (\ref{eq:MPSschmidtGauge}) we have then
\be \label{eq:MPSschmidtGaugeTDVP}\ket{\Psi(A)} = \sum_{q = p_{min}}^{p_{max}}\sum_{\alpha_q = 1}^{D_q}\sqrt{\lambda_{q,\alpha_q}} \ket{\psi_{q,\alpha_q}^{\mathcal{A}_1(2n)}} \ket{\psi_{q,\alpha_q}^{\mathcal{A}_2(2n)}}, \ee
where $\ket{\psi_{q,\alpha_q}^{\mathcal{A}_1(2n)}}$ (resp. $\ket{\psi_{q,\alpha_q}^{\mathcal{A}_2(2n)}}$) are orthonormal unit vectors in the tensor product of the local Hilbert spaces in the region $\mathcal{A}_1(2n)$ (resp. $\mathcal{A}_2(2n)$). The Schmidt values $\lambda_{q,\alpha_q}$, which are non-negative and sum to one, can be obtained as follows: assume $A^{\kappa_1,\kappa_2}$ is brought in a canonical form such that the matrices $r$ and $l$ corresponding to the right and left eigenvectors of the largest eigenvalue of the transfer matrix  \cite{HaegemanBMPS},
\be \sum_{\kappa_1,\kappa_2} A_{\kappa_1,\kappa_2}r[A_{\kappa_1,\kappa_2}]^\dagger = r, \sum_{\kappa_1,\kappa_2} [A_{\kappa_1,\kappa_2}]^\dagger lA_{\kappa_1,\kappa_2} = l, \left( \sum_{\kappa_k} = \sum_{s_k = -1,1}\sum_{p_k = p_{min}(k)}^{p_{max}(k)}, \kappa_k = (s_k,p_k) \right)\ee
are positive definite and diagonal. Here we assume that the largest eigenvalue of the transfer matrix is normalized to one. Then, because $A$ takes the form (\ref{BlockAGI}), $r$ and $l$ will also be degenerate in the eigenvalues of $L(2n)$: $[r]_{(q,\alpha_q);(r,\beta_r)} = r_{q,\alpha_q} \delta_{q,r}\delta_{\alpha_q,\beta_r}$, $[l]_{(q,\alpha_q);(r,\beta_r)} = l_{q,\alpha_q} \delta_{q,r}\delta_{\alpha_q,\beta_r}.$ The Schmidt values $\lambda_{q,\alpha_q}$ are now obtained by multiplying $r$ and $l$: $\lambda_{q,\alpha_q} = r_{q,\alpha_q}l_{q,\alpha_q}$ where $q \in \mathbb{Z}[p_{min},p_{max}]$ labels the eigenvalues of $L(2n)$ and $\alpha_q = 1\ldots D_q$ labels the variational freedom of the matrices $a^{s_1,p_1,s_2}$. 

As can be observed from eq. (\ref{eq:MPSschmidtGaugeTDVP}), truncating to a finite bond dimension thus corresponds to an effective truncation in the Schmidt decomposition of the ground state. Ideally one would want a distribution of $D_q$-values such that the smallest retained Schmidt value is more or less equal for each eigenvalue sector of $L(2n)$. Then if we want a reliable MPS approximation for the ground state, these smallest retained Schmidt values should be sufficiently small, which corresponds to taking $D_q$ sufficiently large. In  practice we do several simulations and adapted $D_q$ until the smallest Schmidt value in each eigenvalue sector of $L(2n)$ was of order $10^{-17}$, i.e. $\min_{\alpha_q}\lambda_{q,\alpha_q} \approx 10^{-17}$.  

\begin{figure}
\begin{subfigure}[b]{.48\textwidth}
\includegraphics[width=\textwidth]{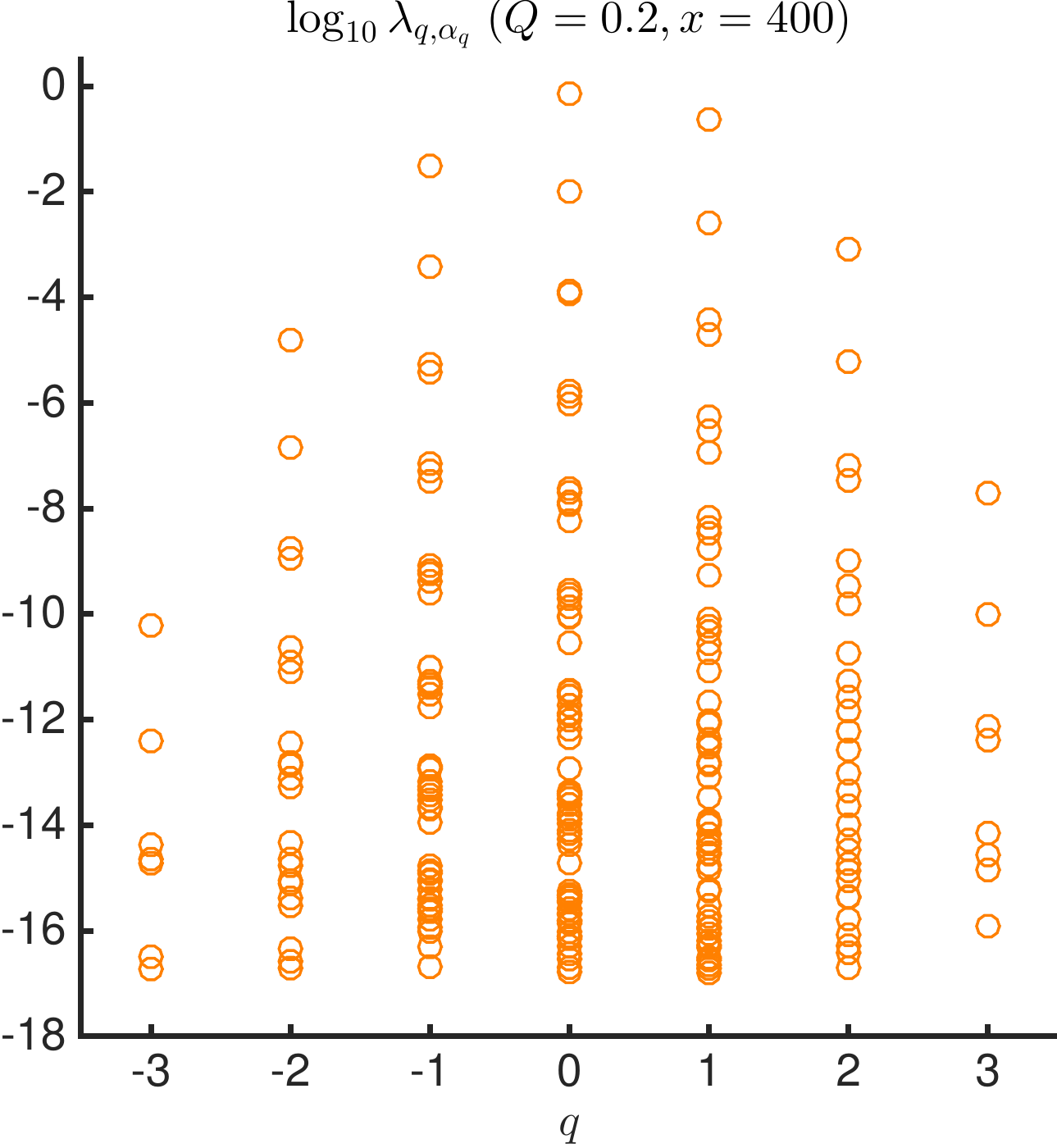}
\caption{\label{fig:STBDa}}
\end{subfigure}\hfill
\begin{subfigure}[b]{.48\textwidth}
\includegraphics[width=\textwidth]{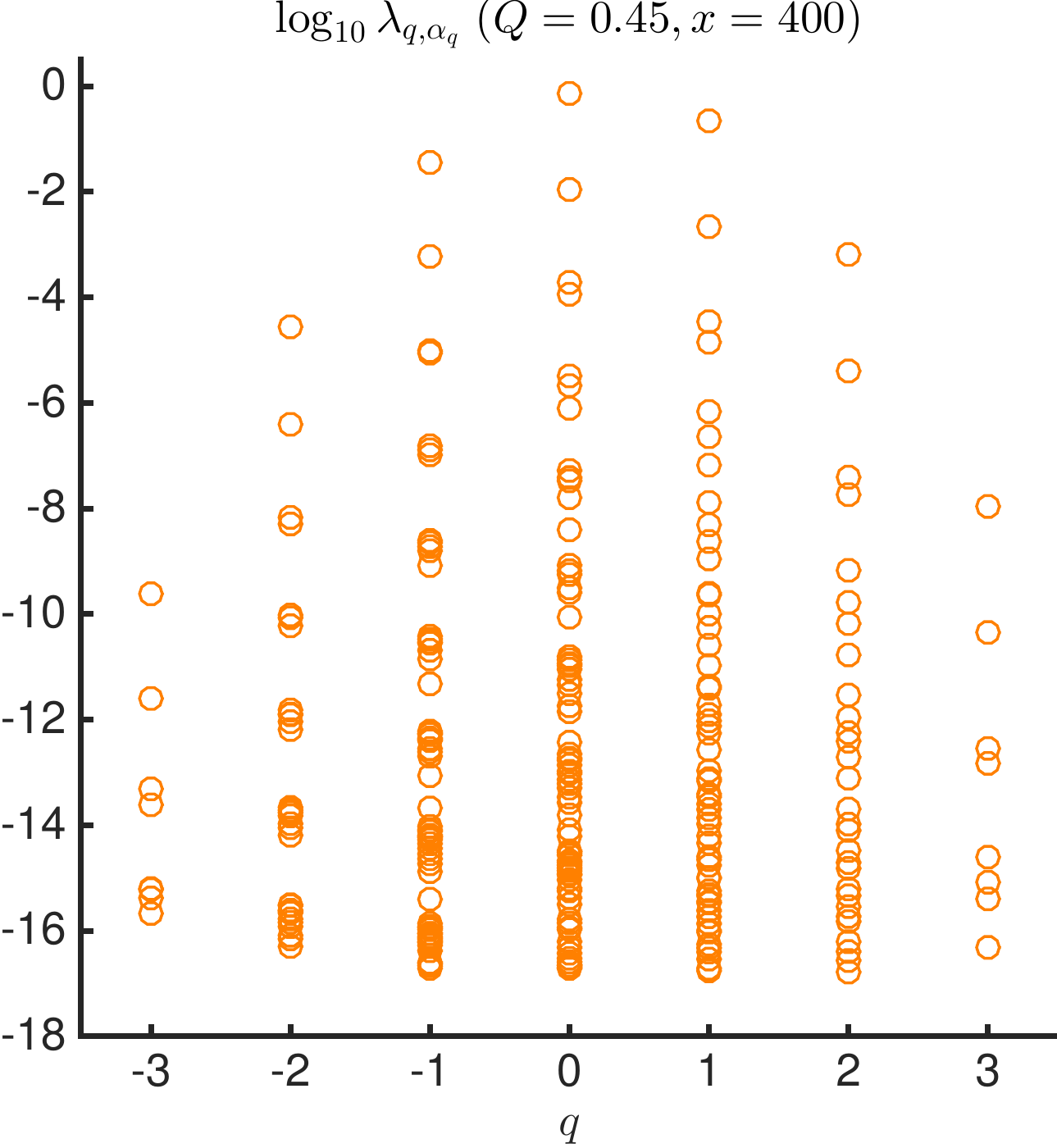}
\caption{\label{fig:STBDb}}
\end{subfigure}\vskip\baselineskip
\begin{subfigure}[b]{.48\textwidth}
\includegraphics[width=\textwidth]{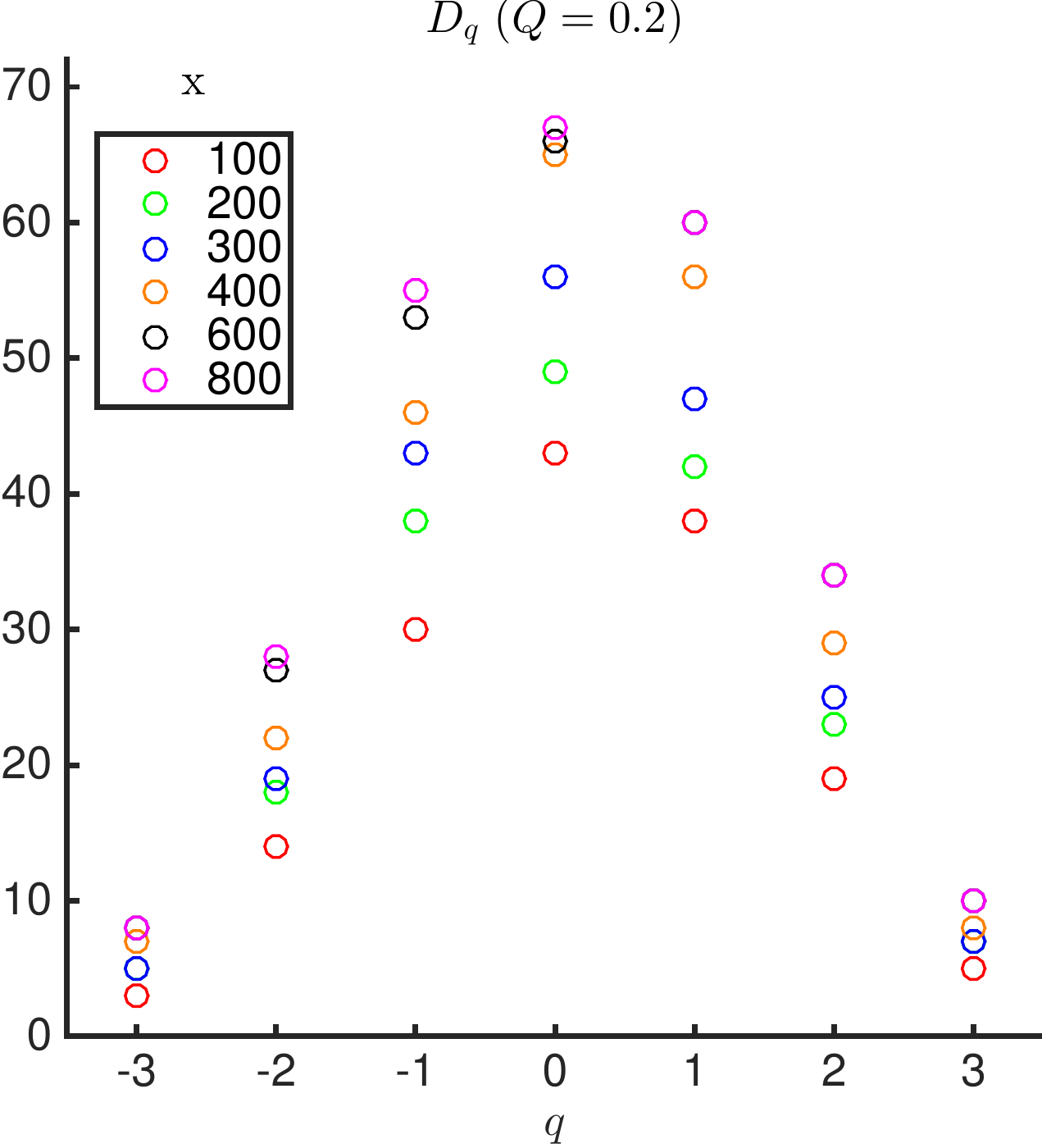}
\caption{\label{fig:STBDc}}
\end{subfigure}\hfill
\begin{subfigure}[b]{.48\textwidth}
\includegraphics[width=\textwidth]{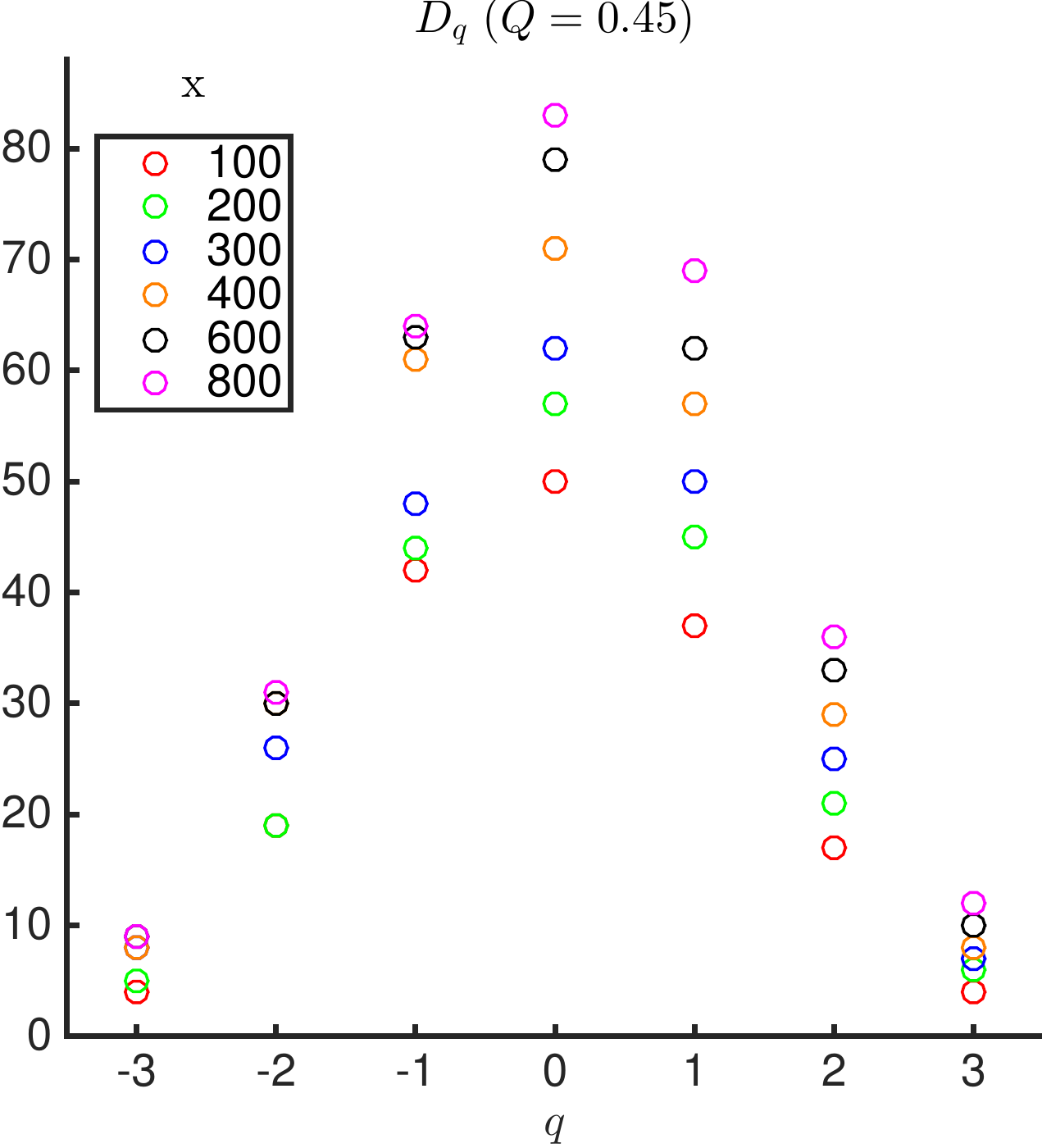}
\caption{\label{fig:STBDd}}
\end{subfigure}\vskip\baselineskip
\captionsetup{justification=raggedright}
\caption{\label{fig:STBD} $m/g = 0.75.$ (a): $Q = 0.2, x = 400.$ Distribution of the 10-base logarithm of the Schmidt values $\lambda_{q,\alpha_q}$ among the eigenvalue sectors $q$ of $L(2n)$. (b): Same as (a) but now for $Q  = 0.45$. (c):  $Q = 0.2.$ Distribution of the bond dimension among the eigenvalue sectors of $L(2n)$ for different values of $x$. (d):  Same as (c) but now for $Q = 0.45$.  }
\end{figure}

In figs. \ref{fig:STBDa} and \ref{fig:STBDb} we plot the distribution of the Schmidt values among the eigenvalue sectors of $L(2n)$ for the final MPS ground-state approximations for $m/g = 0.75, x = 400$ and $Q = 0.2, 0.45$. As in \cite{Buyens}, we observe that the sectors corresponding to $q = 0,-1,1$ are the most dominant ones which justifies our choice of taking $D_q = 0$ for $\vert q \vert > 3$. This can be understood physically from the term proportional to $[L(n) - Q]^2$ in (\ref{Ham02}) which punishes large eigenvalues of $L(n)$. We also display the bond dimensions for each sector and for each simulated value of $x$ in figs. \ref{fig:STBDc} and \ref{fig:STBDd}.  One can observe that as $
 x$ increases we need larger $D_q$ for the same accuracy. This is explained by the fact that the correlation length diverges as we approach the continuum limit ($x \rightarrow + \infty$) and it is well known that critical theories require larger bond dimensions for a good MPS approximation. For the same reason we also need larger $D_q$ when we are getting closer to the phase transition at $m/g = (m/g)_c \approx 0.33$ and $Q = 1/2$ \cite{Coleman,Hamer}. 

\subsection{Continuum extrapolation of the string tension and the electric field}\label{subsec:continuumExtra}
\noindent In the second part of this appendix we discuss how we obtain an estimate for the continuum value of $\sigma_Q$ and $E_Q$. Note that the string tension at $x = 1/g^2a^2$ is obtained from the energy density by:
\be \sigma_Q(x) = \sqrt{x}(\epsilon_Q(x) - \epsilon_0(x))\ee
where $\epsilon_Q(x)$ is the ground-state energy per site of the Schwinger Hamiltonian (\ref{Ham02}). As for $x \rightarrow \infty$, $H/(2g\sqrt{x})$ reduces to the $XY$-model we have that
\be\label{limxED} \lim_{x \rightarrow + \infty}\frac{\epsilon_Q(x)}{2g\sqrt{x}} =  \lim_{x \rightarrow + \infty}\frac{\epsilon_0(x)}{2g\sqrt{x}} = \frac{-1}{\pi} \ee
and it is argued in \cite{Hamer} that $\epsilon_Q(x)/\sqrt{x}$ should behave polynomially as a function of $1/\sqrt{x}$ for large $x$, we have:
\begin{subequations}\label{tensionx}
\be\sqrt{x} \frac{\epsilon_Q(x)}{g} = -\frac{2x}{\pi} + C_Q \sqrt{x} + A_Q + \mathcal{O}\left(\frac{1}{\sqrt{x}}\right) \; (x \gg 1), \ee
\be \sqrt{x}\frac{\epsilon_0(x)}{g} = -\frac{2x}{\pi} + C_0 \sqrt{x} + A_0 + \mathcal{O}\left(\frac{1}{\sqrt{x}}\right) \; (x \gg 1). \ee
\end{subequations}

\begin{figure}
\begin{subfigure}[b]{.48\textwidth}
\includegraphics[width=\textwidth]{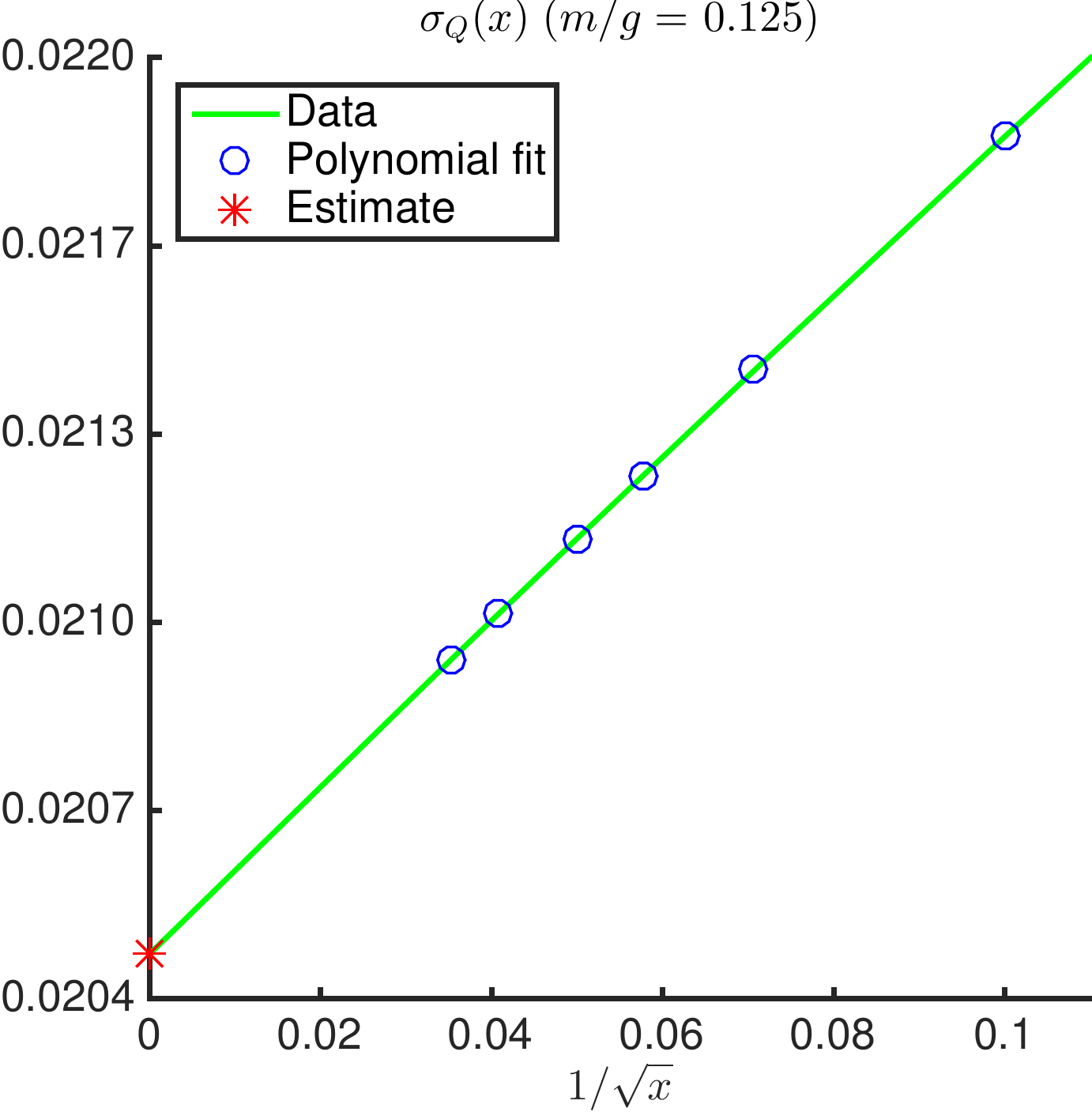}
\caption{\label{fig:ExtrapolationSTa}}
\end{subfigure}\hfill
\begin{subfigure}[b]{.48\textwidth}
\includegraphics[width=\textwidth]{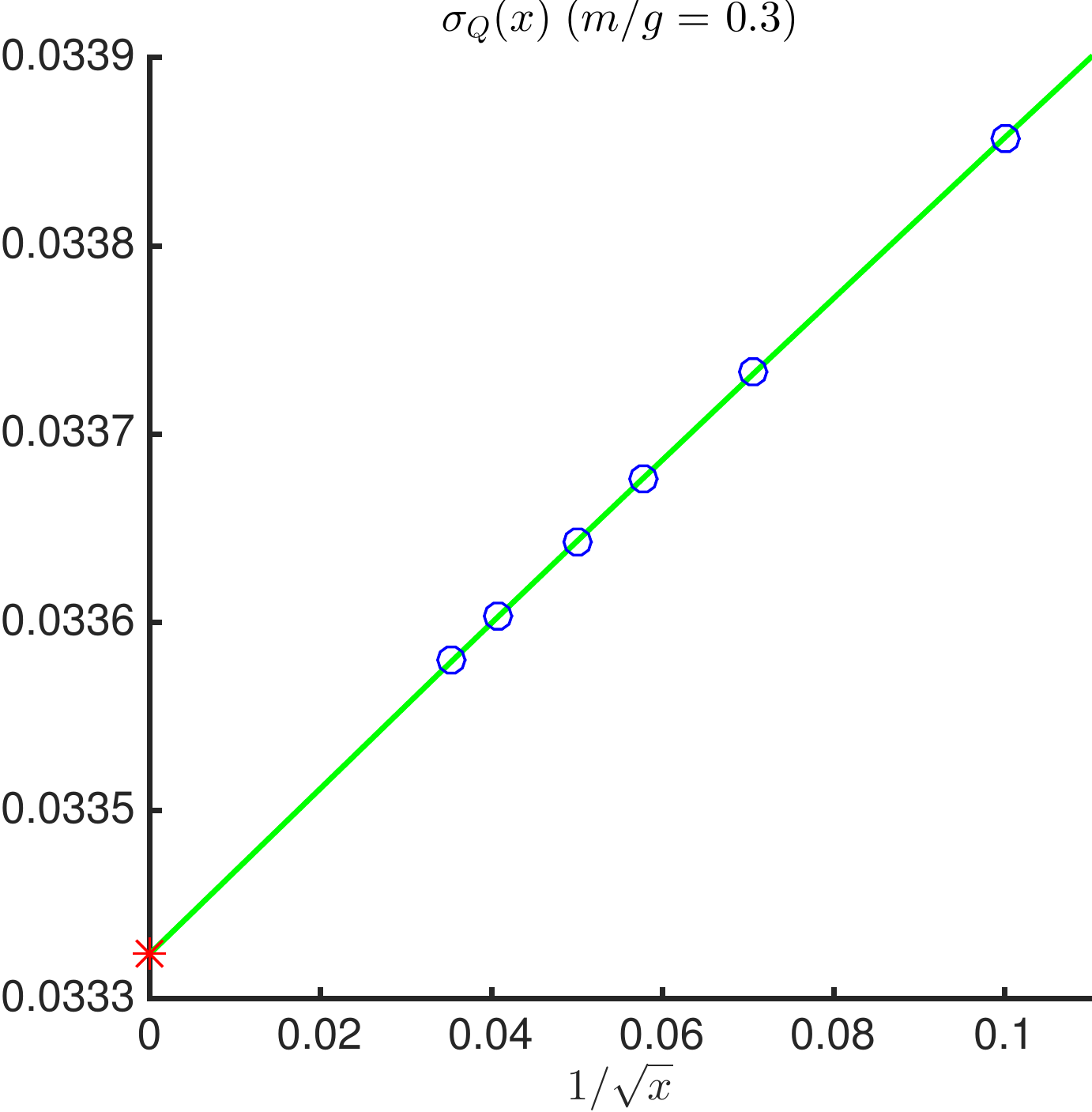}
\caption{\label{fig:ExtrapolationSTb}}
\end{subfigure}\vskip\baselineskip
\begin{subfigure}[b]{.48\textwidth}
\includegraphics[width=\textwidth]{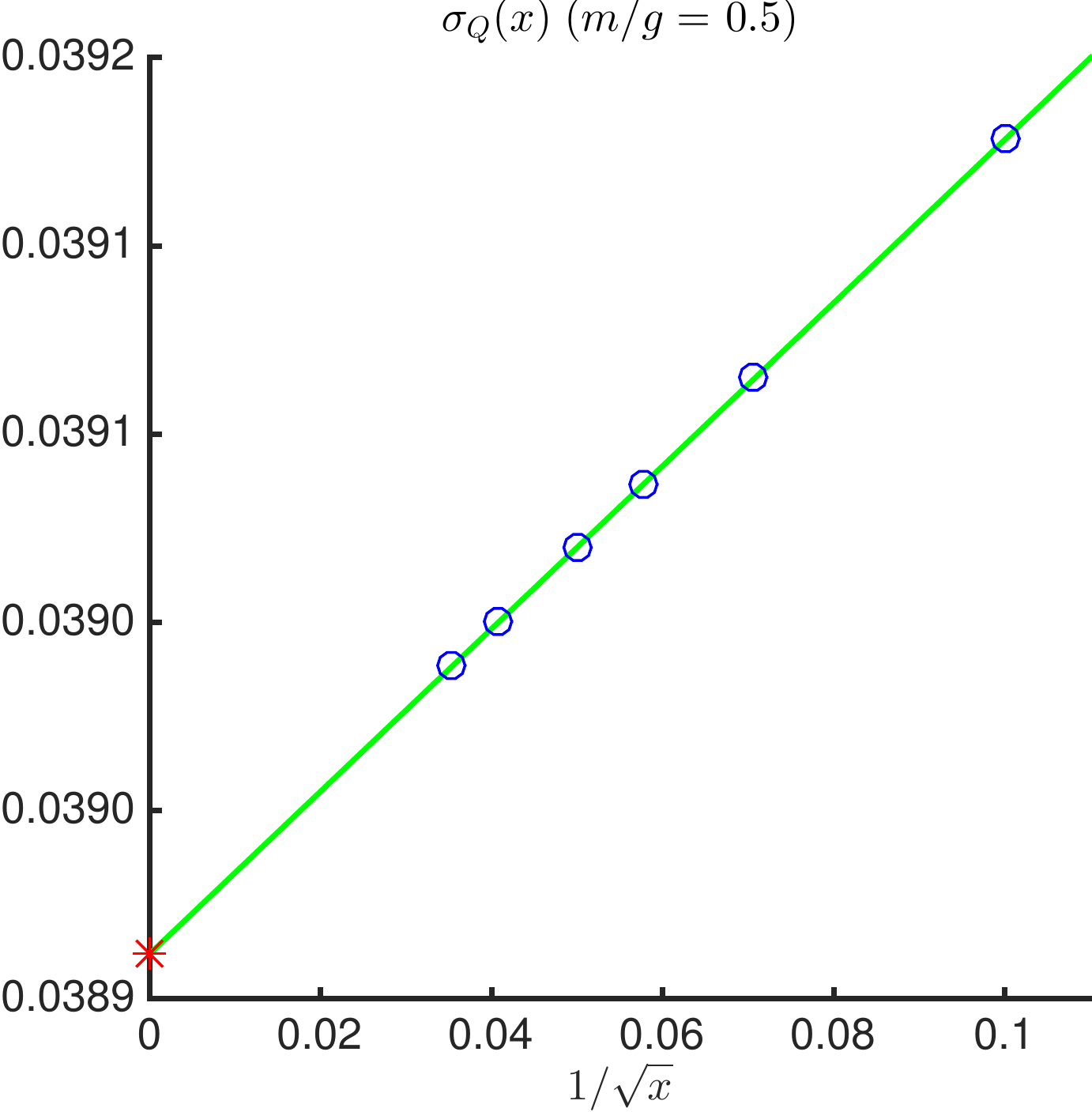}
\caption{\label{fig:ExtrapolationSTc}}
\end{subfigure}\hfill
\begin{subfigure}[b]{.48\textwidth}
\includegraphics[width=\textwidth]{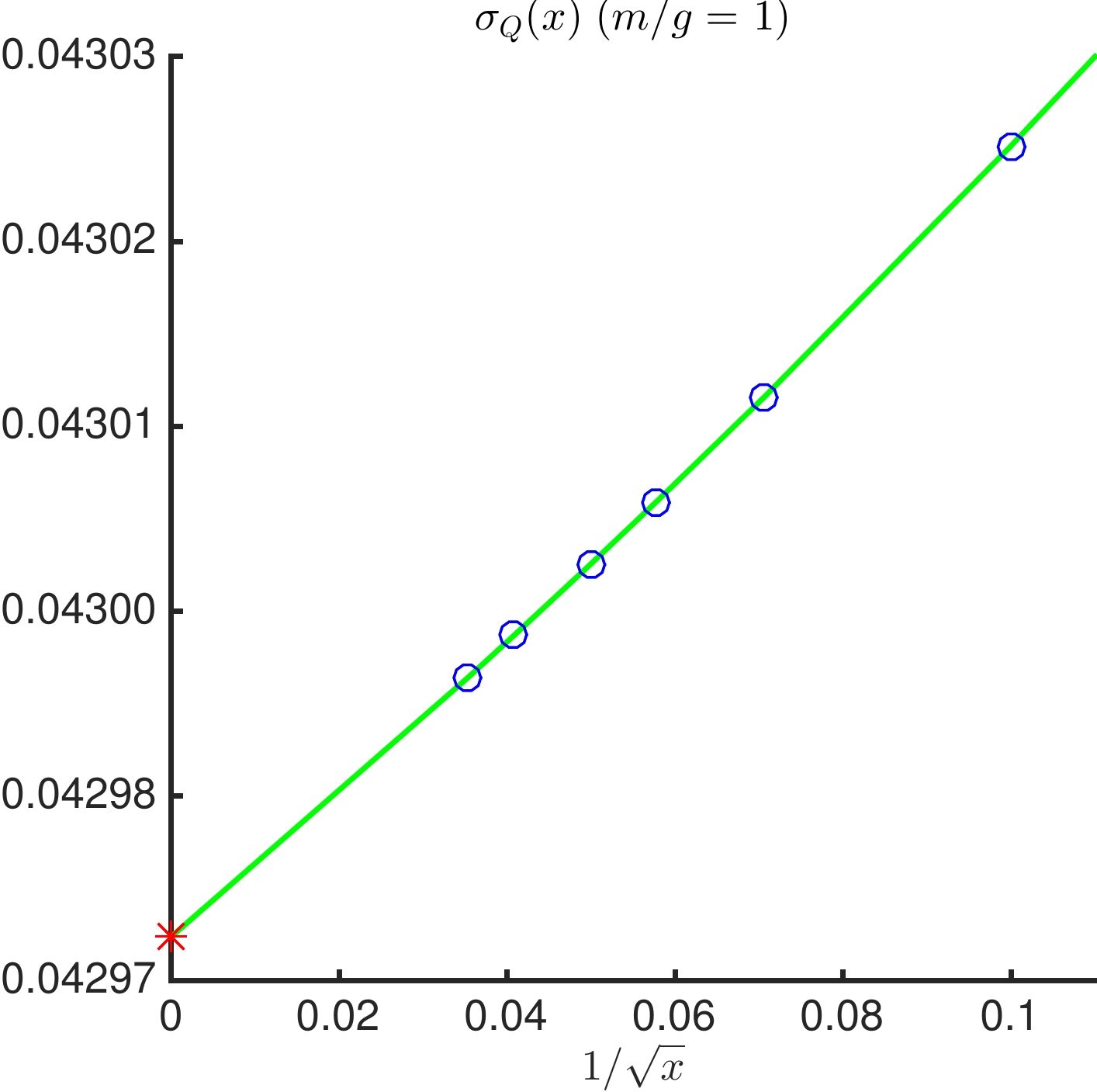}
\caption{\label{fig:ExtrapolationSTd}}
\end{subfigure}\vskip\baselineskip
\captionsetup{justification=raggedright}
\caption{\label{fig:ExtrapolationST} $Q = 0.3:$ Continuum extrapolation of the string tension $\sigma_Q$ for different values of $m/g$. }
\end{figure}

This means that the energy densities $\sqrt{x}\epsilon_Q(x)$ and $\sqrt{x}\epsilon_0(x)$ are UV divergent. But as we see, the string tension which is the difference of these quantities is UV-finite and, thus, we should also have $C_Q = C_0$. However, from the numerical point of view it is clear that small errors in (\ref{limxED}) or/and in $C_Q$ and $C_0$ would lead to large errors in the extrapolated continuum value $\lim_{x\rightarrow \infty} \sigma_Q$. To avoid this problem we first calculate $\epsilon_0$ and subtract it from the Hamiltonian (\ref{Ham02}): $H_Q \leftarrow H_Q - \sum_{n\in \mathbb{Z}}\epsilon_0$. The string tension is then compute as $\sigma_Q(x) = g\sqrt{x}\epsilon_Q(x)$ where $\epsilon_Q(x)$ is the ground state of the renormalized Hamiltonian. As follows from (\ref{tensionx}), for large $x$, $\sigma_Q(x)$ should scale as
\be \frac{\sigma_Q(x)}{g^2} = A_Q + \frac{B_Q}{\sqrt{x}} + \frac{C_Q}{x} + \frac{D_Q}{x^{3/2}} + \frac{E_Q}{x^2} + \mathcal{O}\left(\frac{1}{x^{5/2}}\right). \ee
In our simulations we computed $\sigma_Q(x)$ for $x = 100, 200, 300, 400, 600, 800$. Our estimate $\sigma_Q^{est}$ is obtained by fitting the $\sigma_Q(x)$ corresponding to the five largest $x$ to
\be\label{polfit} f_1(x) = A_Q + \frac{B_Q}{\sqrt{x}} + \frac{C_Q}{x} + \frac{D_Q}{x^{3/2}} \ee
and taking $\sigma_Q^{est} = g^2A_Q$.

In fig. \ref{fig:ExtrapolationST} we plot our results for the string tension as a function of $1/\sqrt{x}$ for $Q = 0.3$ and $m/g = 0.125, 0.3,0.5,1$. The numerical results are represented by circles and our polynomial fit (\ref{polfit}) through the largest five $x$-values is shown by a full line. The star represents our continuum estimate. It is clear that the string tension indeed behaves polynomially as a function of $1/\sqrt{x}$. For larger values of $m/g$ one can also deduce that we are already very close to the continuum limit at $1/\sqrt{x} = 0.1$. Indeed, for $m/g = 1$, the difference of our estimate with  $\sigma_Q(x)$ at $x = 100$ is only of order $10^{-5}$.

\begin{figure}
\begin{subfigure}[b]{.48\textwidth}
\includegraphics[width=\textwidth]{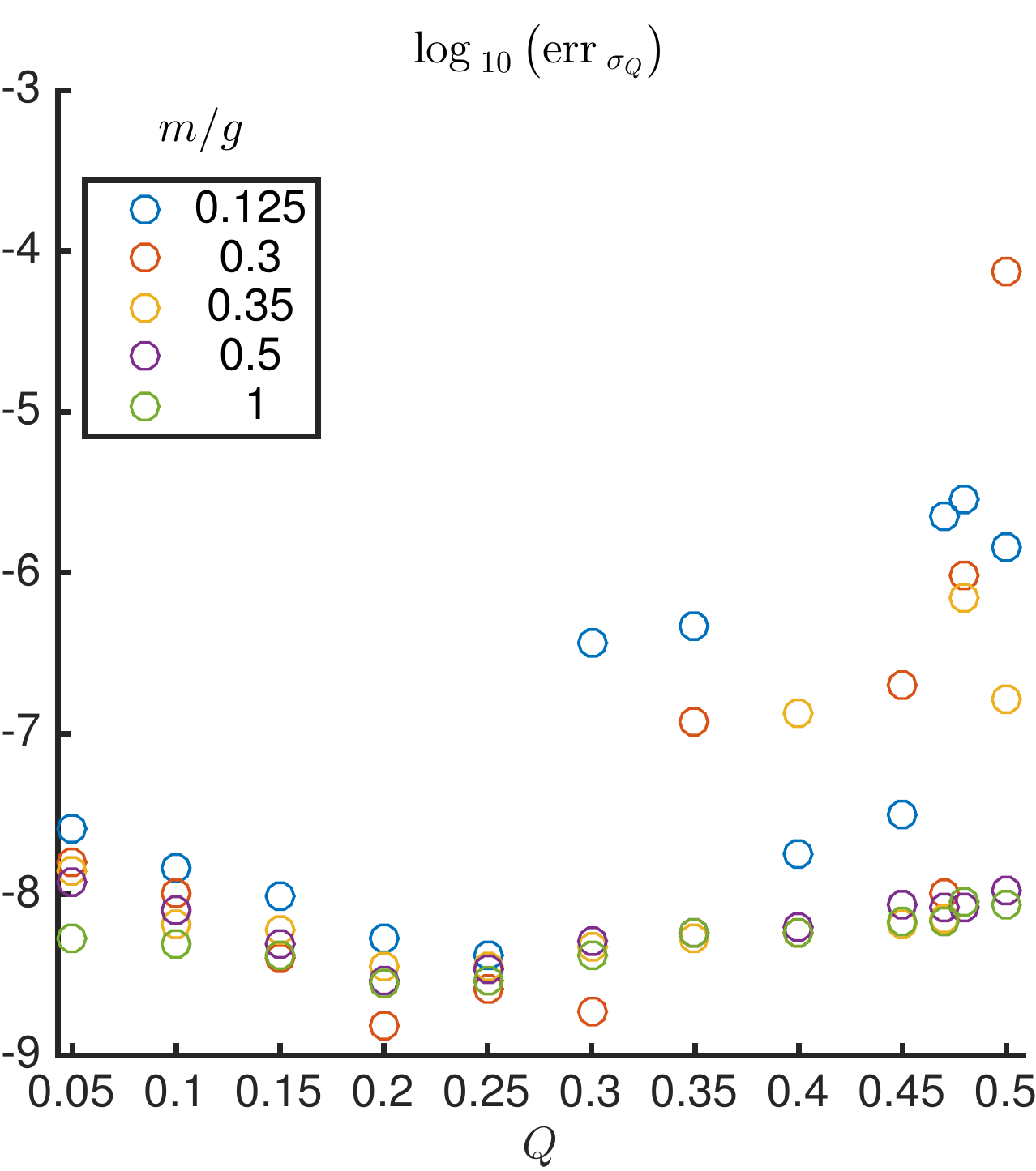}
\caption{\label{fig:ExtrErrora}}
\end{subfigure}\hfill
\begin{subfigure}[b]{.48\textwidth}
\includegraphics[width=\textwidth]{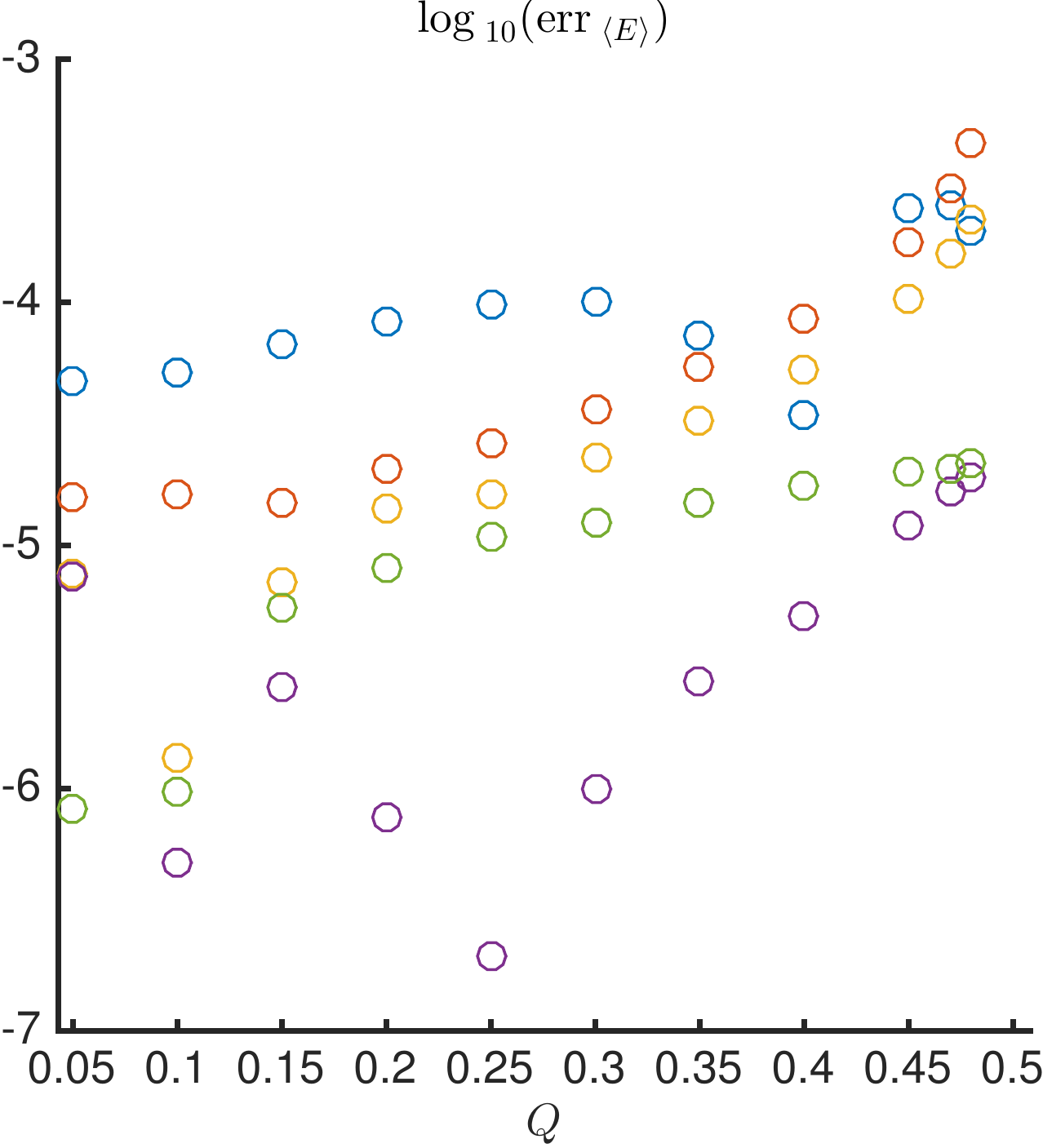}
\caption{\label{fig:ExtrErrorb}}
\end{subfigure}\vskip\baselineskip
\captionsetup{justification=raggedright}
\caption{\label{fig:ExtrError}(a): $\log_{10}\left(\mbox{err}_{\sigma_Q}\right)$ as a function of $Q$. (b):  $\log_{10}\left(\mbox{err}_{\langle E \rangle}\right)$ as a function of $Q$. }
\end{figure}

The continuum extrapolation depends on the chosen interval and the chosen fit. Therefore we also compute the continuum estimates by fitting all the data to $f_1(x)$ (see (\ref{polfit})) and all our data to
\be  f_2(x) = A_Q + \frac{B_Q}{\sqrt{x}} + \frac{C_Q}{x} + \frac{D_Q}{x^{3/2}} + \frac{E_Q}{x^2}. \ee
The error $\mbox{err}_{\sigma_Q}$ is taken to be the maximum of the difference of $\sigma_Q^{est}$ with these two other estimates. In fig. \ref{fig:ExtrErrora} we show the $\log_{10}$ of $\mbox{err}_{\sigma_Q}$ as a function of $Q$ for $m/g = 0.125, 0.3,0.35, 0.5, 1$. It is clear that these errors are quite small. We have the largest error for $m/g = 0.3$ and $Q = 0.5$ which is explained by the fact that the gap is very small there as we are in the vicinity of a phase transition \cite{Byrnes}. As mentioned above, it is well known that for smaller mass gaps, for a given bond dimension, the error on the ground-state MPS approximation will be larger.  

\begin{figure}
\begin{subfigure}[b]{.48\textwidth}
\includegraphics[width=\textwidth]{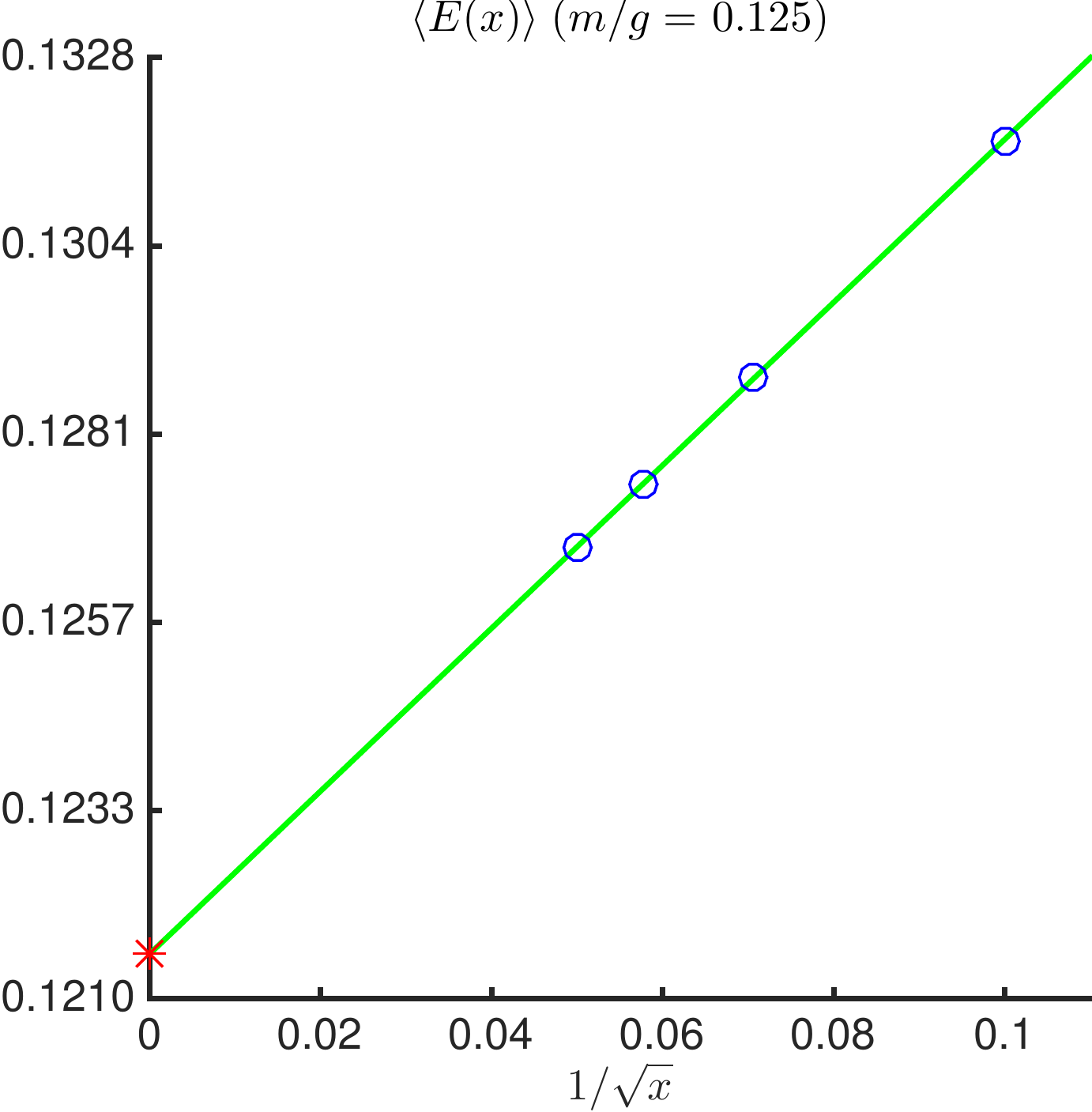}
\caption{\label{fig:ExtrapolationEFa}}
\end{subfigure}\hfill
\begin{subfigure}[b]{.48\textwidth}
\includegraphics[width=\textwidth]{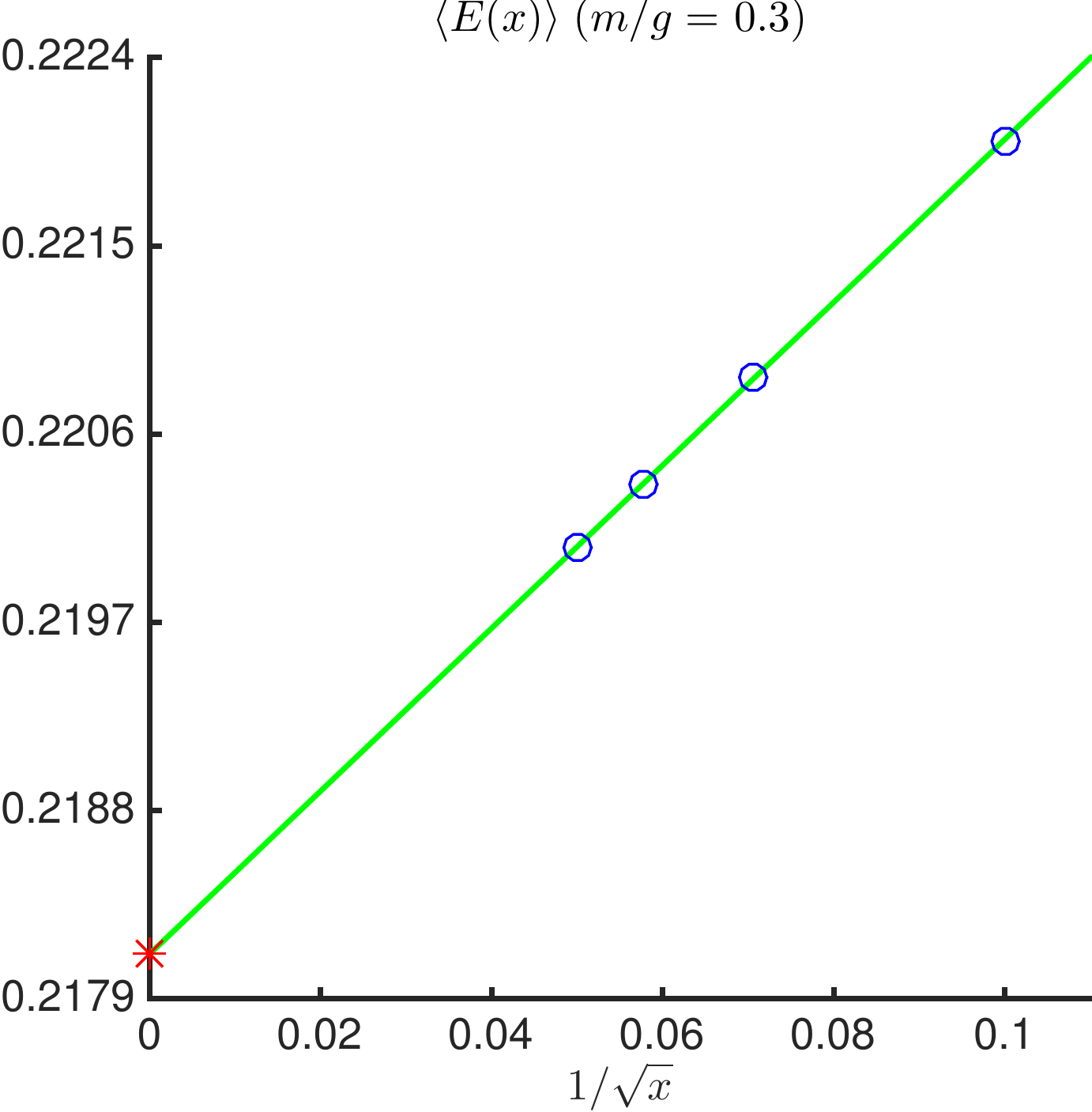}
\caption{\label{fig:ExtrapolationEFb}}
\end{subfigure}\vskip\baselineskip
\begin{subfigure}[b]{.48\textwidth}
\includegraphics[width=\textwidth]{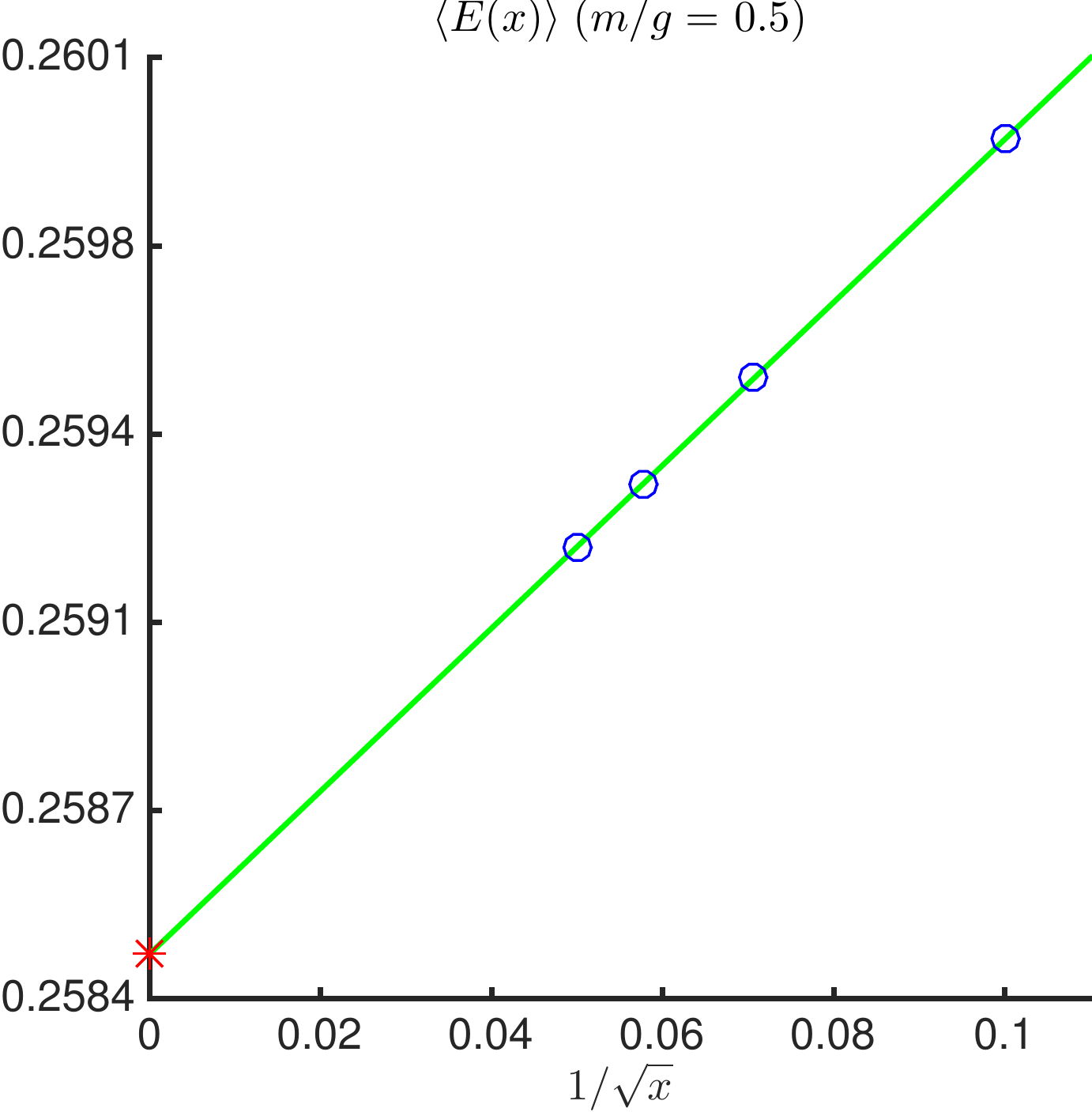}
\caption{\label{fig:ExtrapolationEFc}}
\end{subfigure}\hfill
\begin{subfigure}[b]{.48\textwidth}
\includegraphics[width=\textwidth]{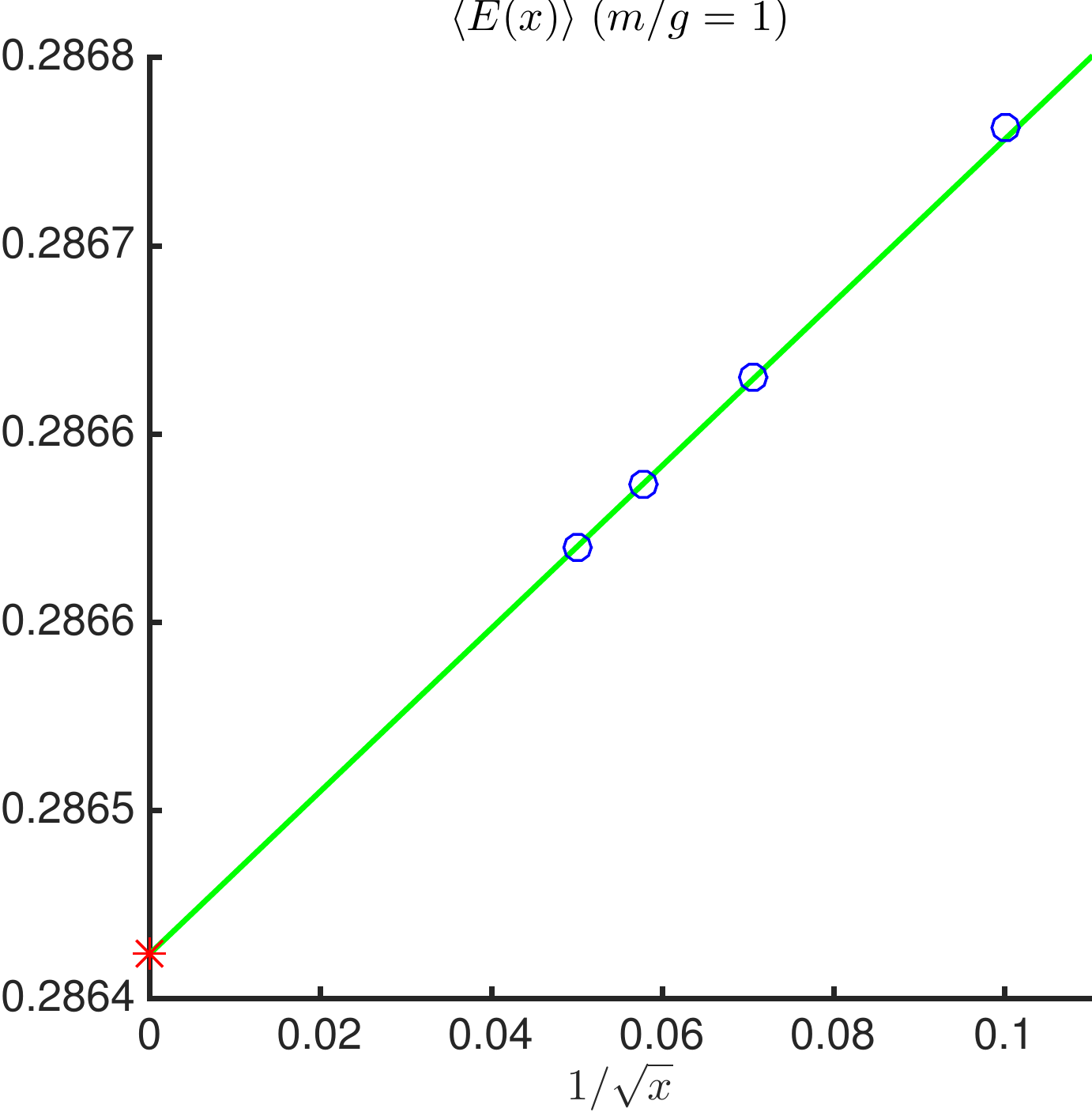}
\caption{\label{fig:ExtrapolationEFd}}
\end{subfigure}\vskip\baselineskip
\captionsetup{justification=raggedright}
\caption{\label{fig:ExtrapolationEF} $Q = 0.3.$ Continuum extrapolation of the electric field $\langle E \rangle$ for different values of $m/g$.}
\end{figure}

\begin{figure}
\begin{subfigure}[b]{.48\textwidth}
\includegraphics[width=\textwidth]{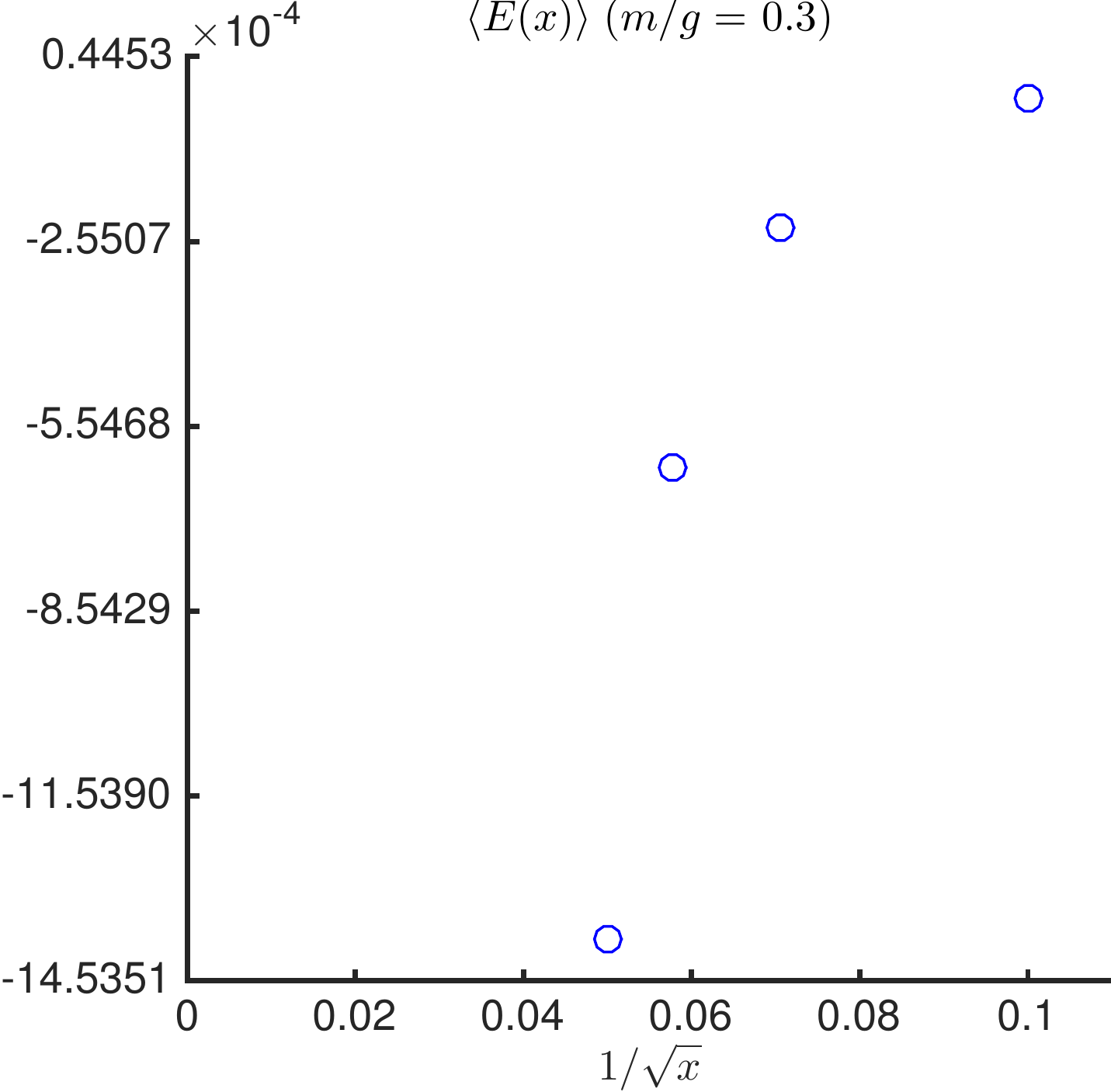}
\caption{\label{fig:SymBra}}
\end{subfigure}\hfill
\begin{subfigure}[b]{.48\textwidth}
\includegraphics[width=\textwidth]{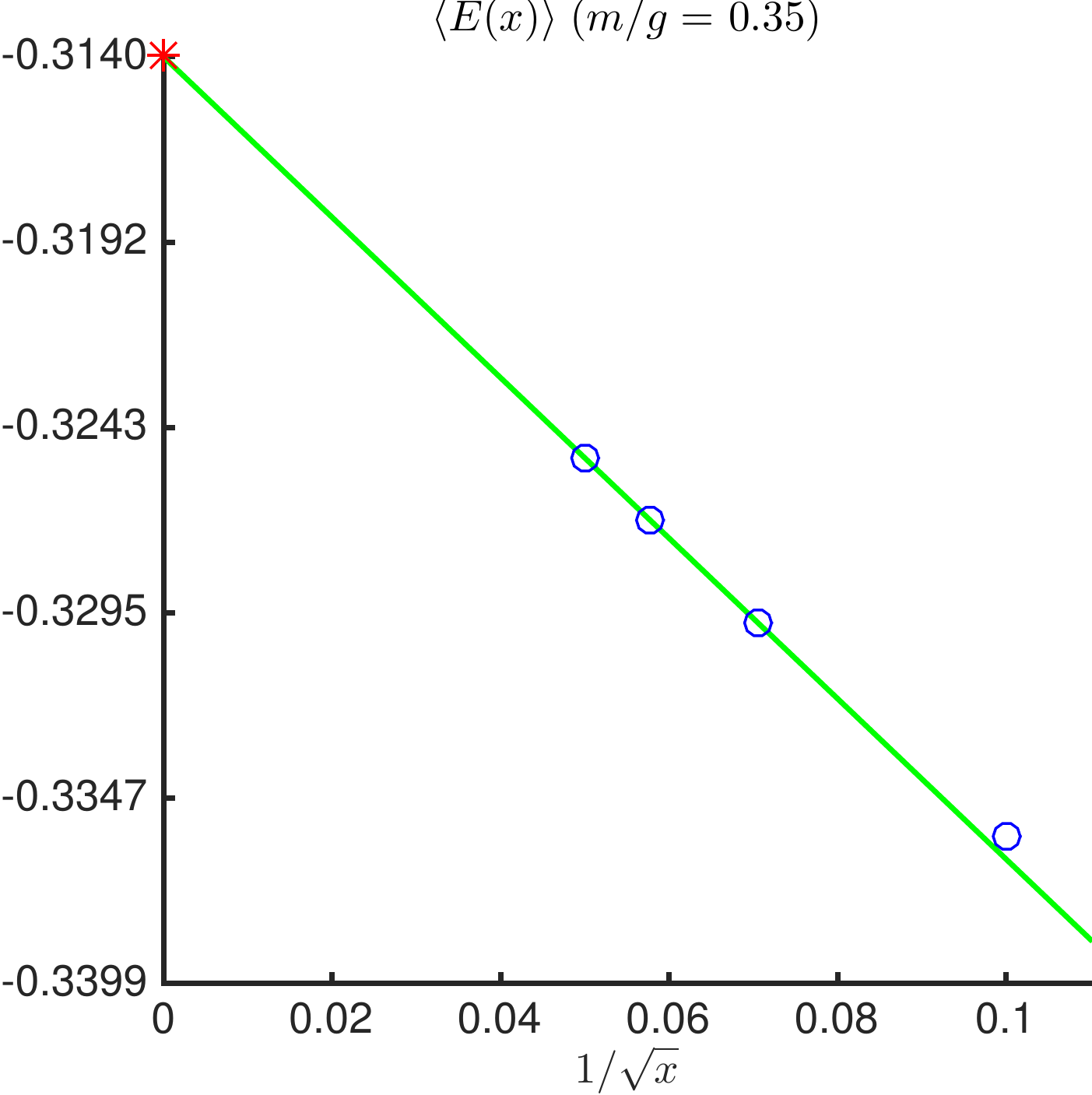}
\caption{\label{fig:SymBrb}}
\end{subfigure}\vskip\baselineskip
\captionsetup{justification=raggedright}
\caption{\label{fig:SymBr} $Q = 1/2$. Continuum extrapolation of the electric field $\langle E \rangle$. (a): $m/g = 0.3$. (b): $m/g = 0.35$.  }
\end{figure}

The continuum extrapolation of the electric field,
\be\frac{\langle E \rangle}{g} =  \frac{1}{2N}\left\langle \sum_{n \in \mathbb{Z}[1,2N]} (L(n) - Q)\right\rangle_Q (N = \vert \mathbb{Z} \vert), \ee
is found in a similar way. Now we use the values computed at $x = 100,200,300,400$ and perform a linear fit,
\be\label{linFit} g_1(x) = A_Q + \frac{B_Q}{\sqrt{x}}, \ee
through the three largest $x-$values. The fact that we again have analytical behavior as a function of $1/\sqrt{x}$ can be observed from fig. \ref{fig:ExtrapolationEF} where we display the electric field as function of $1/\sqrt{x}$. It is also a consequence of the fact that $\langle E(x) \rangle = - d\sigma_Q(x)/dQ$ and we already argued that $\sigma_Q(x)$ is analytical as a function of $x$.
To make our estimate more robust against the choice of the interval and the fitting function we compute estimates by a linear fit (\ref{linFit}) through all the points ($x = 100,200,300,400$) and a quadratic fit,
\be g_2(x) = A_Q + \frac{B_Q}{\sqrt{x}} + \frac{C_Q}{x}, \ee
through all the points. Again, the error $\mbox{err}_{\langle E \rangle}$ is taken to be the maximum of the difference with these two estimates. The $\log_{10}$ of  $\mbox{err}_{\langle E \rangle}$  is displayed in fig. \ref{fig:ExtrErrorb}.  The errors are quite small but become larger again around the phase transition at the critical mass $(m/g)_{c} \approx 0.33$ when going towards $Q = 1/2$.

At $Q = 1/2$ we do not display our error because this is a special case. For $m/g < (m/g)_{c}$ the $CT$ symmetry is not broken and, thus, we should have $\langle E \rangle = 0$, and this for all values of $x$. Therefore a continuum extrapolation of $\langle E \rangle$ is useless, see fig. \ref{fig:SymBra}. To obtain an error bound we take the largest value in magnitude of $\langle E(x) \rangle$ for $x = 100, 200, 300, 400$. It is displayed in table \ref{fig:EFQ}. When $m/g > (m/g)_{c}$ we have two different vacua with opposite sign for the electric field. We will always take the negative sign which comes down to taking the vacuum in the limit $Q \rightarrow 1/2$ for $Q < 1/2$. In this case it is possible to perform a polynomial extrapolation, see fig. \ref{fig:SymBrb}. The results are given in Table \ref{fig:EFQ}.  If possible we compare with \cite{Byrnes}.

\begin{table}
\begin{center}
\begin{tabular}{|r|c|c|}
\hline
$m/g$ & $\langle E \rangle/g$ & $\langle E \rangle/g$  \cite{Byrnes} \\
\hline
\hline
0.125 & 3 $\times 10^{-4}$ & - \\
0.25 & 2 $\times 10^{-4}$ &-\\
0.3 & 0.0014 &0.0(3) \\
0.35 & -0.313(2) & - \\
0.5 & -0.42041(3)& -0.421(1) \\
0.75 & -0.46145(2) & - \\
1 & -0.47692(2) & -0.4769 (5) \\
2 & -0.49364(3) & - \\
4 & -0.49834(3) & - \\
\hline
\end{tabular}
\end{center}
\caption{\label{fig:EFQ} Electric field at $Q = 1/2$ for different value of $m/g$. }
\end{table}

\subsection{Continuum extrapolation of the half-chain von Neumann entropy}\label{subsec:continuumExtraEntropy}
\begin{figure}
\begin{subfigure}[b]{.48\textwidth}
\includegraphics[width=\textwidth]{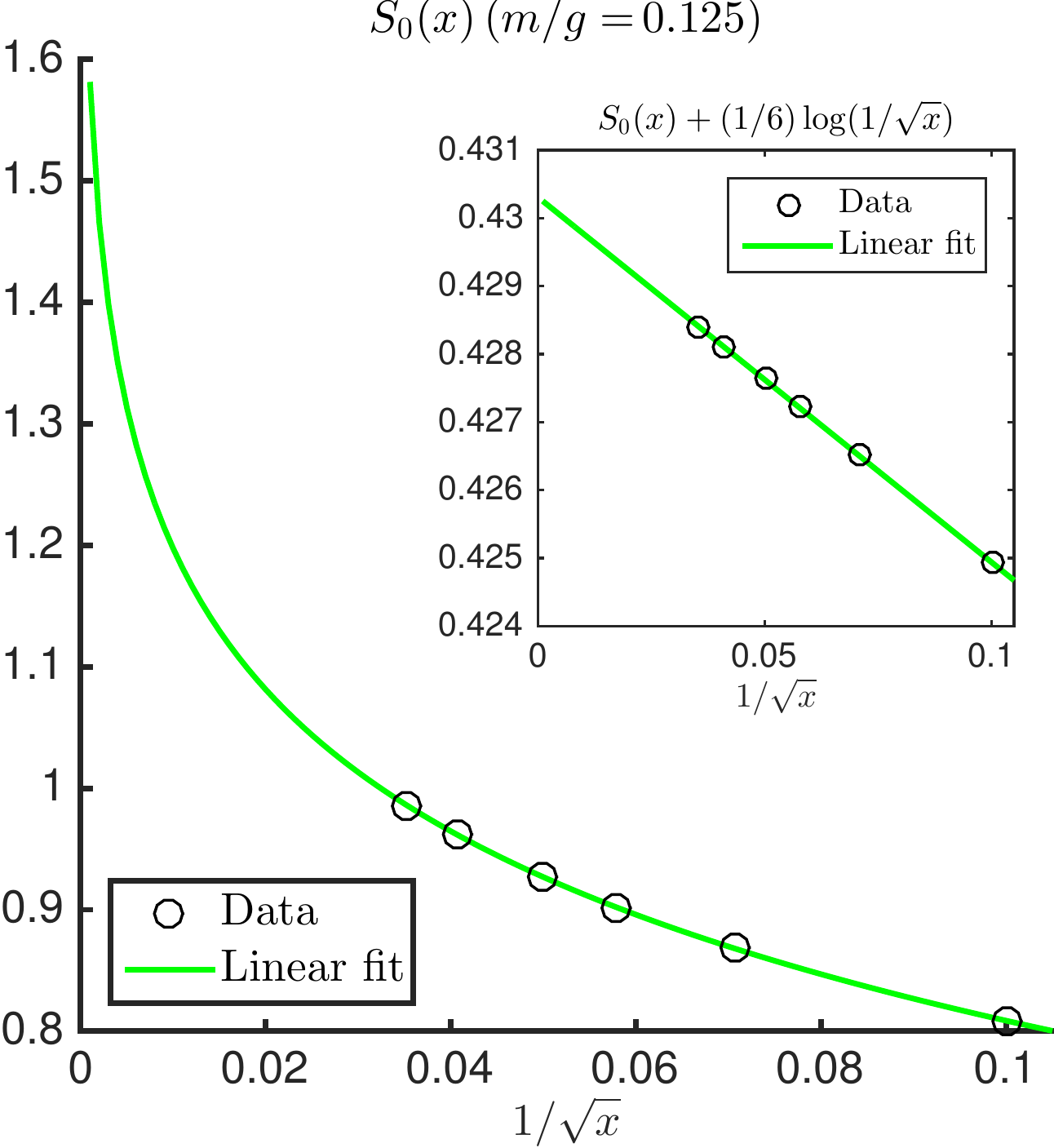}
\caption{\label{fig:EntropyQ0a}}
\end{subfigure}\hfill
\begin{subfigure}[b]{.48\textwidth}
\includegraphics[width=\textwidth]{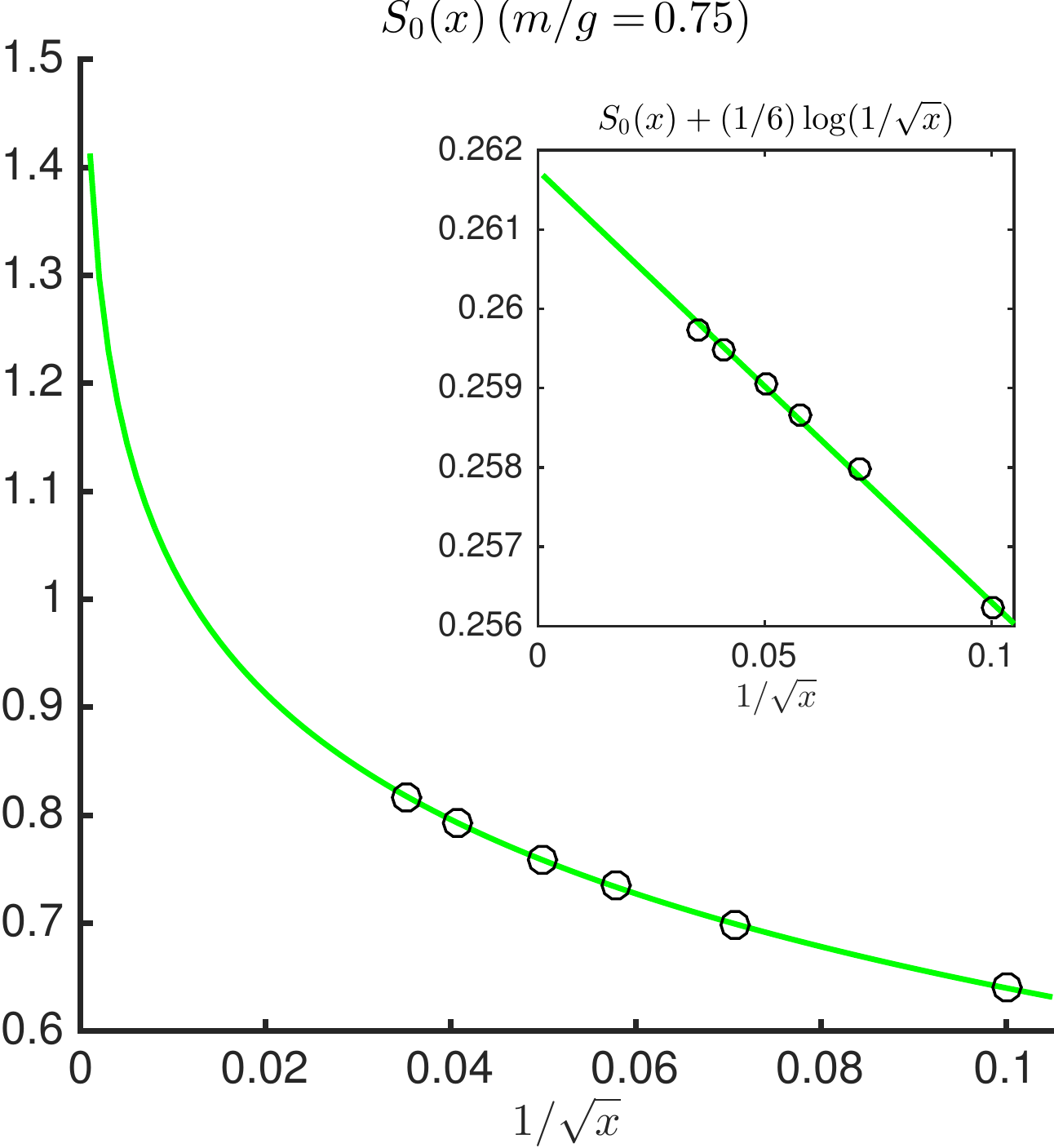}
\caption{\label{fig:EntropyQ0b}}
\end{subfigure}\vskip\baselineskip
\captionsetup{justification=raggedright}
\caption{\label{fig:EntropyQ0} $Q = 0$. (a) $m/g = 0.125.$ Fit of the form $(-1/6)\log(1/\sqrt{x}) + A + C/\sqrt{x}$ through $S_0(x)$. Inset: linear extrapolation of $S_0(x) + (1/6)\log(1/\sqrt{x})$ based on the largest five $x-$values, $x = 200,300,400,600,800$, to obtain the coefficients $A$ and $C$. (b): same as (a) but now for $m/g = 0.75$.}
\end{figure}

\noindent Using the Schmidt values $\lambda_{q,\alpha_q}$, see eq. (\ref{eq:MPSschmidtGaugeTDVP}), we can compute the half-chain von Neumann entropy $S_Q(x)$,
$$S_Q(x) = -\sum_{q = p_{min}}^{p_{max}}\sum_{\alpha_q = 1}^{D_q}\lambda_{q,\alpha_q}\log(\lambda_{q,\alpha_q}),$$
for a particular value of $x$. As already mentioned in the main text, because the Schwinger model is equivalent to a noncritical boson theory \cite{Coleman}, the half-chain von Neumann entropy should diverge as $(-1/6)\log(1/\sqrt{x})$ \cite{Calabrese2004} when $x \rightarrow + \infty$. 

\begin{table}
\begin{center}
\begin{tabular}{| c| | c | c | c | c | c | c | c | }
        \hline
 $m/g$ & 0.125 & 0.25 & 0.3 & 0.35 & 0.5 & 0.75 & 1\\
 \hline 
 $B_0 + 1/6$  &$1 \times 10^{-4}$ & $3 \times 10^{-4}$ & $5 \times 10^{-4}$ & $5 \times 10^{-4}$ & $8 \times 10^{-4}$ &$1.3 \times 10^{-3}$ &$2 \times 10^{-3}$ \\  
  \hline
\end{tabular}
\end{center}
\captionsetup{justification=raggedright}
\caption{\label{fig:tableEntropyQ0} The largest value in magnitude of $B_0 + 1/6$ obtained from the fit (\ref{eq:fitf1Entropy}) through the largest five $x-$values and the fits (\ref{eq:fitf1Entropy}) and (\ref{eq:fitf2Entropy}) through all our data. According to \cite{Calabrese2004} we should have $B_0 + 1/6 = 0$. }
\end{table}

\begin{figure}
\begin{subfigure}[b]{.48\textwidth}
\includegraphics[width=\textwidth]{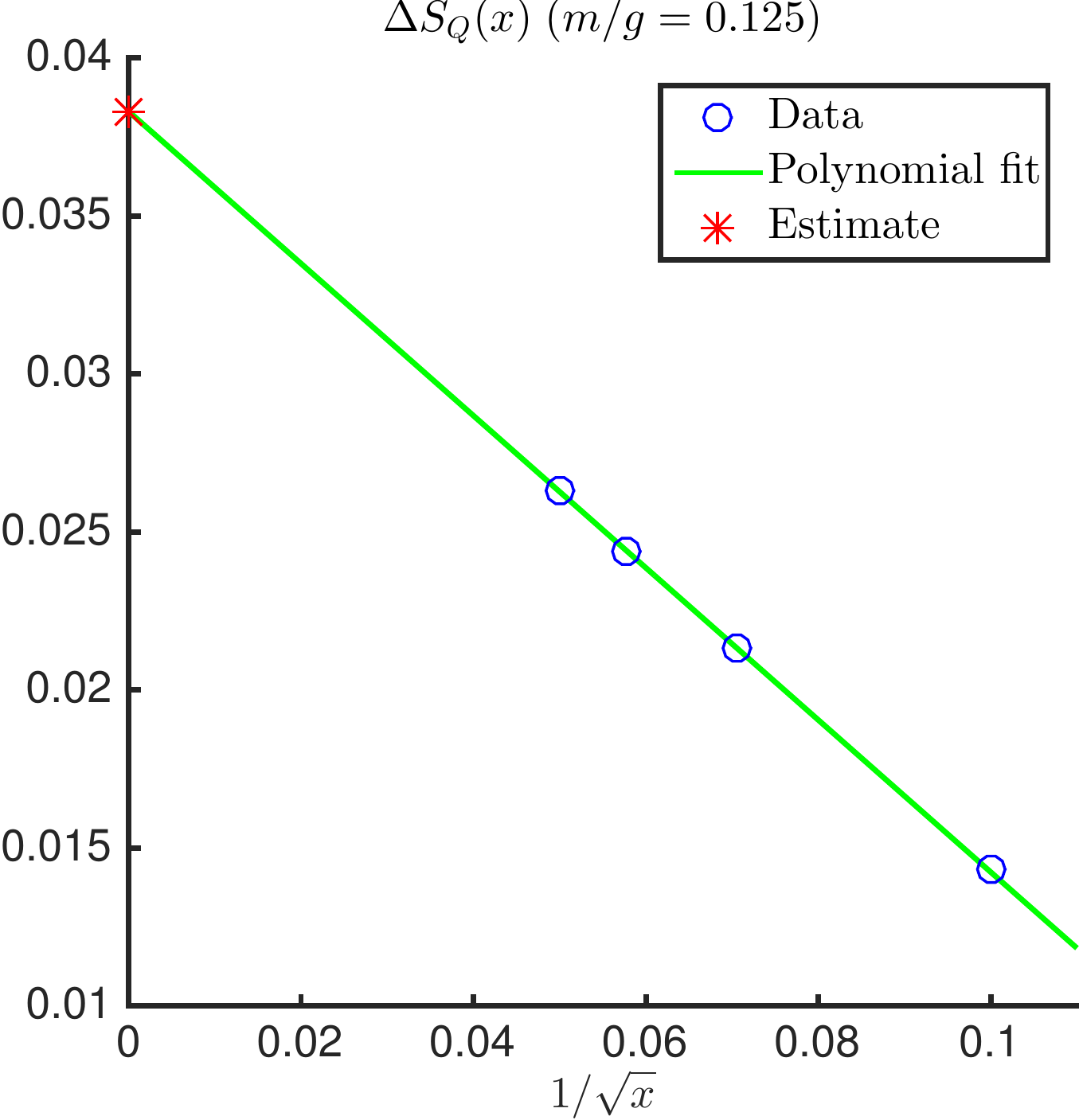}
\caption{\label{fig:ExtrapolationEntra}}
\end{subfigure}\hfill
\begin{subfigure}[b]{.48\textwidth}
\includegraphics[width=\textwidth]{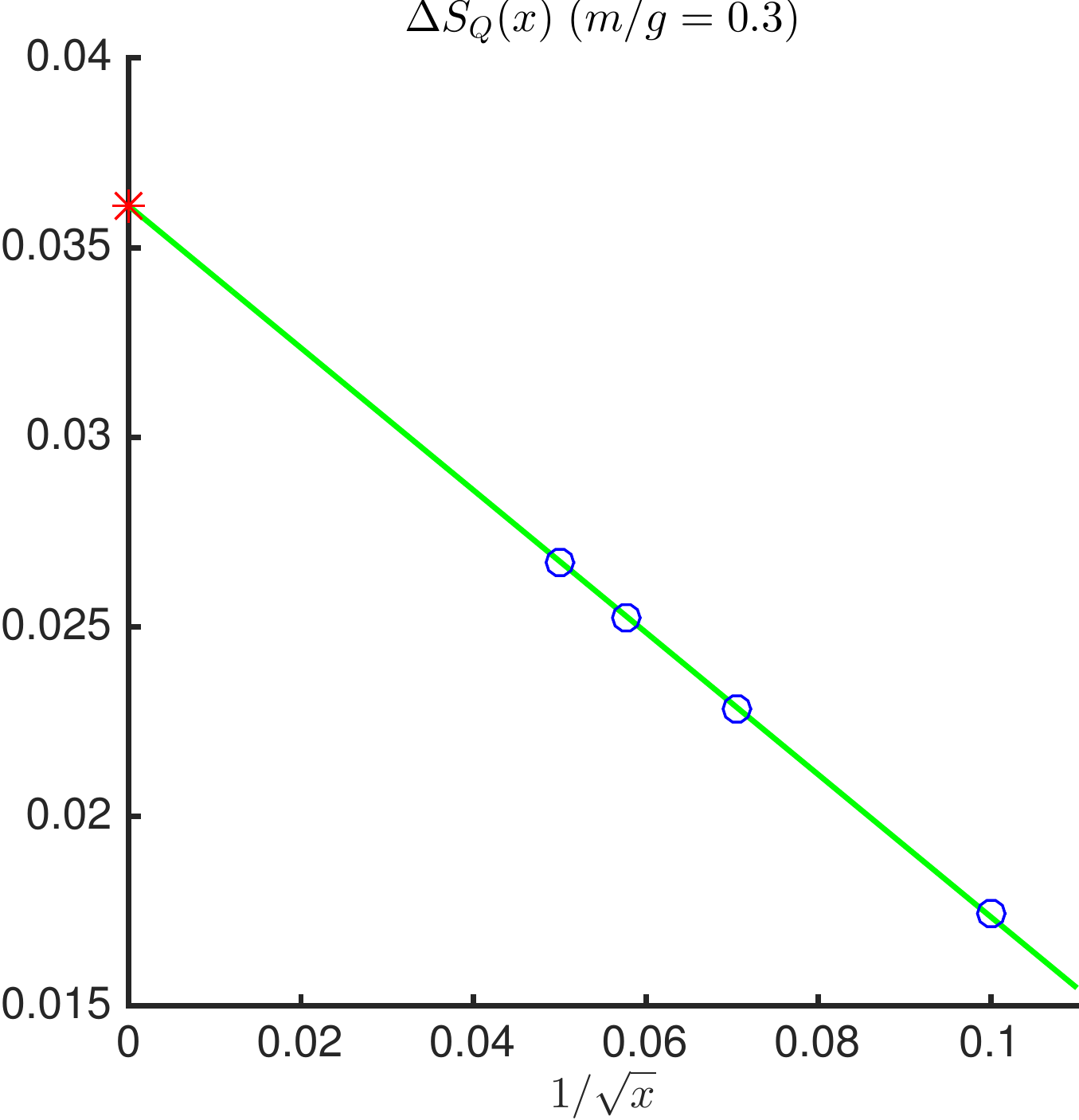}
\caption{\label{fig:ExtrapolationEntrb}}
\end{subfigure}\vskip\baselineskip
\begin{subfigure}[b]{.48\textwidth}
\includegraphics[width=\textwidth]{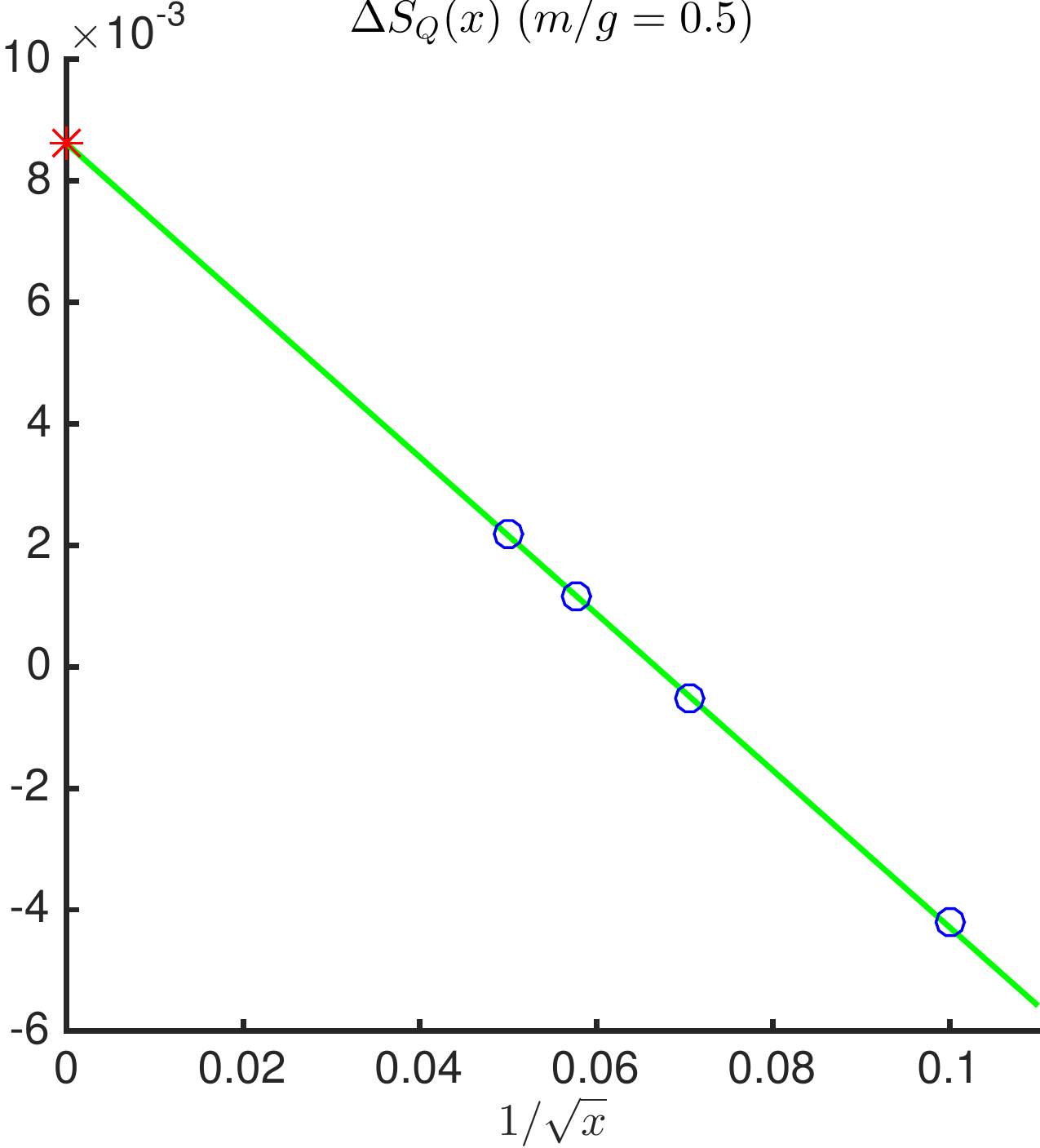}
\caption{\label{fig:ExtrapolationEntrc}}
\end{subfigure}\hfill
\begin{subfigure}[b]{.48\textwidth}
\includegraphics[width=\textwidth]{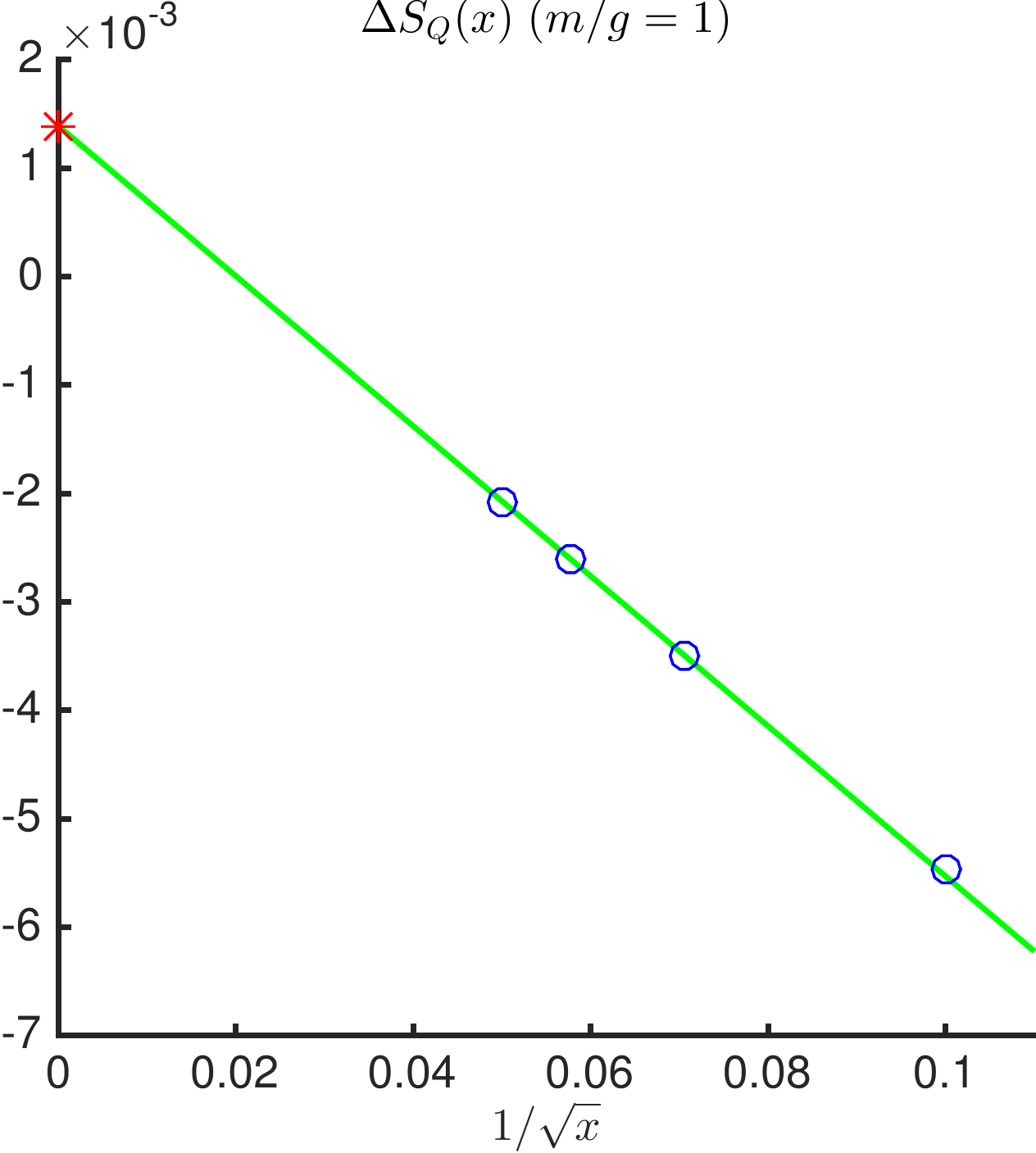}
\caption{\label{fig:ExtrapolationEntrd}}
\end{subfigure}\vskip\baselineskip
\captionsetup{justification=raggedright}
\caption{\label{fig:ExtrapolationEntr} $Q = 0.3.$ Continuum extrapolation of the renormalized half-chain von Neumann entropy for different values of $m/g$.}
\end{figure}

Let us check this for $Q = 0$. Inspecting $S_0(x) + \frac{1}{6}\log(1/\sqrt{x})$ as a function of $x$, we observe that it behaves linear as a function of $1/\sqrt{x}$, see inset figs. \ref{fig:EntropyQ0a} and \ref{fig:EntropyQ0b}. Therefore, we should be able to fit $S_0(x)$ to a function of the form
\be \label{eq:fitf1Entropy} f_1(x) = A_0 + B_0\log\left(\frac{1}{\sqrt{x}}\right) + C_0 \frac{1}{\sqrt{x}} \ee
and find $B_0 = 1/6$. Specifically, we fit our data corresponding to the largest five $x-$values, $x = 200,300,400,600,800$, against $f_1$ to obtain a first estimate for $B_0$. To have some robustness against the choice of fitting interval and the fitting function, we also include our result for $x = 100$ and fitted all our data against $f_1$ and against
\be \label{eq:fitf2Entropy} f_2(x) = A_0 + B_0\log\left(\frac{1}{\sqrt{x}}\right) + C_0 \frac{1}{\sqrt{x}} + D_0\frac{1}{x}. \ee
This gave us two other estimates for $B_0$. In table \ref{fig:tableEntropyQ0} we give the results for $B_0+1/6$. The value that is shown is the largest value for $B_0+1/6$ (in magnitude) from the three fits, i.e. the largest error on the predicted result of \cite{Calabrese2004}. As one observes these errors are at most $2\times 10^{-3}$ and for small values of $m/g$ only of order $10^{-4}$ which is a nice cross-check on our results. In the insets of figs. \ref{fig:EntropyQ0a} and \ref{fig:EntropyQ0b} we show a linear fit of the form  $f(x) = A +  C \frac{1}{\sqrt{x}}$ through $S_0(x) + (1/6)\log(x)$. Here we estimate $A$ and $C$ by taking into account the largest five $x-$values. In the main figures we also show the fit $(-1/6)\log(x) + f(x)$ through $S_0(x)$. As expected, given the results in table \ref{fig:tableEntropyQ0}, this fit matches our data very well. In fig. \ref{fig:EntropyQ0} we show results for $m/g = 0.125$ and $m/g = 0.75$. A similar plot for $m/g = 0.25$ is shown in fig. \ref{fig:UEntropya} (main text).
\\ 
\\ 
Because the coefficient of the logarithmic divergence of the von Neumann entropy is universal, the renormalized entropy $\Delta S_Q = S_Q - S_0$ should be UV-finite. In fig. \ref{fig:ExtrapolationEntr} we plot $\Delta S_Q(x)$ as a function of $1/\sqrt{x}$ and observe that this scales linearly in $1/\sqrt{x}$ to the continuum limit. A continuum result for different values of $Q$ and $m/g$ is obtained in exactly the same way as for the electric field. The results are shown in the main text, see fig. \ref{fig:UEntropyb}. The errors originating of the choice of fitting interval and fitting function are relatively small.

For the electric field and the string tension we have that our results at $x = 100$, or equivalently $ga = 1/\sqrt{x} = 0.1$, only differ from the continuum result by at most 10 percent, see fig. \ref{fig:ExtrapolationST} and fig. \ref{fig:ExtrapolationEF}. In contrast, for the entropy this is not the case at all, see fig. \ref{fig:ExtrapolationEntr}: the result at $x = 100$ and the continuum result differ by a factor of order one and sometimes also have a different sign. The main lesson is that, contrary to other quantities like the electric field and the string tension, we should be careful when extrapolating results at finite $x$ of the renormalized entropy to the continuum limit. In particular, for the nonuniform case, see main text section \ref{sec:potential}, one should always check how the results scale for different values of $x$. 

\section{Perturbative calculation of $\sigma_Q$}\label{pertsigma}
\noindent To compute $\sigma_Q$ in the weak-coupling expansion we start from the Lagrangian (\ref{Lagrangian}) and include a current $j^\mu= g \epsilon^{\mu\nu}\partial_\nu Q$; with $Q$ constant everywhere in the bulk, and $Q\rightarrow 0$ only at the boundaries at infinity (see \cite{ColemanCS}): \bea \mathcal{L} &=& \bar{\psi}\left(\gamma^\mu(i\partial_\mu+g A_\mu) - m\right) \psi - \frac{1}{4} F_{\mu\nu} F^{\mu\nu}-A_\mu j^\mu \nonumber\\
&=&\bar{\psi}\left(\gamma^\mu(i\partial_\mu+g A_\mu) - m\right) \psi - \frac{1}{4} F_{\mu\nu} F^{\mu\nu} - \frac{1}{2} F_{\mu\nu} \bar{F}^{\mu\nu}\,,\label{Lagrangian2} \eea
where on the last line we perform a partial integration and $\bar{F}^{\mu\nu}\equiv \epsilon^{\mu\nu}gQ$. 

\begin{figure}

\includegraphics[width=100mm]{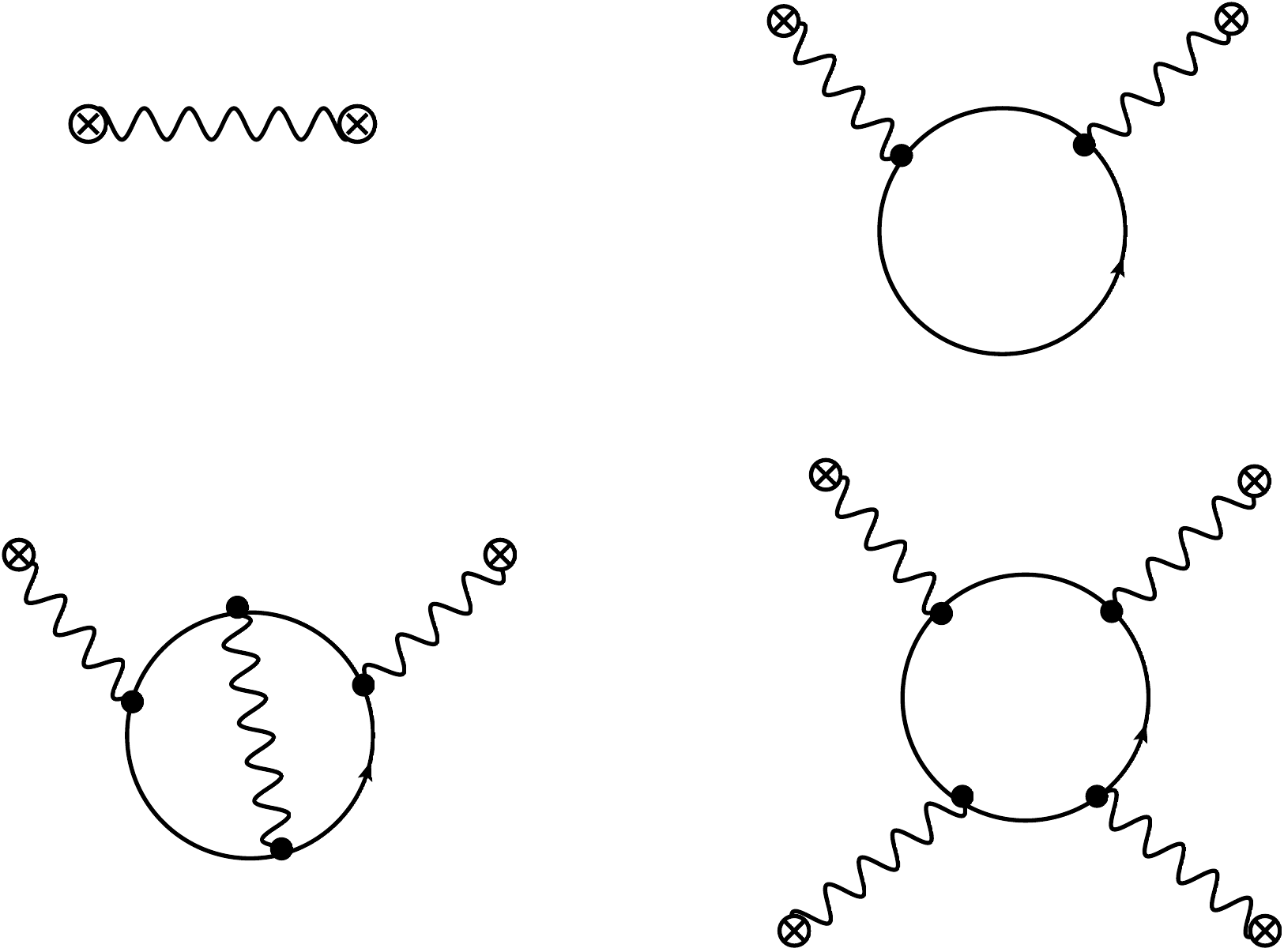}
\captionsetup{justification=raggedright}
\caption{Some diagrams for the effective action from (\ref{Lagrangian2}). On the first line we have the tree-level and the next-to-leading order $g^2/m^2$ contribution to $C_0$ (\ref{C0}).  Evaluation of the first diagram on the second line would give a $g^4/m^4$ correction to $C_0$, while the other diagram would give the leading $g^4/m^4$ contribution to $C_1$.   }
\label{diagram}

\end{figure}

The effective action, obtained by integrating out both the fermion and the gauge fields in the path integral, will then have the general form:
\be S_{eff}=\int\!\! d^2 x\, \,\mathcal{L}_{eff}=\int\!\! d^2\, x\,\,\,C_0(\frac{g}{m})\bar{F}_{\mu\nu}\bar{F}^{\mu\nu}+C_1(\frac{g}{m})\frac{(\bar{F}_{\mu\nu}\bar{F}^{\mu\nu})^2}{m^2}+\ldots\,,\ee where we can exclude derivative terms since $\bar{F}_{\mu\nu}$ is constant. At next-to-leading order we find for the first coefficient $C_0:$
\be C_0=-\frac{1}{4}+\frac{g^2}{24\pi m^2}\,. \label{C0}\ee
The zero-order term here  is the tree-level result while the $g^2/m^2$ term follows from the one-loop Feynman diagram on the first line of fig. \ref{diagram}, which can be calculated with standard techniques (see e.g.\cite{Peskin}). Furthermore one can see that all other nonzero diagrams will lead to contributions to the coefficients $C_i$ that are at least order $g^4/m^4$. Finally, we can then identify $S_{eff}=\int \!\! d^2x\, \sigma _Q$, leading to the result (\ref{tensionwc}). 

\section{Details on the implementation of the DMRG optimization for the nonuniform case}\label{appDMRG} 
\noindent Consider the Schwinger Hamiltonian,
\be\label{Ham2} H =  \frac{g}{2\sqrt{x}}\Biggl( \sum_{n \in \mathbb{Z}} [{L}(n) + \alpha(n)]^2 + \frac{m}{g}\sqrt{x} \sum_{n \in \mathbb{Z}}(-1)^n(\sigma_z(n) + (-1)^{n}) + x \sum_{n \in \mathbb{Z}}(\sigma^+ (n)e^{i\theta(n)}\sigma^-(n + 1) + h.c.)\biggl), \ee
in a nonuniform background field $\alpha(n)$ and assume we already computed a MPS approximation $\Ket{\Psi\bigl(A(1),A(2)\bigl)}$ of the form (\ref{uMPS}) with virtual dimensions $\tilde{D}(1)$ and $\tilde{D}(2)$ for the zero-background  Hamitonian ($\alpha(n) = 0$) \cite{Buyens}, i.e. 
\be\label{uMPSappalph0} \Ket{\Psi\bigl(A(1),A(2)\bigl)} = \sum_{\bm{\kappa}} \bm{v}_L^\dagger \left(\prod_{n \in \mathbb{Z}}A_{\kappa_{2n-1}}(1)A_{\kappa_{2n}}(2n)\right) \bm{v}_R \ket{\bm{\kappa}},\ee
with
\be \kappa_n = (s_n,p_n), s_n \in \{-1,1\}, p_n \in \mathbb{Z}[\tilde{p}_{min}(n+1\mbox{ mod 2}),\tilde{p}_{max}(n+1\mbox{ mod 2})],\ket{\bm{\kappa}} = \ket{\{\kappa_n\}_{n \in \mathbb{Z}}},\ee 
is a ground state of (\ref{Ham2}) with $\alpha(n) = 0$. Note that the tensors $A(n)$ take the form
\bea
{[A_{s,p}(n)]}_{(q,\alpha_q),(r,\beta_r)} &=& {[a_{s,p}(n)]}_{\alpha_q,\beta_r}\delta_{q+(s_n+(-1)^n)/2,r}\delta_{r,p}, \alpha_q = 1\ldots \tilde{D}_q(n \mbox{ mod 2}), \beta = 1 \ldots \tilde{D}_r(n+1 \mbox{ mod 2}),\label{AnGI}
\eea
$q \in \mathbb{Z}[\tilde{p}_{min}(n \mbox{ mod 2}),\tilde{p}_{max}(n \mbox{ mod 2})];$ $\tilde{p},\tilde{r} \in \mathbb{Z}[\tilde{p}_{min}(n+1 \mbox{ mod 2}),\tilde{p}_{max}(n+1 \mbox{ mod 2})]$ in order to enforce Gauss' law, $G(n)\Ket{\Psi\bigl(A(1),A(2)\bigl)} = 0$ with 
\be G(n) = L(n) - L(n-1) - \frac{\sigma_z(n) + (-1)^n}{2} = 0, \ee
to the state (see eq. (\ref{gaugeMPS})). We now consider a constant background electric field $\alpha(n)$ which has compact support: $\alpha(n) = \alpha \in \mathbb{R}$ for $n \in \mathbb{N}[0,k]$, $\alpha(n) = 0$ for $n \notin \mathbb{N}[0,k]$. The MPS trial state as ansatz for the ground state of this Hamiltonian that we will consider is, see eq. (\ref{nonuMPS}),
\be\label{DMRGAn}\ket{\Phi(\bm{B})} =  \sum_{\bm{\kappa}} \bm{v}_L^\dagger \left(\prod_{n < r_L}A_{\kappa_{n}}(n)\right) \left(\prod_{n = r_L}^{r_R-1}B_{\kappa_{n}}(n)\right)  \left(\prod_{n \geq r_R}A_{\kappa_{n}}(n)\right) \bm{v}_R  \ket{\bm{\kappa}}, \ee
where $r_L \ll 0 \leq k \ll r_R$ and $A_\kappa(n) = A_\kappa(n \mbox{ mod 2})$ corresponds to the MPS approximation (\ref{uMPSappalph0}) of the ground state of the zero-background Hamiltonian. We take $r_L$ and $r_R$ odd. To enforce Gauss' law, $G(n)\ket{\Phi(\bm{B})} = 0$, the $B(n)$ must take the form (\ref{AnGI}):
\bea
{[B_{s,p}(n)]}_{(q,\alpha_q),(r,\beta_r)} &=& {[b_{s,p}(n)]}_{\alpha_q,\beta_r}\delta_{q+(s_n+(-1)^n)/2,r}\delta_{r,p}, \alpha_q = 1\ldots D_q(n), \beta = 1 \ldots D_r(n+1), \label{BnGI}
\eea
$ q \in \mathbb{Z}[p_{min}(n),p_{max}(n)];$  $p,r \in \mathbb{Z}[p_{min}(n+1),p_{max}(n+1)]$ where $D_q(r_L)= \tilde{D}_q(1)$, $D_q(r_R) = \tilde{D}_q(1)$, $p_{min/max}(r_L) =p_{min/max}(r_R) =  \tilde{p}_{min/max}(1)$. The formal virtual dimensions of this MPS are $D(n) = \sum_{q = p_{min}(n)}^{p_{max}(n)} D_q(n)$. Later in this appendix we will come back to the issue of which values to take for $D_q(n)$ and $p_{min/max}(n)$. \\
\\
To obtain the best approximation within this class of states of the ground state of the Hamiltonian (\ref{Ham2}) we have to minimize
\be H(\bar{\bm{b}}, \bm{b}) = \frac{\bra{\Phi(\bar{\bm{B}}[\bar{\bm{b}}])} H \ket{\Phi(\bm{B}[\bm{b}])}}{\langle \Phi(\bar{\bm{B}}[\bar{\bm{b}}]) \ket{\Phi(\bm{B}[\bm{b}])}}\ee
with respect to $b(r_L),\ldots, b(r_R - 1)$. This is a perfect problem to tackle with the DMRG \cite{White}. We briefly sketch how this works in our case.

The DMRG first minimizes $H(\bar{\bm{b}}, \bm{b})$  with respect to $b(r_L)$ while keeping $b(r_L+1),\ldots, b(r_R - 1)$ fixed, then minimizes $H(\bar{\bm{b}}, \bm{b})$  with respect to $b(r_L + 1)$ while keeping $b(r_L),b(r_L + 2),\ldots, b(r_R - 1)$ fixed and so on until $b(r_R - 1)$. After this sweep, it will sweep back: minimizing $H(\bar{\bm{b}}, \bm{b})$  with respect to $b(r_R - 1)$ while keeping $b(r_L),\ldots, b(r_R - 2)$ fixed, then minimizing $H(\bar{\bm{b}}, \bm{b})$  with respect to $b(r_R - 2)$ while keeping $b(r_L),\ldots, b(r_R - 3), b(r_R - 1)$ fixed and so on until $b(r_L)$. The algorithm keeps sweeping until convergence of the quantity $H(\bar{\bm{b}}, \bm{b})$ is reached.

Let us now discuss how to minimise $H(\bar{\bm{b}}, \bm{b})$  with respect to $b(m)$ ($r_L \leq m \leq r_R - 1$). It is convenient to use the gauge freedom \footnote{This is the gauge freedom inherent to MPS, not be confused with the gauge freedom of QED.} of the matrices: \be A_\kappa(1) \rightarrow U(1)A_\kappa(1) U(2)^{-1}, A(2) \rightarrow U(2)A_\kappa(2) U(1)^{-1}, B_\kappa(n) \rightarrow V(n) B_\kappa(n) V(n+1)^{-1}, V(r_L)= U(1), V(r_R) = U(1),\ee
to bring  (\ref{DMRGAn}) in the following form
\be\label{DMRGAnN}\ket{\Phi(\bm{B})} =  \sum_{\bm{\kappa}} \bm{v}_L^\dagger \left(\prod_{n < r_L}L_{\kappa_{n}}(n)\right) \left(\prod_{n = r_L}^{m - 1}B_{\kappa_{n}}^{(L)}(n)\right) B_{\kappa_{m}}(m)\left(\prod_{n = m+1}^{r_R-1}B_{\kappa_n}^{(R)}(n)\right) \left(\prod_{n \geq r_R}R_n^{\kappa_{n}}\right) \bm{v}_R  \ket{\bm{\kappa}}\ee
where $L(n)$, $B^{(L)}(n)$ are in the left-canonical form: $\sum_\kappa(L_\kappa(n))^\dagger L_\kappa (n) = \idm,$ $\sum_\kappa (B_{\kappa}^{(L)}(n))^\dagger B_{\kappa}^{(L)}(n) = \idm,$ and $R_n$, $B^{(R)}(n)$ are in the right-canonical form: $\sum_\kappa R_n^\kappa (R_n^\kappa)^\dagger = \idm,$ $\sum_\kappa B_{\kappa}^{(R)}(n) (B_{\kappa}^{(R)}(n))^\dagger = \idm$ ($\sum_{\kappa} = \sum_{s = -1,1} \sum_{p = p_{min}(n)}^{p_{max}(n)}, \kappa = (s,p)$). In this case, the norm of the state is $N[B(m)] = \sqrt{\langle \Phi(\bar{\bm{B}}) \ket{\Phi(\bm{B})}} = \sqrt{\sum_\kappa \mbox{tr}[(B_\kappa(m))^\dagger B_\kappa(m)]}$ which can be put to one by rescaling $B(m)$: $B(m) \rightarrow B(m)/N[B(m)]$. Note that $L(n)$ and $R(n)$ depend only on the parity of $n$. Furthermore, we can use the remaining gauge freedom in the matrices to find positive-definite diagonal matrices $\Lambda(n)$ with $\mbox{tr}[\Lambda(n)] = 1$, such that
\begin{subequations}\label{SmC}
\begin{multline}\sum_\kappa L_\kappa(1)\Lambda(r_L) (L_\kappa(1))^\dagger = \Lambda(r_L-1),\sum_\kappa L_\kappa(2)\Lambda(r_L-1) (L_\kappa(2))^\dagger = \Lambda(r_L), \\ \sum_\kappa B_{\kappa}^{(L)}(n)\Lambda(n)(B_{\kappa}^{(L)}(n))^\dagger = \Lambda(n-1) (r_L \leq n \leq m-1)\end{multline}
and
\begin{multline} \sum_\kappa (B_{\kappa}^{(R)}(n))^\dagger \Lambda(n-1)B_{\kappa}^{(R)}(n) = \Lambda(n) (m+1 \leq n \leq r_R), \sum_\kappa (R_\kappa(1))^\dagger \Lambda(r_R-1)R_\kappa(1) = \Lambda(r_R), \\ \sum_\kappa (R_\kappa(2))^\dagger \Lambda(r_R) R_\kappa(2) = \Lambda(r_R+1). \end{multline}
\end{subequations}
Because the tensors $R,L$ and $B$ take the form (\ref{AnGI}) and (\ref{BnGI}), the diagonal elements of $\Lambda_n$ can be labeled by the eigenvalues $q$ of $L(n)$:
\begin{multline} [\Lambda(n)]_{[(q\alpha_q),(r\beta_r)]} = \delta_{\alpha_q,\beta_r}\delta_{q,r}\lambda_{q,\alpha_r}(n) \\ \alpha_q, \beta_q = 1\ldots D_q(n+1), 0\leq \lambda_{q,D_q(n+1)}(n) \leq \lambda_{q,D_q(n+1) - 1}(n) \leq \ldots \leq \lambda_{q,1}(n) \leq 1, \sum_{q = p_{min}(n+1)}^{p_{max}(n+1)}\sum_{\alpha_q=1}^{D_p(n+1)} \lambda_{q,\alpha_q}(n) = 1,\label{eq:MPSschmidtGaugeNonu} \end{multline}
$q,r \in \mathbb{Z}[p_{min}(n+1),p_{max}(n+1)].$ These diagonal elements of $\Lambda(n)$ are the Schmidt values associated with the bipartition $\{\mathcal{A}_1(n)=\mathbb{Z}[-\infty,n],\mathcal{A}_2(n) = \mathbb{Z}[n+1, + \infty]\}$ of the lattice. More specifically, we have that the Schmidt decomposition with respect to this bipartition reads, see eq. (\ref{eq:MPSschmidtGauge}),
\be \label{eq:MPSschmidtGaugeAp} \ket{\Phi(\bm{B})} = \sum_{q = p_{min}(n+1)}^{p_{max}(n+1)} \sum_{\alpha_q=1}^{D_q(n+1)} \sqrt{\lambda_{q,\alpha_q}(n)} \ket{\psi_{q,\alpha_q}^{\mathcal{A}_1(n)}}\ket{\psi_{q,\alpha_q}^{\mathcal{A}_2(n)}} \ee
where $\ket{\psi_{q,\alpha_q}^{\mathcal{A}_1(n)}}$ (resp. $\ket{\psi_{q,\alpha_q}^{\mathcal{A}_2(n)}}$) are orthonormal unit vectors in the tensor product of the local Hilbert spaces in the region $\mathcal{A}_1(n)$ (resp. $\mathcal{A}_2(n)$). At the boundaries ($n < r_L$ and $n \geq r_R$)  the Schmidt values depend only on the parity of $n$, more specifically:
\begin{subequations}\label{eq:SchmidtValueBoundary}
\be \lambda_{q,\alpha_q}(2n-1) = \lambda_{q,\alpha_q}(r_L)\;(\forall n: 2n-1 \leq r_L), \lambda_{q,\alpha_q}(2n) = \lambda_{q,\alpha_q}(r_L-1) (\forall n: 2n \leq r_L - 1,) \ee
\be \lambda_{q,\alpha_q}(2n-1) = \lambda_{q,\alpha_q}(r_R)\;(\forall n: 2n-1 \geq r_R), \lambda_{q,\alpha_q}(2n) = \lambda_{q,\alpha_q}(r_R+1) (\forall n: 2n \leq r_R + 1). \ee
\end{subequations}
These Schmidt values correspond to those of the ground state (\ref{uMPSappalph0}) of the zero-background Hamiltonian and were already computed before (see appendix \ref{appasymptotic} subsection \ref{subsec:MPSansatz} for the details).

With the MPS in the form (\ref{DMRGAnN}) $H(\bar{\bm{b}}, \bm{b})$ is minimized with respect to $b_{m}$ by finding the smallest eigenvalue $\mathcal{E}_0$ and the corresponding eigenvector of the matrix $H(m)$ with components
\be [H(m)]_{(s_1,p_1,\alpha_1,\beta_1);(s_2,p_2,\alpha_2,\beta_2)} = \frac{\partial}{\partial [\bar{b}_{s_1,p_1}(m)]_{\alpha_1\beta_1}}\frac{\partial}{\partial [b_{s_2,p_2}(m)]_{\alpha_2\beta_2}}\bra{\Phi(\bar{\bm{B}}[\bar{\bm{b}}])} H \ket{\Phi(\bm{B}[\bm{b}])}, \ee
where $s_k \in \{-1,1\}, p_k \in \mathbb{Z}[p_{min}(m+1), p_{max}(m+1)], \alpha_k = 1\ldots D_{p_k - (s_k + (-1)^k)/2}(m),\beta_k = 1\ldots D_{p_k}(m+1)$ $(k = 1,2)$. Because we are only interested in the smallest eigenvalue $\mathcal{E}_0$ and its eigenvector we can use the Lanczos iteration \cite{Lanczos}. For this we only need the action of $H(m)$ on $\bm{b}(m)$. Exploiting the gauge-invariant structure of the tensors $R,L$ and $B$, see (\ref{AnGI}) and (\ref{BnGI}), the computation time of every sweep scales as $$\mathcal{O}\left(\sum_{n=r_L-1}^{r_R + 1} \sum_{q = p_{min}(n)}^{p_{max}(n)} (D_q(n))^3\right) \sim \mathcal{O}\left((r_R - r_L+2) \max_n(p_{max}(n) - p_{min}(n)) \left[\max_{n,q}(D_q(n))\right]^3\right).$$

\noindent We conclude this appendix by discussing how to fix $p_{min/max}(n)$ and $D_q(n)$. As choosing finite values for these quantities means an effective truncation in the Schmidt decomposition (\ref{eq:MPSschmidtGaugeAp}) we need to look at the weight of the Schmidt values $\lambda_{q,\alpha_q}(n)$ over the sectors $q$ corresponding to the eigenvalues of $L(n)$ for any $n$ with $r_L \leq n \leq r_R$. Assuming that the ground-state approximations for $n < r_L$ and $n > r_R$ are accurate (see appendix \ref{appasymptotic} subsection \ref{subsec:MPSansatz}), we don't have to care about the Schmidt values (\ref{eq:SchmidtValueBoundary}) at the boundary.   

In practice we start with a certain distribution of $D_q$-values for each $n$, anticipating that the dominant eigenvalue sector of $L(n)$ would shift from $q=0$ at large $n$ to $q\approx Q$ at the center. After a first full DMRG-optimization, the initial $D_q$ values are updated: increased in case that the minimal retained Schmidt value in the particular eigenvalue sector is larger than $\lambda_{min}=10^{-18}$, decreased in case that the minimal retained Schmidt value is smaller. This is repeated a few times until all retained minimal Schmidt values are smaller or of the same order as $\lambda_{min}$. As for the choice of $r_L$ and $r_R$, we verify a posteriori that the inhomogeneous interval of the MPS (\ref{DMRGAnN}) is taken to be large enough, by verifying the convergence of local observables at large distances to their value for the homogeneous ground state.

\begin{figure}
\begin{subfigure}[b]{.48\textwidth}
\includegraphics[width=\textwidth]{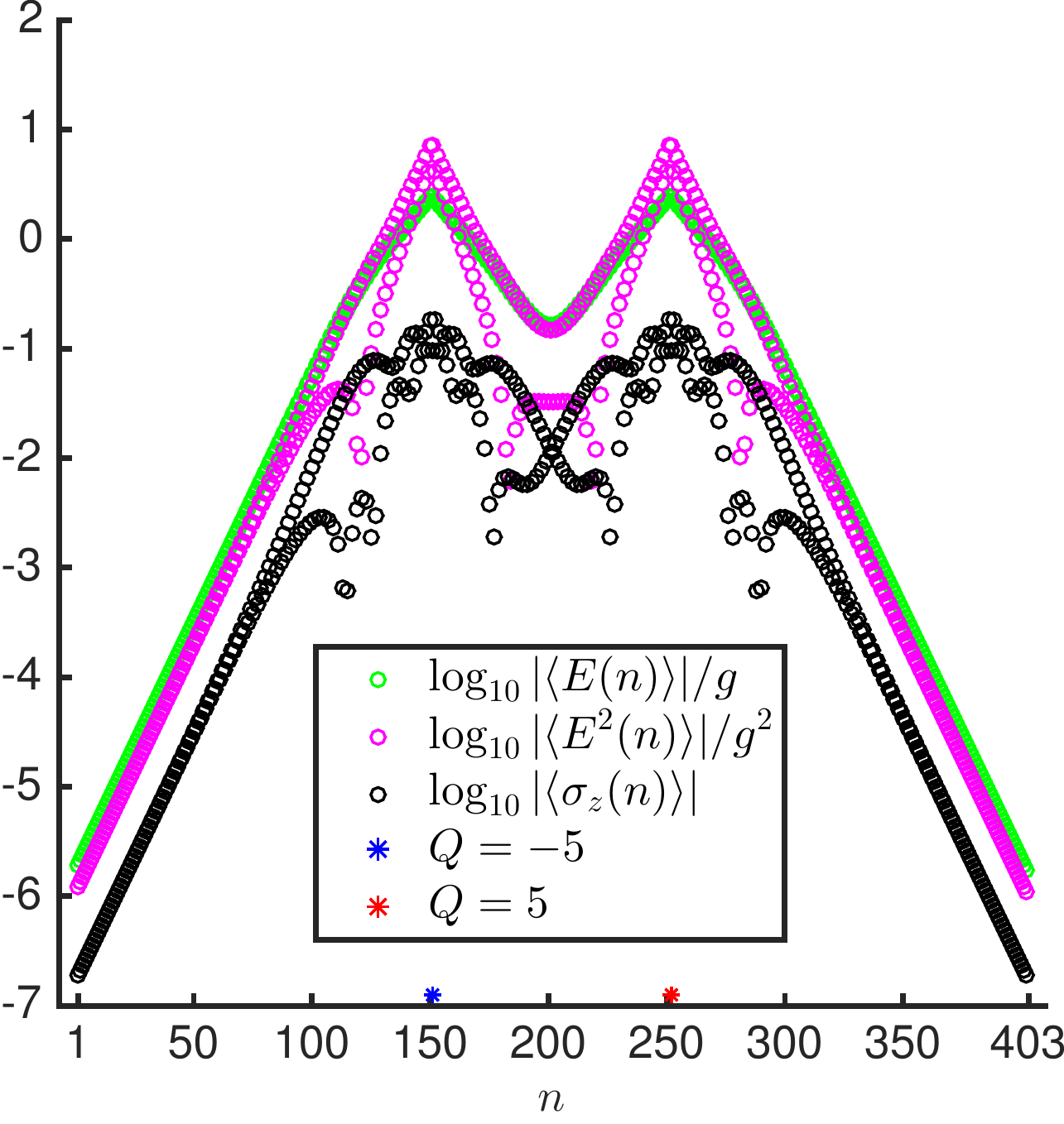}
\caption{\label{fig:maxChargea}}
\end{subfigure}\hfill
\begin{subfigure}[b]{.48\textwidth}
\includegraphics[width=\textwidth]{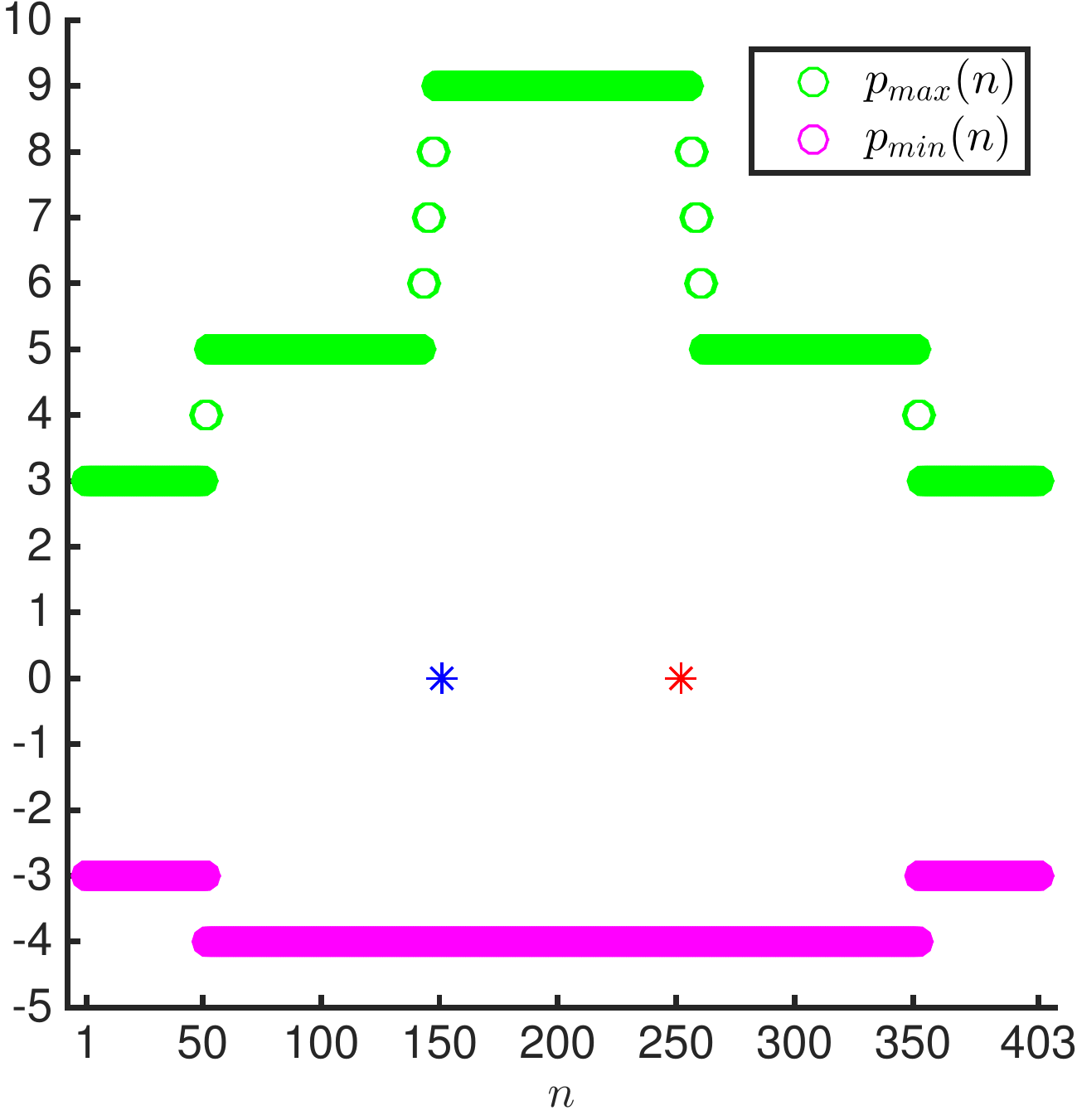}
\caption{\label{fig:maxChargeb}}
\end{subfigure}\vskip\baselineskip
\begin{subfigure}[b]{.48\textwidth}
\includegraphics[width=\textwidth]{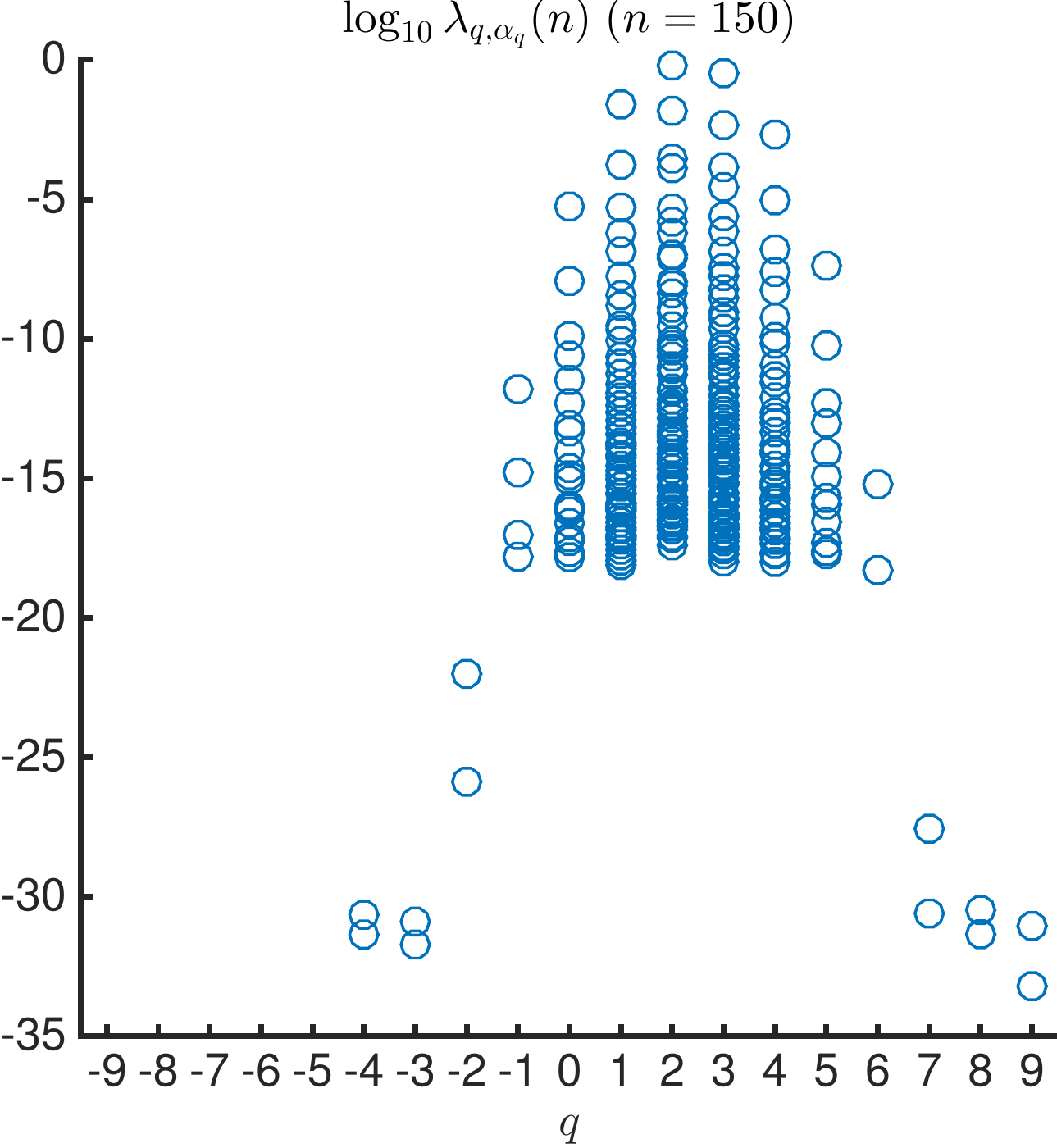}
\caption{\label{fig:maxChargec}}
\end{subfigure}\hfill
\begin{subfigure}[b]{.48\textwidth}
\includegraphics[width=\textwidth]{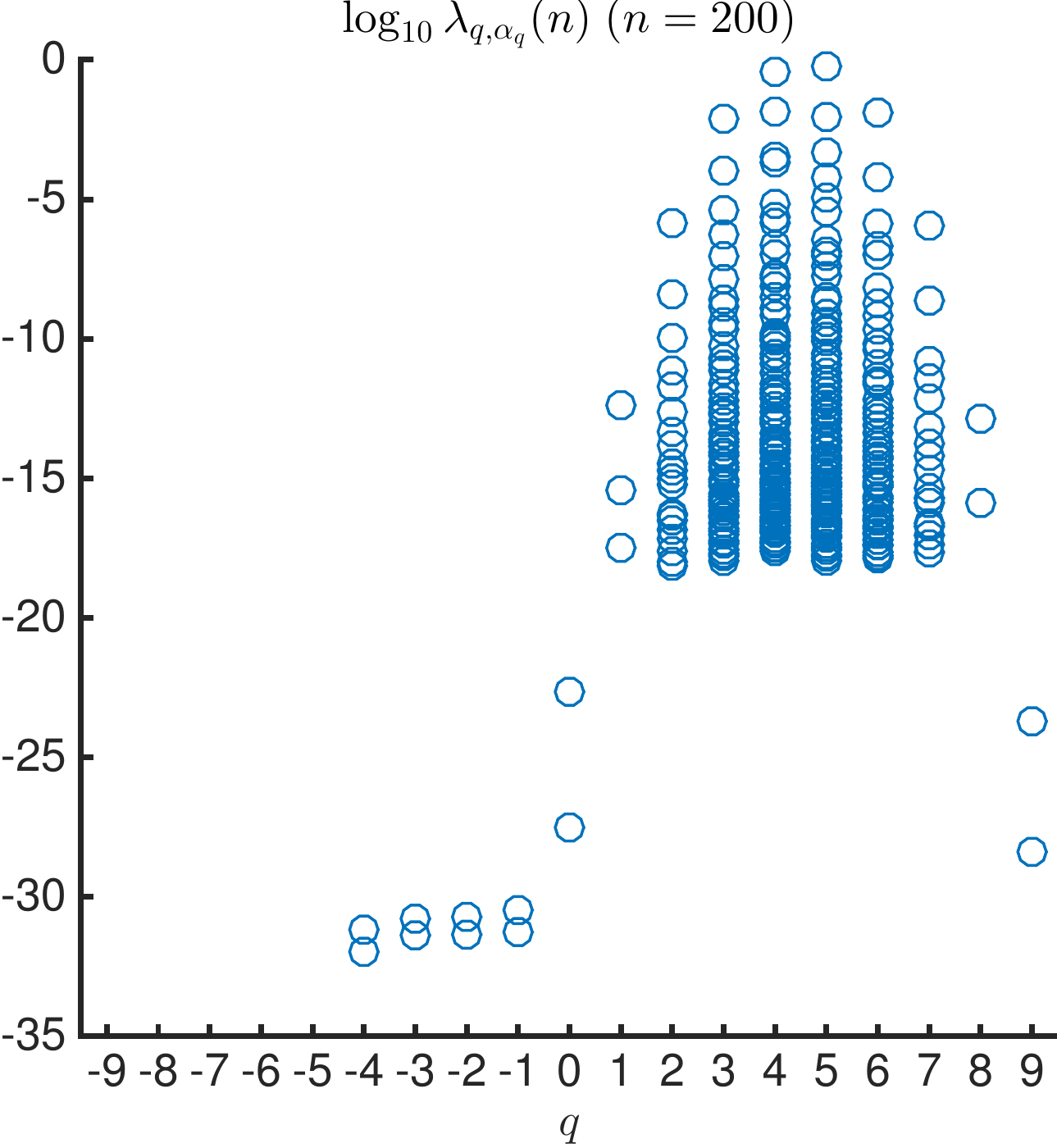}
\caption{\label{fig:maxCharged}}
\end{subfigure}\vskip\baselineskip
\captionsetup{justification=raggedright}
\caption{\label{fig:maxCharge} $m/g = 0.25, x = 100, Q = 5, Lg = 10.1.$ The stars represent the external charges. Between them the electric background field $-Q = -5$ is applied. (a): 10-base logarithm of the expectation values of some local quantities with the Schwinger vacuum-value subtracted. At the boundaries one observes that they are sufficiently small indicating that we took the nonuniform range wide enough. (b): maximum and minimum eigenvalues $p_{max}(n)$ and $p_{min}(n)$ of $L(n-1)$ we took into account in our numerical scheme on every site. (c): Distribution of the $10-$base logarithm of the Schmidt values $\lambda_{q,\alpha_q}(n)$ among the eigenvalue sectors $q$ of $L(n)$ for $n = 150$. (d): Distribution of the of the $10-$base logarithm of the Schmidt values $\lambda_{q,\alpha_q}(n)$ among the eigenvalue sectors $q$ of $L(n)$ for $n = 200$. }
\end{figure}

\begin{figure}
\begin{subfigure}[b]{.48\textwidth}
\includegraphics[width=\textwidth]{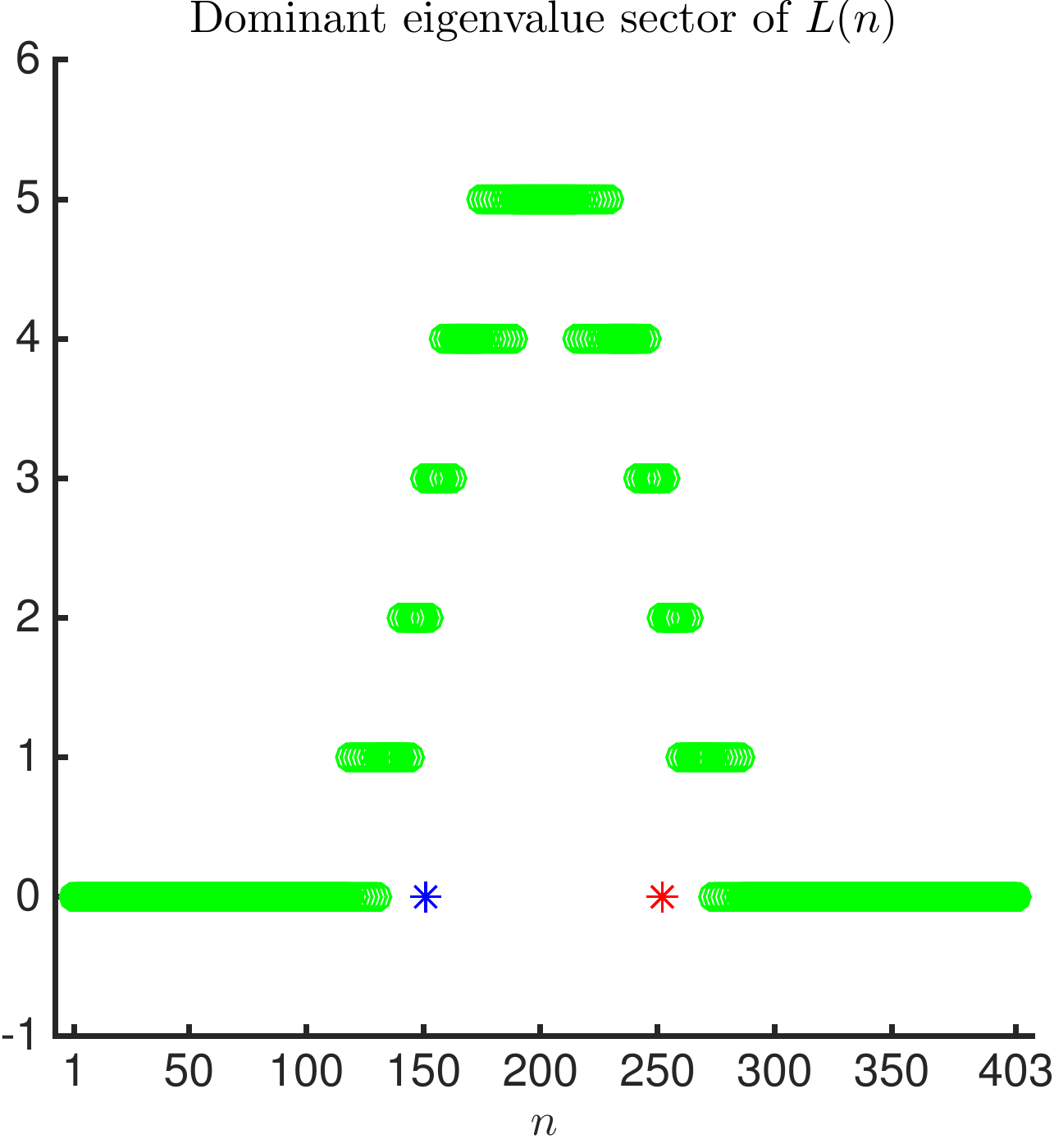}
\caption{\label{fig:maxBDa}}
\end{subfigure}\hfill
\begin{subfigure}[b]{.48\textwidth}
\includegraphics[width=\textwidth]{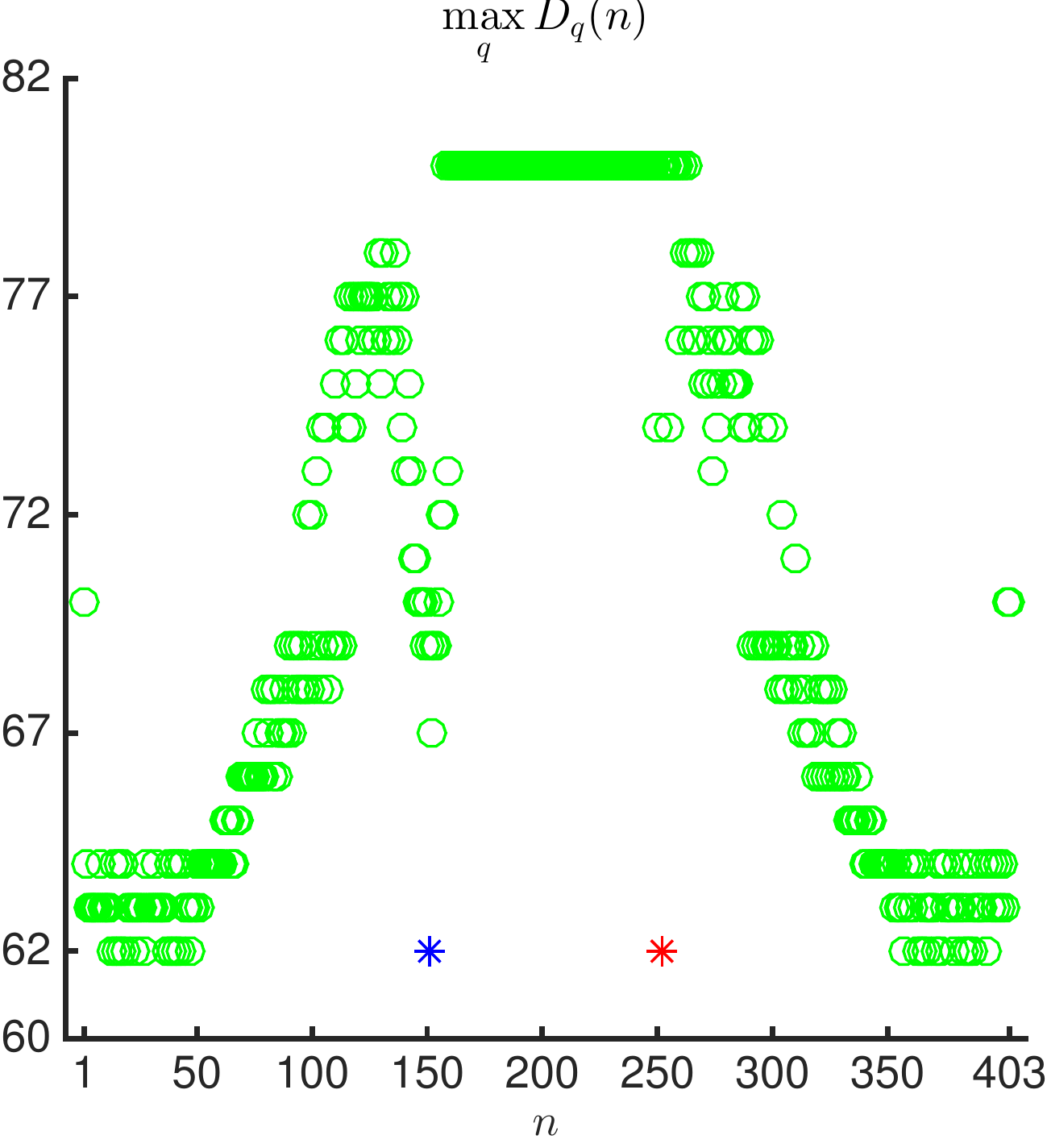}
\caption{\label{fig:maxBDb}}
\end{subfigure}\vskip\baselineskip
\captionsetup{justification=raggedright}
\caption{\label{fig:maxBD}$m/g = 0.25, x = 100, Q = 5, Lg = 10.1.$ The stars represent the external charges. Between them the electric background field $-Q = -5$ is applied.(a): Dominant eigenvalue sector of $L(n)$, i.e. eigenvalue $q$ of $L(n)$ with largest $\sum_{\alpha_q = 1}^{D_q(n)}\lambda_{q,\alpha_q}(n)$. $D_q(n)$ is taken such that smallest Schmidt value is around $10^{-18}$. (b): $\max_q D_q(n):$ Largest bond dimension among the eigenvalue sectors of $L(n)$ at every site $n$. }
\end{figure}

Let us give a specific example. In fig. \ref{fig:maxCharge} and fig. \ref{fig:maxBD} we show some details of the simulation of the ground state for $m/g = 0.25, Q = 5, x = 100, Lg = 10.1$. In our setup with lattice spacing $1/g^2\sqrt{x} = 0.1/g$ this corresponds to a distance of 101 sites between the external antiquark with charge $-gQ$  and the external quark with charge $gQ$. Specifically, we put the antiquark at site 151 and the quark at site $252$. And we reserve $150$ sites on the left of the antiquark and $150$ sites on the right for the nonuniform part of our MPS ansatz. In total we thus have $151 + 101 + 150 = 402$ tensors $B_n$ that need to be optimized. By looking at the 10-base logarithm of the expectation value of some local quantities with respect to the Schwinger vacuum, see fig. \ref{fig:maxChargea}, we observe that we take the range of the nonuniform part large enough: the errors by taking a finite range for the nonuniform part are of order $10^{-6}$.\\
In fig. \ref{fig:maxChargeb} we show the distribution of the minimum charge $p_{min}(n)$ and maximum charge $p_{max}(n)$ we used. For $q< p_{min}(n)$ and $q > p_{max}(n)$ we thus put $D_q(n) = 0$. The $p_{min}(n)$ and $p_{max}(n)$ we took at the boundaries, i.e. $n \gtrsim 1$ and $n \lesssim 402$ correspond to the $p_{min}$ and $p_{max}$ of the Schwinger vacuum, i.e. the vacuum without external charges, that we simulated in \cite{Buyens}. Between the boundaries and the external charges we anticipate the increasing electric field and raised $p_{max}(n)$ to $4 + Q = 9$ anticipating the dominant eigenvalue sector $p_0\approx Q$ at the center.

In figs. \ref{fig:maxChargec} and \ref{fig:maxCharged} we plot the distribution of the Schmidt values among the eigenvalues sector $q$ of $L(n)$ at the sites $n = 150$ (c) and $n = 200$ (d). As we explained above, we adapted the bond dimensions such that for each site $n$ and at each eigenvalue sector $p$ of $L(n)$: $\min_{\alpha_q} \lambda_{q,\alpha_q}(n)  \lesssim 10^{-18}$. Comparing with figs. \ref{fig:STBDa} and \ref{fig:STBDb} we observe that the dominant eigenvalue sector is shifted to $q = 2$ for $n = 150$ and to $q = 5$ for $n = 200$. One can also see that our $p_{min}(n)$ and $p_{max}(n)$ are not entirely optimal: for certain charge sectors the largest Schmidt-value is still well below $10^{-18}$, and these sectors could have been discarded altogether.  As we can see by looking at fig. \ref{fig:maxBDa} the most dominant eigenvalue sector of $L(n)$, i.e. the eigenvalue sector $q$ with the largest value for $\sum_{\alpha_q = 1}^{D_q(n)}\lambda_{q,\alpha_q}(n)$ shifts from $q = 0$ to $q = 5$ as we go from the left boundary to the middle and then decreases to $q = 0$ as we go to the right boundary. \\
We also show the maximum bond dimension $\max_{q}D_q(n)$ in fig. \ref{fig:maxBDb}. The largest bond dimension is required in the region where the electric background field is applied.

\section{The string state and broken-string state for $Q=1$ in the weak-coupling limit}\label{apppert}
\noindent As we mention in the introduction, a major difference between QCD and QED$_2$ is that the latter theory is already confining at the perturbative level, as the Coulomb potential is linear in 1+1 dimensions.  This also allows us to understand the transition from the confining state to the broken-string two-meson state in the weak-coupling limit $m/g\gg 1$. In this nonrelativistic limit one can obtain the ground state by diagonalizing the Hamiltonian in subspaces of the different (fermion) particle sectors. The zero-particle sector simply consists of the Fock vacuum of the free Dirac field and corresponds to the confining string state with an energy \be \mathcal{E}_{string}=g^2 L/2 \label{Estring}\,,\ee for probe charge $Q=1$ and separation length $L$. The broken-string state will correspond to the ground state in the subspace of all states containing one (light) quark-antiquark pair. For this state the light antiquark will bind to the external probe quark and vice versa. We can make this
  more quantitative, by considering the effective Hamiltonian in the nonrelativistic limit for this particle sector: \be H_{q\bar{q}}= 2m -\frac{\nabla_A^2}{2m}-\frac{\nabla_B^2}{2m}+\frac{g^2}{2}|x_A+L/2|+\frac{g^2}{2}|x_B-L/2|+\frac{g^2}{2}|x_A-x_B|-\frac{g^2}{2}|x_A-L/2|-\frac{g^2}{2}|x_B+L/2|+\frac{g^2}{2}L\,.\ee
Here $x_A$ and $x_B$ are the coordinates for the light antiquark and quark, and we put the probe quark at $x=-L/2$ and the probe antiquark at $x=L/2$. Anticipating binding of the light fermions to the probe charges for large $L$, we can assume $x_A<x_B$, $x_A< L/2$ and $x_B>-L/2$ leading to a cancellation of the last four potential terms $H_{q\bar{q}}\approx H_A+H_B$ with  \bea H_{A}&=&m-\frac{\nabla_A^2}{2m}+\frac{g^2}{2}|x_A+L/2|\,,\nonumber\\
 H_{B}&=&m-\frac{\nabla_B^2}{2m}+\frac{g^2}{2}|x_B-L/2|\,.\label{Schrodinger}\eea
A ground-state solution will therefore be of the form $\Psi(x_A,x_B)=\phi_A(x_A)\phi_B(x_B)\,$ where now $\phi_A(x_A)$ and $\phi_B(x_B)$ are both ground states of the nonrelativistic one-particle problem for a linear potential. All eigenstates for this nonrelativistic Hamiltonian $H_A$ (and similar for $H_B$) can be written in terms of the so-called Airy function $Ai$ \cite{Airy}:
\be \phi^{(n)}_A(x_A)=\mathcal{N}Ai\left((g^2m)^{1/3}  |x_A+L/2|-2\mathcal{E}_n \frac{m^{1/3}}{g^{4/3}}\right)\,,\ee
where $\mathcal{N}$ is the normalization factor and $\mathcal{E}_n$ is the (kinetic) eigenenergy of the eigenstate. These energies follow from the continuity requirement on $\phi_A$ and $\phi_A'$ at $x_A=-L/2$, leading to either even or odd $\phi_A$ under $x_A+L/2\rightarrow -(x_A+L/2)$. The ground-state wave-function is even and the ground-state energy $\mathcal{E}_0$ is related to the first zero of the first derivative of the Airy function, $Ai'(x_1)=0$, at $x_1\approx-1.0188$: $\mathcal{E}_0=-\frac{x_1}{2}\frac{g^{4/3}}{m^{1/3}}$. So in the nonrelativistic approximation we find: \be \mathcal{E}_{2meson}=2m + 1.0188 \frac{g^{4/3}}{m^{1/3}}\,.\label{} \ee Notice that relativistic corrections to this approximation will necessarily involve quantum field contributions from other particle sectors. The relativistic one-particle Dirac equation has no bound state solutions for a linear (vector) potential \cite{Galic:1987rp,Capri:1986fs}. 

Finally, in the nonrelativistic approximation we can then understand the transition from the string state to the broken-string state as a level crossing at the critical length $L$, where $\mathcal{E}_{string}=\mathcal{E}_{2meson}$.

\end{document}